\documentclass[11pt, a4paper]{article}

\usepackage{setspace}
\usepackage{amssymb}
\usepackage[polutonikogreek, english]{babel}
\usepackage{cite}
\usepackage{amsmath}
\usepackage{dsfont}
\usepackage{amsfonts}
\usepackage{epsfig}
\usepackage{latexsym}
\usepackage{slashed}
\usepackage{xcolor}
\usepackage{cancel}
\usepackage{verbatim}
\usepackage{nicefrac}
\usepackage{amsbsy}
\usepackage{graphicx}
\usepackage[heightrounded]{geometry}
\usepackage{comment}

\allowdisplaybreaks[1]

\def\nb{n_B}
\def\nbb{n_{\bar{B}}}
\def\ng{n_\gamma}

\newcommand{\rmeq}{{\rm eq}}
\newcommand{\barphi}{\bar{\phi}}
\newcommand{\barell}{\bar{\ell}}
\newcommand{\ml}{\mathcal{L}}

\newcommand{\bh}[1]{\hat{\bf #1}}

\newcommand{\rmd}{{\rm d}}
\newcommand{\rmi}{{\rm i}}
\newcommand{\rme}{{\rm e}}
\newcommand{\tr}{{\rm tr}}

\newcommand{\CP}{{C\!P}}
\newcommand{\hc}{{\rm h.c.}}
\newcommand{\gb}{\boldsymbol{\gamma}}

\newcommand{\D}{\Delta}

\def\openone{\leavevmode\hbox{\small1\kern-3.8pt\normalsize1}}
\def\a{\alpha}

\def\c{\chi}
\def\d{\delta}
\def\e{\epsilon}

\def\g{\gamma}
\def\h{\eta}

\def\j{\psi}

\def\l{\lambda}

\def\o{\omega}
\def\p{\pi}
\def\q{\theta}

\def\D{\Delta}

\def\G{\Gamma}

\def\P{\Pi}

\def\S{\Sigma}

\def\cl{{\cal L}}

\def\bo{{\raise-.3ex\hbox{\large$\Box$}}}               % D'Alembertian
\def\face{{\raise.2ex\hbox{$\displaystyle \bigodot$}\mskip-2.2mu \llap {$\ddot
        \smile$}}}                                      % happy face
\def\leftrightarrowfill{$\mathsurround=0pt \mathord\leftarrow \mkern-6mu
        \cleaders\hbox{$\mkern-2mu \mathord- \mkern-2mu$}\hfill
        \mkern-6mu \mathord\rightarrow$}       % <--> double differential
\def\dvec#1{\vbox{\ialign{##\crcr
        \leftrightarrowfill\crcr\noalign{\kern-1pt\nointerlineskip}
        $\hfil\displaystyle{#1}\hfil$\crcr}}}           % <--> accent
\def\beq{\begin{equation}}
\def\eeq{\end{equation}}

\def\beqx{\begin{displaymath}}
\def\eeqx{\end{displaymath}}

\def\bea{\begin{eqnarray}}
\def\eea{\end{eqnarray}}

\def\bs{\begin{split}}
\def\ensp{\end{split}}

\def\ba{\begin{align}}
\def\ea{\end{align}}

\def\bef{\begin{figure}}
\def\eef{\end{figure}}

\def\bec{\begin{center}}
\def\eec{\end{center}}

\begin{document}

\begin{flushright}
MPP-2011-62\\[7mm]
\end{flushright}

\begin{center}
{\bf \Large Hard-Thermal-Loop Corrections in Leptogenesis II:\\ Solving the Boltzmann Equations}\\[4mm]

%\today\\[4mm]

Clemens P.~Kie\ss ig%$^\star$
\footnote{E-mail: \texttt{ckiessig@mpp.mpg.de}},
%Alois S.~Kabelschacht
%\footnote{E-mail: \texttt{kabel@mpp.mpg.de}} ,
Michael Pl\"umacher%$^\star$
\footnote{E-mail: \texttt{pluemi@mpp.mpg.de}}
%Markus H.~Thoma$^\dagger$
%\footnote{E-mail: \texttt{mthoma@mpe.mpg.de}}, 

\vspace*{0.5cm}
%$^\star$ 
\it
Max-Planck-Institut f\"{u}r Physik (Werner-Heisenberg-Institut),\\
F\"ohringer Ring 6, D-80805 M\"unchen, Germany\\[0.1cm]
%$^\dagger$ \it Max-Planck-Institut f\"ur extraterrestrische Physik,\\
%Giessenbachstra\ss e, D-85748 Garching, Germany

\vspace*{0.4cm}
\end{center}

\begin{abstract}

  We investigate hard-thermal-loop (HTL) corrections to the final
  lepton asymmetry in leptogenesis. To this end we solve the Boltzmann
  equations with HTL-corrected rates and $C\!P$ asymmetries, which we
  calculated in paper I of this series. We pay special attention to
  the influence of the two leptonic quasiparticles that arise at
  non-zero temperature. We include only decays and inverse decays and
  allow for the lepton modes to be either decoupled from each other,
  or to be in chemical equilibrium by some strong interaction,
  simulating the interaction with gauge bosons. In two additional
  cases, we approximate the full HTL lepton propagators with
  zero-temperature propagators, where we replace the zero-temperature
  mass by the thermal mass of the leptons $m_\ell(T)$ or the
  asymptotic mass $\sqrt{2} \, m_\ell(T)$. We compare the final lepton
  asymmetries of the four thermal cases and the zero-temperature case
  for zero, thermal and dominant initial neutrino abundance. The final
  lepton asymmetries of the thermal cases differ considerably from the
  vacuum case and from each other in the weak washout regime for zero
  initial neutrino abundance and in the intermediate regime for
  dominant initial neutrino abundance. In the strong washout regime,
  the final lepton asymmetry can be enhanced by a factor of two in the
  case of strongly interacting lepton modes.

\end{abstract}

\section{Introduction}
\label{sec:introduction}

The question of the origin of all things was always essential to
mankind and has driven them to search for answers in science, among
others. Physics as the science of nature 
%(\greektext <h fusik'h
%\latintext ``nature'') 
and within physics, cosmology 
%(\greektext <o
%k`osmoc \latintext ``order'') 
as the science of the order and the
evolution of the universe, address this question and have their own
formulation of it. What is the origin of the matter that is the
building block of all things we observe, including ourselves?  

The matter in nature consists of leptons
%(\greektext lept'os
%\latintext ``small''),
and the much heavier baryons, 
%(\greektext bar'uc
%\latintext ``heavy'') 
which are in turn made up of quarks. According
to the standard model of particle physics (SM), matter particles,
quarks or leptons, can only be created in pairs together with their
antiparticles, that is, antiquarks and antileptons, at least in
perturbation theory. If we assume that the early universe was indeed
without form and void
%\footnote{We know that the early universe was by
%  no means without form and void, but rather a vibrant soup of all
%  particles of the SM and your favourite extensions thereof,
%  interacting rapidly with each other.}~\cite{KingJames:1611aa}
, that
is in the language of particle physics, there was no excess of one
particle species over the other, there would have to be an equal
amount of particles and antiparticles today. More specifically, since
annihilation of particles and antiparticles proceeds at fast rates, no
structures like atoms, molecules, galaxies, stars, planets, DNA, cells
and finally living organisms could have formed and we would
observe
%\footnote{Or rather, not observe.}
 a universe populated almost
exclusively by photons and the slowly interacting neutrinos. This
scenario is obviously not realised. If we believe in inflation, we
cannot assume a sizeable matter-antimatter asymmetry as an initial
condition of the universe because that asymmetry would be diluted by
inflation, not even mentioning the highly unsatisfactory character of
such an approach from a scientific point of view. Therefore we have to
employ a baryogenesis theory, a mechanism that creates a baryon
asymmetry dynamically and explains the value of
\begin{align}
  \label{eq:12}
  \h \equiv \left. \frac{\nb - \nbb}{\ng} \right |_0 =(6.16 \pm 0.16)
  \times 10^{-10} \, .
\end{align}
This value has been inferred from the WMAP seven-year cosmic microwave
background (CMB) anisotropy data~\cite{Komatsu:2010fb}, where $\nb$,
$\nbb$, and $\ng$ are the number densities of baryons, antibaryons,
and photons, respectively, and the subscript 0 implies present cosmic
time. The value agrees with the abundance of light elements inferred
from big bang nucleosynthesis.

Leptogenesis~\cite{Fukugita:1986hr} is a very attractive baryogenesis
theory, since it simultaneously explains the creation of the baryon
asymmetry and the smallness of neutrino masses via the seesaw
mechanism~\cite{Minkowski:1977sc,Yanagida:1979as,GellMann:1980vs,
  Mohapatra:1980yp,Schechter:1980gr,Schechter:1981cv}.  We add three
heavy right-handed neutrinos $N_i$ to the SM, which are assumed to
have rather large Majorana masses $M_i$, close to the scale of some
possibly underlying grand unified theory (GUT)~\cite{Ross:1985ai
%,Antusch:2009gu
}, $E_{\rm GUT} \sim 10^{15 \ldots 16} \, {\rm GeV}$. The interaction
with the SM neutrinos suppresses their mass when we integrate out the
heavy neutrinos. In the early universe, the heavy neutrinos decay into
leptons and Higgs bosons and create a lepton asymmetry, which is
lateron converted to a baryon asymmetry by the anomalous sphaleron
processes~\cite{Klinkhamer:1984di,Kuzmin:1985mm}.  The three Sakharov
conditions~\cite{Sakharov:1967dj} that are necessary for a
baryogenesis theory are fulfilled, that is lepton number $L$ and $B-L$
are violated, $C\!P$ symmetry is violated in the decays and the decays
can be out of equilibrium.

Ever since the development of the theory 25 years ago, the
calculations of leptogenesis dynamics have become more refined and
many effects and scenarios that have initially been neglected have
been considered\footnote{For an excellent review of the development in
  this field, we refer to reference\cite{Davidson:2008bu}.}. Notably
the question how the hot and dense medium of SM particles influences
leptogenesis dynamics has received increasing attention over the last
years~\cite{Covi:1997dr,Giudice:2003jh,Anisimov:2010dk,Anisimov:2010gy,
  Garny:2010nz,Beneke:2010dz,Beneke:2010wd,Garbrecht:2010sz}. At high
temperature, particles show a different behaviour than in vacuum due
to their interaction with the medium: they acquire thermal masses,
modified dispersion relations and modified helicity properties. All
these properties can be summed up by viewing the particles as thermal
quasiparticles with different behaviour than their zero-temperature
counterparts, much like the large zoo of single-particle and
collective excitations that are known in high density situations in
solid-state physics. At high temperature, notably fermions can in the
hard-thermal-loop-limit (HTL) occur in two distinct states with a
positive or negative ratio of helicity over chirality and different
dispersion relations than at zero temperature, where these dispersion
relations do not break the chiral symmetry as a zero-temperature mass
does.

Thermal effects have been considered by
references~\cite{Covi:1997dr,Giudice:2003jh,Anisimov:2010dk,Anisimov:2010gy,
  Garny:2010nz,Beneke:2010dz,Beneke:2010wd,Garbrecht:2010sz}. Notably
reference~\cite{Giudice:2003jh} performs an extensive analysis of the
effects of thermal masses that arise by resumming propagators using
the HTL resummation within thermal field theory (TFT). However, the
authors approximated the two fermionic helicity modes with one
simplified mode that behaves like a vacuum particle with its
zero-temperature mass replaced by a thermal mass\footnote{Moreover, an
  incorrect thermal factor for the $\CP$-asymmetry was obtained, as
  has been pointed out in reference~\cite{Garny:2010nj}.}. Due to
their chiral nature, there are serious consequences to assigning a
chirality breaking mass to fermions, hence the effects of abandoning
this property should be examined. Moreover, it seems questionable to
completely neglect the negative-helicity fermionic state which,
according to TFT, will be populated at high temperature. We argue in
this study that one should include the effect of the fermionic
quasiparticles in leptogenesis calculations and possibly in other
early universe dynamics, since they behave differently from
zero-temperature states with thermal masses, both conceptually and
regarding their numerical influence on the final lepton asymmetry. We
do this by analysing the dynamics of a leptogenesis toy model that
includes only decays and inverse decays of neutrinos and Higgs bosons,
but takes into account all HTL corrections to the leptons and Higgs
bosons, paying special attention to the two fermionic
quasiparticles. In a slightly different scenario, we assume chemical
equilibrium among the two leptonic modes, thereby simulating a
scenario where the modes interact very fast. As a comparison, we
calculate the dynamics for two models where we approximate the lepton
modes with ordinary zero-temperature states and modified masses, the
thermal mass $m_\ell(T)$ and the asymptotic mass of the
positive-helicity mode, $\sqrt{2} \, m_\ell(T)$.

This paper is the second part of a two-paper series, where we have
calculated the HTL-corrections to $C\! P$-asymmetries in the first
part~\cite{Kiessig:2011fw}. The topic of this work is solving the
Boltzmann equations with HTL-corrected rates and
$C\!P$-asymmetries. It is structured as follows: In
section~\ref{sec:prop-at-finite-t}, we briefly review the imaginary
time formalism of thermal field theory (TFT) and discuss the hard
thermal loop (HTL) resummation. In
section~\ref{sec:hard-thermal-loop}, we summarise and present our
previous calculations of interaction rates and $C\!P$-asymmetries in
references\cite{Kiessig:2010pr},~\cite{Kiessig:2010zz}
and~\cite{Kiessig:2011fw}. Section~\ref{sec:boltzmann-equations} deals
with the evaluation of the Boltzmann equations. We derive the
equations and compare our four thermal scenarios, wich are the
decoupled and strongly coupled two-mode approach and the one-mode
approach with thermal and asymptotic mass, to the vacuum case. We show
the evolution of the abundances for three different initial conditions
for the neutrinos, that is, zero, thermal and dominant abundance. We
explain the dynamics of the different cases in detail and find
considerable differences both of the thermal approaches to the vacuum
case and of the two-mode cases to the one-mode cases. We summarise the
main insights of this work in the conclusions and give an outlook on
future work and prospects. In appendix~\ref{sec:boltzm-equat-at}, we
derive Boltzmann equations at zero temperature, while in
appendix~\ref{sec:subtr-shell-prop}, we explicitly perform the
subtraction of on-shell propagators for our cases.

\section{Propagators at Finite Temperature}
\label{sec:prop-at-finite-t}

When going to finite temperature~\cite{LeBellac:1996}, one has to
employ ensemble weighted expectation values of operators rather than
the vacuum expectation values, so for an operator $\hat{A}$ we get
\begin{align}
  \label{eq:1}
  \langle 0 | \hat{A} | 0 \rangle \rightarrow \langle \hat{A} \rangle_\rho \equiv {\rm tr} (\rho \hat{A}) \, .
\end{align}
There are two formalisms for calculating Green's functions at finite
temperature, the imaginary time formalism and the real time
formalism. Both are equivalent and we employ the imaginary time
formalism, where the $k_0$-integration is replaced by a sum over
discrete energies, the so-called Matsubara frequencies.

\begin{figure}
  \centering
  \includegraphics[scale=0.7]{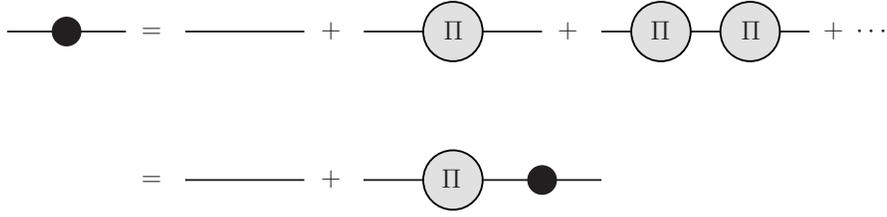}
  \caption{The resummed scalar propagator.}
  \label{fig:resummedpi}
\end{figure}

Naive perturbation theory at finite temperature can lead to serious
conceptual problems, such as infrared
divergent~\cite{Braaten:1991jj,Braaten:1991we} and gauge
dependent~\cite{Kalashnikov:1979cy, Gross:1980br} results and results
that are not complete to leading order. In order to cure these
shortcomings, the hard thermal loop (HTL) resummation technique has
been invented~\cite{Braaten:1989mz, Braaten:1990az}. One distinguishes
between hard momenta of order $T$ and soft momenta of order $gT$,
where $g$ is the coupling constant of the corresponding theory. In a
strict sense, this is only possible in the weak coupling limit where
$g \ll 1$. If all external momenta are soft, then the bare thermal
propagators have to be replaced by resummed propagators. The
self-energies that are resummed are the HTL self-energies, for which
all internal momenta are hard. For a scalar field with a
HTL-self-energy $\Pi$, the resummed effective HTL-propagator $\D^*$
follows from the Dyson-Schwinger equation in
figure~\ref{fig:resummedpi} as
\begin{align}
  \label{eq:2}
    \rmi \, \D^* & = \rmi \,\D + \rmi \, \D \left( - \rmi \, \P \right) \rmi \, \D +
  \dots \nonumber \\
  & = \frac{\rmi}{\D^{-1}-\P} = \frac{\rmi}{K^2-m_0^2-\P} \, ,
\end{align}
where $\D$ is the bare propagator, $K$ the momentum and $m_0$ the
zero-temperature mass of the scalar.
The dispersion relation for this effective excitation is given by the pole of the propagator as
\begin{align}
  \label{eq:3}
  k_0^2 = k^2+m_0^2+\Pi \, ,
\end{align}
so we get an effective mass of $m_{\rm eff}^2 = m_0^2 + m_S^2$ where
the thermal mass of the scalar is given by the self-energy, which is proportional to $gT$, $m_S^2 =
\P \propto (g T)^2$. It is possible to neglect the zero-temperature mass if $m_S \gg
m_0$.

For fermions with negligible zero-temperature mass, the general
expression for the self-energy in the rest frame of the thermal bath
is given by~\cite{Weldon:1982bn}
\begin{align}
  \label{eq:30}
  \Sigma(P)= -a(P) \slashed{P}-b(P) \slashed{u} \, ,
\end{align}
where $u^\alpha=(1,0,0,0)$ is the four-velocity of the heat bath. The
factors $a$ and $b$ are given by
\begin{align}
  \label{eq:31}
  a(P) &= \frac{1}{4 p^2} \left[ \tr\left(\slashed{P} \Sigma\right) -
    p_0 \tr \left( \g_0 \S \right)\right] \, , \\ \nonumber
  b(P) &= \frac{1}{4 p^2} \left[P^2 \tr\left(\g_0 \Sigma\right) -
    p_0 \tr \left( \slashed{P} \S \right) \right] \, .
\end{align}
In the HTL limit, the traces are given by~\cite{LeBellac:1996}
\begin{align}
  \label{eq:32}
  T_1 \equiv \tr \left(\slashed{P} \S\right) &= 4 \, m_F^2 \, ,
  \nonumber \\
  T_2 \equiv \tr\left(\g_0\S\right) &= 2 \, m_F^2 \frac{1}{p} \ln
  \frac{p_0+p+\rmi \, \e}{p_0-p+\rmi \, \e} \, ,
\end{align}
where the effective thermal fermion mass $m_F \propto gT$ depends on the
interaction that gives rise to the fermion self-energy.
%\begin{align}
%  \label{eq:33}
%  m_F^2 =
%\begin{cases}
%  e^2 T^2/8 & \text{for QED} \\
%  g^2 T^2/16 & \text{for a Yukawa interaction } \mathcal{L}_Y = - g
%  \overline{\j} \j \f \, .
%\end{cases}
%\end{align}

The resummed fermion propagator is then written as
\begin{align}
  \label{eq:4}
    S^*(K)=\frac{1}{\slashed{K}-\Sigma_{\rm HTL}(K)}\, .
\end{align}
It is convenient to rewrite this propagator in the helicity-eigenstate
representation~\cite{Braaten:1990wp,Braaten:1992gd},
\begin{align}
  \label{eq:35}
  S^*(K)=\frac{1}{2} \Delta_+(K) (\gamma_0-\hat{\bf k} \cdot
\boldsymbol{\gamma}) +\frac{1}{2} \Delta_-(K) (\gamma_0+\hat{\bf k} \cdot
\boldsymbol{\gamma}),
\end{align}
where $\hat{\bf k}={\bf k}/k$, and
\begin{align}
\label{eq:42}
\Delta_\pm(K)=\left [ -k_0 \pm  k + \frac{m_F^2}{k} \left ( \pm 1 - 
\frac{\pm k_0 - k}{2k} \ln \frac{k_0+k}{k_0-k}  \right ) \right ]^{-1} \, .
\end{align}

\begin{figure}
  \centering
  \includegraphics[scale=0.8]{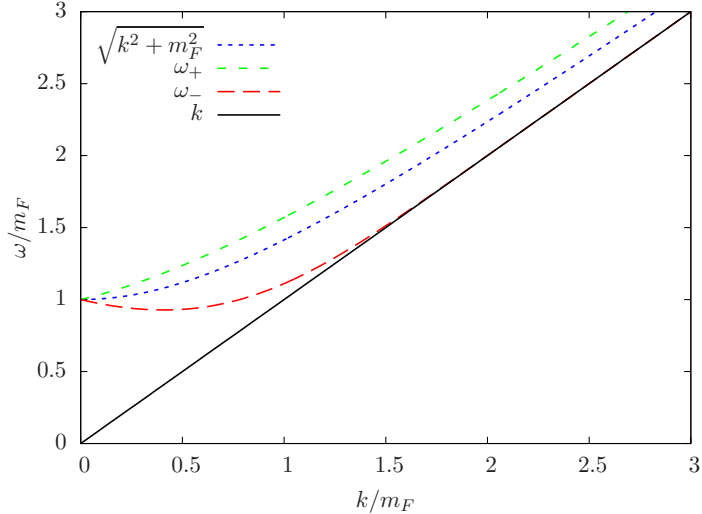}
  \caption[Fermionic dispersion relations.]{The two dispersion laws
    for fermionic excitations compared to the standard dispersion
    relation $\o^2=k^2+m_F^2$.}
  \label{fig:omega}
\end{figure}

This propagator has two poles, the zeros of the two denominators
$\D_\pm$. The poles can be seen as the dispersion relations of
single-particle excitations of the fermions that interact with the hot
plasma,
\begin{align}
  \label{eq:40}
  k_0=\o_\pm(k) \, .
\end{align}
We have presented an analytical expression for the two dispersion
relations making use of the Lambert $W$ function in
reference~\cite{Kiessig:2010pr}. The dispersion relations are shown in
figure~\ref{fig:omega}.

Note that even though the dispersion relations resemble the behaviour
of massive particles and $\o =m_F$ for zero momentum $k$, the
propagator $S^*(K)$ \eqref{eq:35} does not break chiral invariance
like a conventional mass term. Both the self energy $\S(K)$
\eqref{eq:30} and the propagator $S^*(K)$ anticommute with $\g_5$. The
Dirac spinors that are associated with the pole at $k_0=\o_+$ are
eigenstates of the operator $(\g_0-\bh{k} \cdot \gb)$ and they have a
positive ratio of helicity over chirality, $\c=+1$. The spinors
associated with $k_0=\o_-$, on the other hand, are eigenstates of
$(\g_0+\bh{k} \cdot \gb)$ and have a negative helicity-over-chirality
ratio, $\c=-1$. At zero temperature, fermions have $\c=+1$. The
introduction of a thermal bath gives rise to fermionic modes which
have $\c=-1$. These modes have been called plasminos since they are
new fermionic excitations of the plasma and have first been noted in
references~\cite{Weldon:1982bn,Klimov:1981ka}.

% \noindent \bof{Spinor eigenstates of the HTL fermion propagator}
% {\it Optional later}

We can introduce a spectral representation for the two parts of the
fermion propagator \eqref{eq:42} \cite{Pisarski:1989cs},
\begin{align}
  \label{eq:41}
  \D_\pm(K)=\int_{-\infty}^\infty \rmd
   \o \frac{\rho_\pm(\o,k)}{\o-k_0-\rmi \, \e} \, ,
\end{align}
where the spectral density $\rho_\pm(\o,k)$
\cite{Braaten:1990wp, Kapusta:1991qp} has two contributions, one from
the poles,
\begin{align}
  \label{eq:43}
  \rho^{\rm pole}_\pm(\o,k)= Z_\pm(\o,k) \, \d(\o-\o_\pm(k)) +Z_\mp(\o,k) 
  \, \d(\o+\o_\mp(k)) \, ,
\end{align}
and one discontinuous part,
\begin{align}
  \label{eq:44}
  \rho^{\rm disc}_\pm(\o,k)= \frac{\frac{1}{2}\, m_F^2(k\mp \o)}
  {\left\{k(\o\mp k)-m_F^2\left[Q_0(x) \mp Q_1(x)\right]\right\}^2 +
    \left[\frac{1}{2}\, \p\, m_F^2 (1 \mp x) \right]^2} \times
  \q(k^2-\o^2) \, ,
\end{align}
where $x=\o/k$, $\q(x)$ is the heaviside function and $Q_0$ and $Q_1$
are Legendre functions of the second kind,
\begin{align}
  \label{eq:45}
  Q_0(x)=\frac{1}{2} \ln \frac{x+1}{x-1} \, , \hspace{1cm} Q_1(x) = x \,
  Q_0(x) -1 \, .
\end{align}
The residues of the quasi-particle poles are given by
\begin{align}
  \label{eq:46}
  Z_\pm(\o,k)=\frac{\o_\pm^2(k)-k^2}{2 \, m_F^2} \, , \quad {\rm where}
  \quad Z_+ + Z_- = 1 \, .
\end{align}
One can describe the non-standard dispersion relations $\o_\pm$ by
momentum-dependent effective masses $m_\pm(k)$ which are given by
\begin{align}
  \label{eq:47}
  m_\pm(k)=\sqrt{\o_\pm^2(k)-k^2}=\sqrt{2 \, Z(\o,k)} \, m_F \, .
\end{align}
\begin{figure}
  \centering
  \includegraphics[scale=0.8]{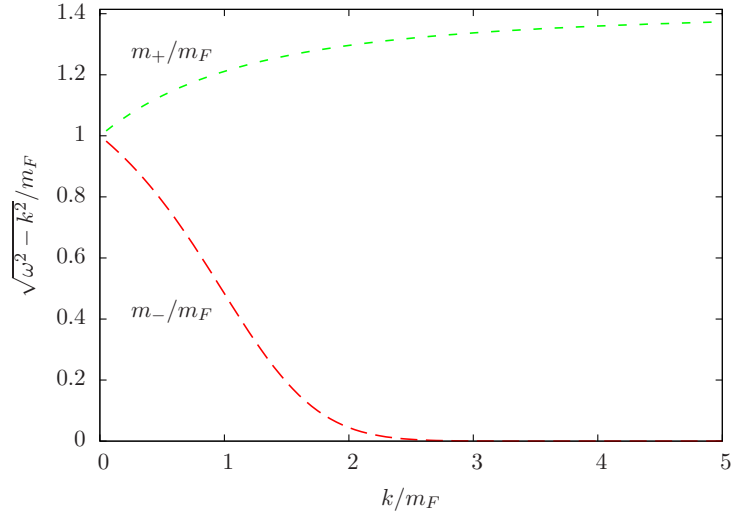}
  \caption{The momentum-dependent effective masses $m_\pm$.}
  \label{fig:mpm}
\end{figure}
These masses are shown in figure~\ref{fig:mpm}.

Considering gauge theories, one might also have to use HTL-corrected
effective vertices that are related to the propagators by Ward
identities~\cite{LeBellac:1996}. We do not consider these vertices
since we are only looking at Yukawa vertices. In the HTL framework, it
is sufficient to use bare propagators if at least one of the external
legs is hard. However, it is always possible to resum self-energies
and thus capture effects which arise from higher-order loop diagrams
and take into account the appearance of thermal masses and modified
dispersion relations in a medium. In fact, since the effective masses
we encounter do typically not satisfy the condition $m_{\rm eff} \ll
T$ but are rather in the range $m_{\rm eff}/T \sim 0.1$ -- $1$, the
effect of resummed propagators is noticeable even when some or all
external momenta are hard. In summary, we always resum the propagators
of particles that are in equilibrium with the thermal bath, which are
the Higgs bosons and the leptons in our case, in order to capture the
effects of thermal masses, modified dispersion relation and modified
helicity structures. This approach is justified a posteriori by the
sizeable corrections it reveals, similar to the treatment of meson
correlation fuctions in reference~\cite{Karsch:2000gi}.

In leptogenesis, the leptons and Higgs bosons acquire thermal masses
that have been calculated in
references~\cite{Weldon:1982bn,Klimov:1981ka,Cline:1993bd,
  Elmfors:1993re} and are given by
\begin{align}
m_\phi^2(T)&=\left (\frac{3}{16} g_2^2+ \frac{1}{16} g_Y^2 +
\frac{1}{4} y_t^2 + \frac{1}{2} \lambda \right) T^2 \, , \nonumber \\
m_\ell^2(T)&=\left (\frac{3}{32} g_2^2+ \frac{1}{32} g_Y^2 \right ) T^2.
\end{align}
The couplings denote the SU(2) coupling $g_2$, the U(1) coupling
$g_Y$, the top Yukawa coupling $y_t$ and the Higgs self-coupling
$\lambda$, where we assume a Higgs mass of about $115$ GeV. The other
Yukawa couplings can be neglected since they are much smaller than
unity and the remaining couplings are renormalised at the first
Matsubara mode, $2 \pi T$, as explained in
reference~\cite{Giudice:2003jh} and in
reference~\cite{Kajantie:1995dw} in more detail. The heavy neutrinos
$N_1$ do acquire a thermal mass, but since the Yukawa couplings are
much smaller than unity, this effective mass can be neglected compared
to the zero-temperature mass.

\section{HTL Corrections to Decays and $\boldsymbol{C \! P}$-Asymmetries}
\label{sec:hard-thermal-loop}

\subsection{Decay and inverse decay rates}
\label{sec:decay-inverse-decay}

The additional terms for the right-handed neutrinos in the Lagrangian are
\begin{align}
\label{eq:15}
\cl={\rm i} \; \bar{N}_i \partial_\mu \g^\mu N_i - \l_{i \alpha}
\bar{N_i} (\phi^a \e_{ab} \ell_\alpha^b)- \frac{1}{2} \sum_i M_i
\bar{N}_i N_i^c + \hc \, ,
\end{align}
where the Higgs doublet $\phi$ is normalised such that its vacuum
expectation value (vev) in
\begin{equation}
  \langle \phi \rangle =
\begin{pmatrix}
0 \\ v
\end{pmatrix}
\end{equation}
is $v \simeq 174 \, {\rm GeV}$ and $\l_{i \a}$ is the Yukawa coupling
connecting the Higgs doublet, the lepton doublet and the heavy
neutrino singlet. The indices $a$ and $b$ denote doublet indices and
$\e_{ab}$ is the two-dimensional total antisymmetric tensor that
ensures antisymmetric $SU(2)$-contraction. 

\begin{figure}
\begin{center}
\includegraphics[scale=0.5]{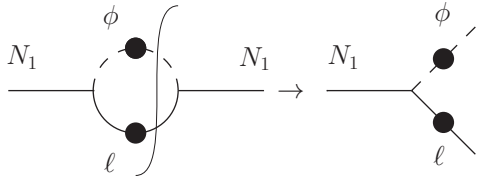}
\caption[The optical theorem in neutrino decays]{\label{cut} $N_1$ decay
  via the optical theorem with dressed propagators denoted by a blob.}
\end{center}
\end{figure}

We have discussed the HTL corrections to neutrino decays $N_1 \to H L$
in detail in reference~\cite{Kiessig:2010pr}. When the temperature is
so high that $m_\phi > M_1$, the neutrino decay is kinematically
forbidden in the HTL-approximation\footnote{It has been shown in
  reference~\cite{Anisimov:2010gy}, that the decay is still allowed if
  one considers the effect of collinear external momenta.}, but the
decay of Higgs bosons into neutrinos and leptons becomes
possible\footnote{The lepton decay is not possible, since $m_\phi >
  m_\ell$ for all temperatures}~\cite{Kiessig:2011fw}. We have briefly
shown the Higgs boson decay rate in reference~\cite{Kiessig:2010zz}
and explain the calculation in more detail in a separate
work~\cite{Kiessig:2011fw}. The basic idea is to calculate the rate
via the optical theorem by cutting the neutrino self-energy with
resummed propagators as in figure~\ref{cut}.
Using this method, we describe the particles that are affected by
thermal corrections, the Higgs boson and the lepton, by thermal
propagators, whereas the external particle is not affected
thermally since the couplings are small.  According to
finite-temperature cutting rules \cite{Weldon:1983jn,Kobes:1986za},
the interaction rate for a neutrino with momentum $P$ reads
\begin{align}
\label{eq:5}
\Gamma(P) = - \frac{1}{2 p_0} \; {\rm tr} [ (\slashed{P}+M_1) \; {\rm Im} \; 
\Sigma(p_0 + \rmi \, \e,{\bf p})].
\end{align}
Integrating over neutrino momenta $\bf p$, we can write the decay
density in a familiar form that is corrected for the statistical
distribution of the particles,
\begin{align}
\label{eq:6}
\gamma(N_1 \to H L) = \int \rmd \tilde{p}_{N_1} \rmd \tilde{p}_H \rmd
\tilde{p}_L (2 \pi)^4 \delta^4(p_{N_1} - p_L - p_H) \left| \mathcal{M}_h
\right|^2 f_{N_1}^\rmeq (1-f_L^\rmeq) (1+f_H^\rmeq)
\end{align}
and in the same way for the Higgs boson decays
\begin{align}
\label{eq:7}
\gamma(H \to N_1 L) = \int \rmd \tilde{p}_{N_1} \rmd \tilde{p}_H \rmd
\tilde{p}_L (2 \pi)^4 \delta^4(p_{N_1} + p_L - p_H) \left| \mathcal{M}_h
\right|^2 (1-f_{N_1}^\rmeq) (1-f_L^\rmeq) f_H^\rmeq \, ,
\end{align}
where $\rmd \tilde{p} = \rmd^3 p / [(2 \pi)^3 2 E]$, we have $E_H^2 =
p_H^2 + m_\phi^2$, $E_L = \omega_h(p_L)$ and $h=\pm 1$ denotes the
helicity-over-chirality ratio of the lepton. The inverse processes $HL
\to N_1$ and $N_1L \to H$ can be written in the same way with the
appropriate statistical factors. The matrix element for neutrino and
Higgs boson decays turns out to be the same, like at zero temperature,
\begin{align}
  \label{eq:8}
  \left| \mathcal{M}_h(P,K) \right|^2 = 4 \, (\l^\dagger \l)_{11}
  P_\mu K_h^\mu = 4 \, (\l^\dagger \l)_{11} Z_h \omega_h (p_0 -h \,
  {\bf p} \cdot \hat{\bf k}) ,
\end{align}
where we have introduced a chirally invariant four-momentum $K_h^\mu =
Z_h(k) \o_h(k) (1,h \, \hat{\bf k})$ for the lepton and $\hat{\bf k} =
{\bf k}/ k$. From this matrix element, one can derive a multiplication rule for
the HTL lepton spinors,
\begin{align}
\label{eq:9}
u_\ell^\pm(K) \overline{u}_\ell^\pm(K) = Z_\pm
\omega_\pm (\gamma_0 \mp \hat{\bf k} \cdot
\boldsymbol{\gamma})\, ,
\end{align}
and the antiparticle spinors,
  \begin{align}
\label{eq:10}
    v_\ell^\pm(K) \overline{v}_\ell^\pm(K) = - Z_\pm
\omega_\pm (\gamma_0 \pm \hat{\bf k} \cdot
\boldsymbol{\gamma}) \, .
\end{align}
These rules make it easy to calculate processes that involve
HTL-corrected leptons as external particles.

\begin{figure}
\centering
\includegraphics[width=0.75 \textwidth]{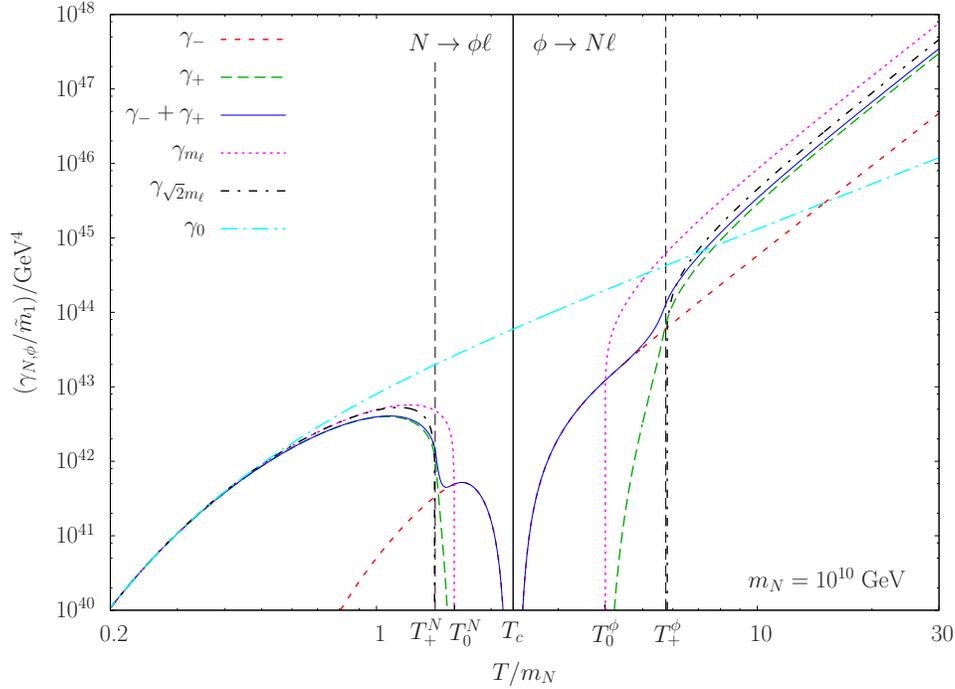}
\caption[Neutrino and Higgs boson decay densities]{The decay densities
  for the neutrino and the Higgs boson decay. We show the one-mode
  approach with the thermal mass as $\gamma_{m_{\ell}}$ and with
  the asymptotic mass as $\gamma_{\sqrt{2} m_{\ell}}$; Also the $T=0$ rate
  $\gamma_0$ and our two modes $\gamma_\pm$. The temperature
  thresholds are explained in the text.}
\label{comp} 
\end{figure}

In figure~\ref{comp}, we compare our consistent HTL calculation to the
one-mode approximation adopted by reference~\cite{Giudice:2003jh},
while we add quantum-statistical distribution functions to their
calculation, which equals the approach of using an approximated lepton
propagator $1/(\slashed{K}-m_\ell)$~\cite{Kiessig:2009cm}.  In
addition, we show the one-mode approach for the asymptotic mass
$\sqrt{2} \, m_\ell$. We evaluate the decay rates for $M_1=10^{10}$
GeV and normalise the rates by the effective neutrino mass
$\widetilde{m}_1 = (\l^\dagger \l)_{11} v^2 / M_1$, where $v=174$ GeV
is the vacuum expectation value of the Higgs field. This effective
mass is often taken as $\widetilde{m}_1 = 0.06 \; {\rm eV}$, inspired
by the mass scale of the atmospheric mass splitting.

The decay densities are analysed in detail in
reference~\cite{Kiessig:2011fw}.  Summarising, we can distinguish five
different thresholds for the thermal decay rates. Going from low
temperature to high temperature, these are given by the following
conditions:
\begin{align}
T_+^N: \quad M_1&=\sqrt{2} \, m_{\ell}+m_\phi \, , \nonumber \\
T_0^N: \quad M_1&=m_{\ell}+m_\phi \, , \nonumber \\
 T_c: \quad M_1&=m_\phi \, , \nonumber \\
 T_0^\phi: \quad m_\phi&=m_{\ell}+M_1 \, , \nonumber \\
 T_+^\phi: \quad m_\phi&=\sqrt{2} \, m_{\ell}+M_1 \, .
\end{align}
These thresholds correspond to three different thermal lepton masses,
the asymptotic mass of the $(+)$-mode, $\sqrt{2} \, m_\ell$, the naive
thermal mass $m_\ell$, which is the effective mass for zero lepton
momentum, and the vanishing asymptotic mass of the $(-)$-mode.

\subsection{$\boldsymbol{C \!P}$-asymmetries}
\label{sec:cp-asymm}

\begin{figure}
\begin{center}
\includegraphics[width=\textwidth]{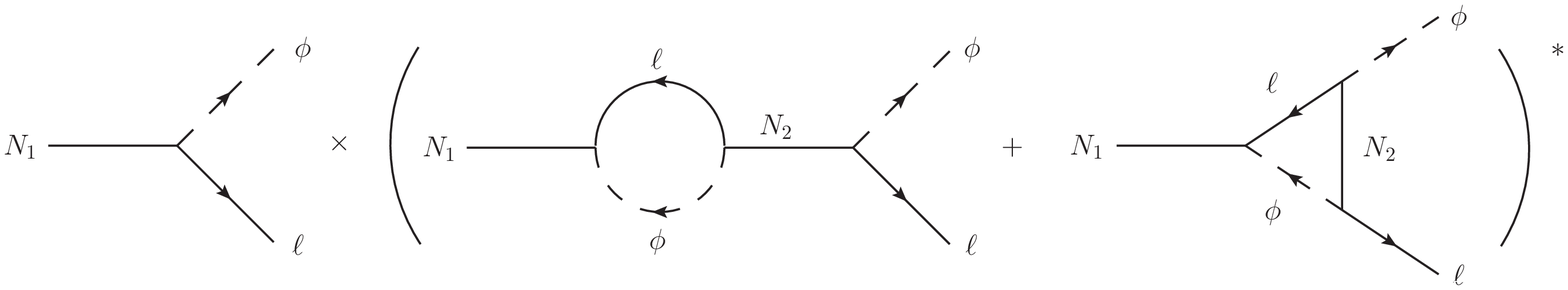}
\caption[$\CP$-asymmetry in neutrino decays]{The $\CP$-asymmetry
  in neutrino decays. The graph in the middle 
is the self-energy contribution, the graph on the right the
vertex contribution.}
\label{fig:cpasfigure}
\end{center}
\end{figure}

\begin{figure}
\begin{center}
\includegraphics[width=\textwidth]{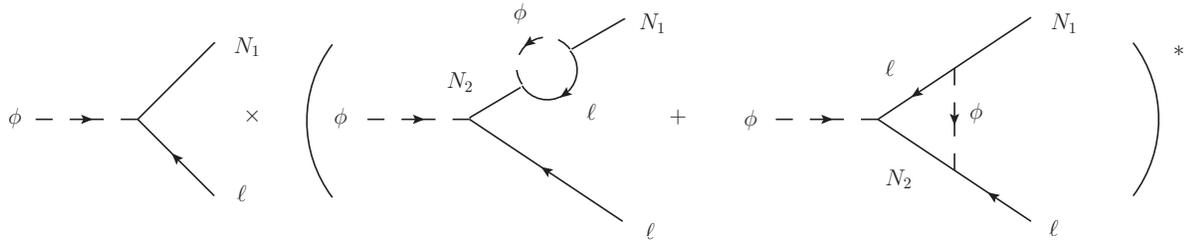}
\caption[$\CP$-asymmetry in Higgs boson decays]{The $\CP$-asymmetry
  in Higgs boson decays at high temperature. Again, the graph in the middle 
is the self-energy contribution, the graph on the right the
vertex contribution.}
\label{fig:cpasphi}
\end{center}
\end{figure}

The $C \!P$-asymmetry in neutrino decays at zero temperature is
defined as
\begin{align}
\label{epsilonGamma}
\epsilon_0=\frac{\Gamma(N \rightarrow \phi \ell) - 
\Gamma(N \rightarrow \bar{\phi} \bar{\ell})}
{\Gamma(N \rightarrow \phi \ell) + 
\Gamma(N \rightarrow \bar{\phi} \bar{\ell})} \, .
\end{align}
At finite
temperature, we have to calculate the $\CP$-asymmetry via the
integrated decay rates,
\begin{align}
\label{epsilongamma}
\epsilon_h(T)=\frac{\gamma^{T>0}(N \rightarrow \phi \ell_h) - 
\gamma^{T>0}(N \rightarrow \bar{\phi} \bar{\ell_h})}
{\gamma^{T>0}(N \rightarrow \phi \ell_h) + 
\gamma^{T>0}(N \rightarrow \bar{\phi} \bar{\ell_h})},
\end{align}
where we define the $\CP$-asymmetry for each lepton mode, denoted by
$h$. The $C\!P$-asymmetry arises as an interference between tree level
and one-loop diagrams in neutrino and, at high temperature, Higgs
boson decays, shown in figures~\ref{fig:cpasfigure} and \ref{fig:cpasphi}.
In order to calculate the interference, one has to take the imaginary
part of the relevant one-loop diagram. At zero temperature, this can
be done by cutting through the diagram and determining the
discontinuity via the optical theorem and the Cutkosky
rules. Kinematically, it is only possible to put the lepton propagator
and the Higgs boson propagator in the loop on-shell, so there is one
possible cutting for the self-energy graph and one possible cutting
for the vertex correction graph.

At finite temperature, there exist cutting rules for the RTF. In the
ITF, it is possible to isolate terms with certain momentum relations
that correspond to certain cuttings, while a direct relation to the
RTF cutting rules is not straightforward. Since we can exchange energy
with the heat bath, also terms that correspond to cutting through the
$N_2$ in the loop are
possible~\cite{Giudice:2003jh,Garny:2010nj,Garbrecht:2010sz,Kiessig:2011fw}. In
the hierarchical limit $M_2 \gg M_1$, which we assume here, these
terms are suppressed, so again the cuts through the lepton and the
Higgs boson in the loop survive. Due to the two possibilities for
lepton and Higgs doublets in the loop, the $C \! P$-asymmetry from the
self-energy graph is exactly twice as large as the $C\!P$-asymmetry
from the vertex correction graph in the hierarchical limit. We can
therefore distinguish four contributions to the $C\!P$-asymmetry,
taking into account the two lepton modes in the loop and two modes for
the external leptons.

\begin{figure}[h]
  \centering
  \includegraphics[width=0.8 \textwidth]{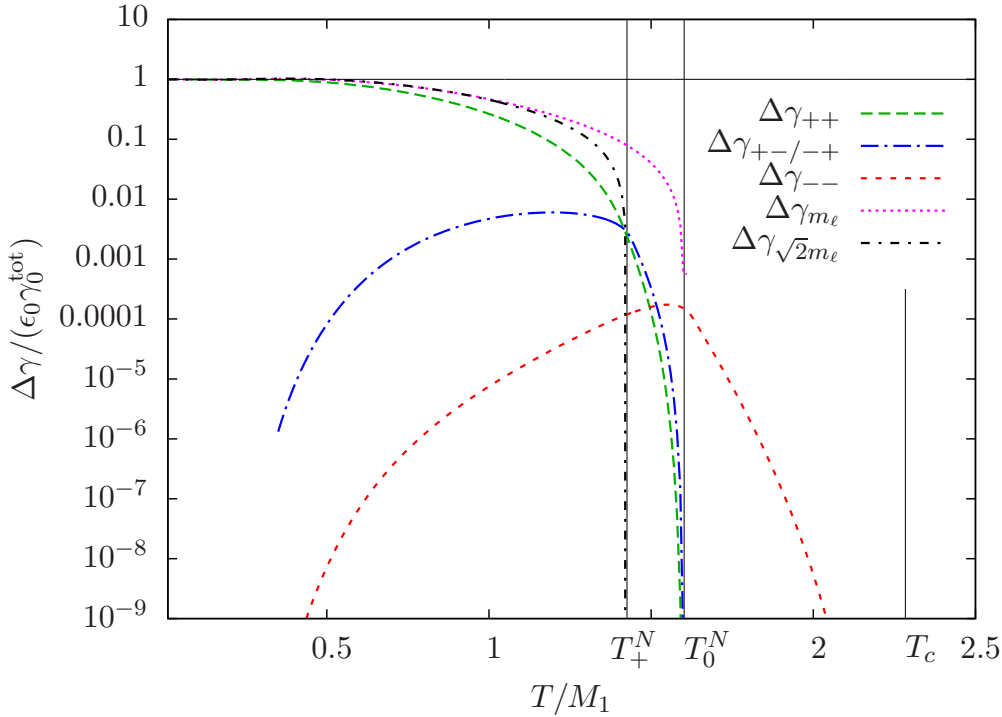}
  \caption[The $\CP$-asymmetries in $N$ decays in units of $\e_0
  \g_0$]{The $\CP$-asymmetries in neutrino decays normalised by the
    $\CP$-asymmetry in vacuum and the total decay density in vacuum,
    $\D \g/(\gamma^{\rm tot}_0 \e_0)$. We choose $M_1=10^{10} \, {\rm
      GeV}$ and $M_2 \gg M_1$. The term $\Delta \gamma_{h_1 h_2}$
    denotes the difference between the decay rate and its $\CP$
    conjugated rate, which is proportional to the
    $\CP$-asymmetry. Here, $h_1$ denotes the mode of the external
    lepton, while $h_2$ denotes the mode of the lepton in the
    loop. For example, $\Delta \gamma_{+-} = \gamma(N \to \phi
    \ell_+)- \gamma(N \to \barphi \barell_+)$, where a minus-mode
    lepton is present in the loop. $\Delta \gamma_{m_\ell}$ and
    $\Delta \gamma_{\sqrt{2} \, m_\ell}$ denote the rate differences
    for the one-mode approach with a thermal mass $m_\ell$ and an
    asymptotic thermal mass $\sqrt{2} \, m_\ell$.}
  \label{fig:cpn0}
\end{figure}

\begin{figure}
  \centering
\includegraphics[width=0.75 \textwidth]{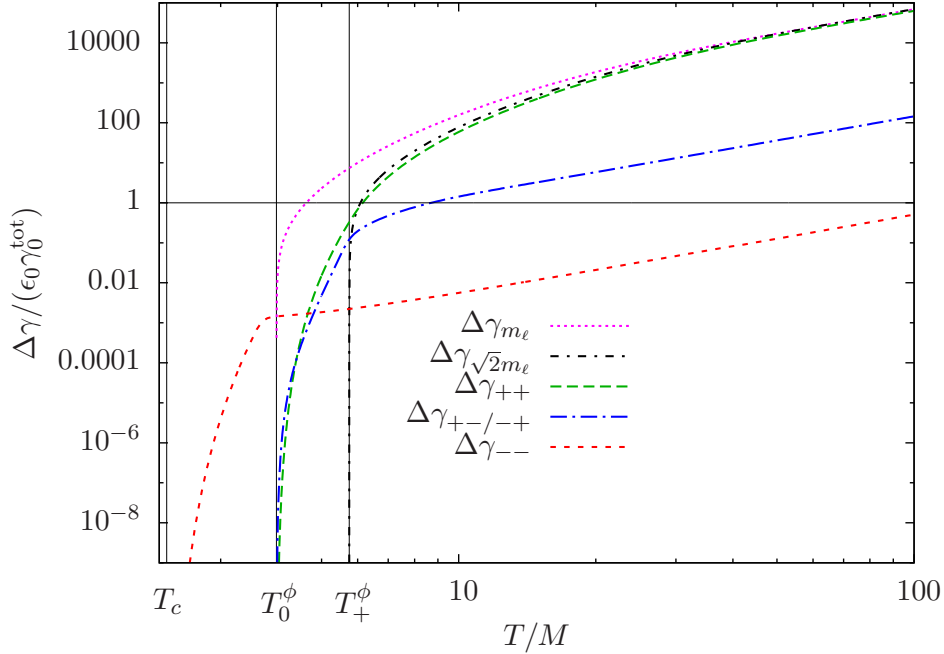}
\caption[The $\CP$-asymmetries in $\phi$ decays in units of $\e_0
\g_0$]{The $\CP$-asymmetries in Higgs boson decays normalised by the
  $\CP$-asymmetry in vacuum and the total decay density in vacuum, $\D
  \g/(\gamma^{\rm tot}_0 \e_0)$, where the asymmetries $\D \g$ are
  explained in figure~\ref{fig:cpn0}. We choose $M_1=10^{10} \, {\rm
    GeV}$ and $M_2 \gg M_1$.}
  \label{fig:cpphi0}
\end{figure}

The differences in neutrino decay rates, that is $\Delta \gamma = \g(N
\to \phi \ell) - \g(N \to \bar{\phi} \bar{\ell})$, are shown in
figure~\ref{fig:cpn0} and compared to the two one-mode approaches with
masses $m_\ell$ and $\sqrt{2} \, m_\ell$.
The same is shown for Higgs boson decays in figure~\ref{fig:cpphi0}.
We discuss these $C\!P$-asymmetries in detail in reference~\cite{Kiessig:2011fw}.

\section{Boltzmann Equations}
\label{sec:boltzmann-equations}

%\subsection*{Preliminary}
%\label{sec:preliminary}

We calculate the Boltzmann equations for leptogenesis. We include
decays and inverse decays involving neutrinos, leptons and Higgs
bosons. We neglect all scatterings, expect the on-shell contribution
of the $\Delta L=2$ scatterings, which we have to take into account
for consistency reasons. We take into account thermal dispersion
relations, but assume distributions close enough to equilibrium that
we can use Boltzmann equations. For the distribution functions, we use
the full quantum statistics, that is, Fermi-Dirac and Bose-Einstein
statistics, but assume the kinetic equilibrium approximation $f_i =
n_i/n_i^\rmeq f_i^\rmeq$. It has been shown by
reference~\cite{HahnWoernle:2009qn} that this is a good
approximation. The Boltzmann equations at zero temperature are derived
in appendix~\ref{sec:boltzm-equat-at}.

\subsection{Low Temperature}
\label{sec:low-temperature}

\subsubsection*{Neutrino evolution}
\label{sec:neutrino-evolution}

The formulation of the Boltzmann equations we need is given by
equation~\eqref{eq:b2} in appendix~\ref{sec:particle-kinematics}. The
equation for the evolution of the lightest right-handed neutrino reads
\begin{align}
  \label{eq:b6}
  \frac{\rmd n_{N_1}}{\rmd z}= - \frac{z}{HM_{N_1})}
    & \left[ \gamma(N_1 \rightarrow \phi \ell_+) + \gamma(N_1 \to \barphi
    \barell_+) 
    - \gamma(\phi \ell_+ \to N_1) - \gamma(\barphi \barell_+ \to N_1) 
  \right. \nonumber \\
+   & \left. \gamma(N_1 \rightarrow \phi \ell_-) + \gamma(N_1 \to \barphi
    \barell_-) 
    - \gamma(\phi \ell_- \to N_1) - \gamma(\barphi \barell_- \to N_1) 
  \right]
\end{align}
where $\ell_\pm$ denote the two lepton modes.  We neglect scatterings
since they are of higher order in the coupling constant.  We will from
now on omit the subscript 1 for the neutrino and write $N$. The
$CP$-asymmetry in the matrix element is not relevant for neutrino
decay, so we calculate the matrix element, which is the same for the
above processes and define
\begin{align}
  \label{eq:b68}
  \left|\mathcal{M}^0_\pm \right|^2 \equiv \left| \mathcal{M} (N \to
    HL_\pm) \right|^2 =  \left| \mathcal{M}
    (HL_\pm \to N) \right|^2, 
\end{align}
where now $HL_\pm$ denotes the sum of leptons and Higgs doublets,
$\ell_\pm$ and $\phi$, and their charge conjugated states
$\bar{\ell}_\pm$ and $\bar{\phi}$. The matrix elements are, however,
different for the different lepton modes and also the
momentum-conserving delta functions differ from each other. The
subscript $\pm$ means ($+$) or ($-$), not the sum. When summing an
expression $A_\pm$ that is dependent on the kind of lepton dispersion
relation over the lepton modes, we write $\sum_\pm A_\pm$. We have
\begin{align}
  \label{eq:b69}
    \frac{\rmd n_{N}}{\rmd z}= - \frac{z}{H(M_N)} \sum_\pm
    \left[ \gamma(N \rightarrow HL_\pm) - \gamma(HL_\pm \to N)  \right]
\end{align}
For each of the two lepton modes, we have now
\begin{align}
  \label{eq:b7}
  \gamma(N \rightarrow H L_\pm) - \gamma(HL_\pm \rightarrow N) 
& =  \int
  \rmd \tilde{p}_{N_1} \rmd \tilde{p}_{L \pm} \rmd \tilde{p}_H (2 \pi)^4 \delta^4(p_{N} -p_H-p_{L \pm}) \nonumber \\
& \times
  \left[ \left|\mathcal{M}(N \rightarrow H L_\pm)\right|^2 f_{N}
    (1+f_H)(1-f_{L \pm}) \right. \nonumber \\ 
&- \left. \left|\mathcal{M}(HL_\pm \rightarrow N)\right|^2
    (1-f_{N}) f_H f_{L \pm} \right].
\end{align}
The term in square brackets in equation~\eqref{eq:b7} reduces to
\begin{equation}
  \label{eq:b10}
  \left|\mathcal{M}^0_\pm \right|^2  \left[ f_{N}
    (1+f_H)(1-f_{L \pm})- (1-f_{N}) f_H f_{L \pm} \right] =
  \left|\mathcal{M}^0_\pm \right|^2 \left[ c_{N \rightarrow HL\pm}- c_{HL\pm
      \rightarrow N} \right],
\end{equation}
where
\begin{align}
  \label{eq:b11}
  c_{N \rightarrow HL\pm} &= f_N (1+f_H)(1-f_{L \pm}) \, , \nonumber \\
  c_{HL\pm \rightarrow N} &= (1-f_N) f_H f_{L \pm} \, .
\end{align}
Throughout this section, we make the kinetic equilibrium assumption,
that is, the phase space densities can be written as
\begin{equation}
  \label{eq:b9}
  f_i=\frac{n_i}{n_i^{\rm eq}} f_i^{\rm eq}= x_i f_i^\rmeq,
\end{equation}
where $x_i \equiv n_i/n_i^\rmeq$, and $n^\rmeq$ and $f^\rmeq$ are the
equilibrium number densities and distributions.  For the neutrino
evolution, we can assume that the Higgs bosons are in equilibrium
since they couple very strongly to the thermal bath,
$f_H=f_H^\rmeq$. The lepton distributions of the two modes are,
strictly speaking, out of equilibrium since leptons and antileptons
are created asymmetrically. However, the leptons are much closer to
equilibrium than the neutrinos, so the neutrino evolution is not
influenced by the lepton asymmetry. Therefore we approximate the
lepton densities with their equilibrium density, $f_{L \pm}=f_{L
  \pm}^\rmeq$. We will relax this assumption in the section on the
lepton asymmetry evolution.

Using the relation
\begin{equation}
  \label{eq:b13}
  f_N^\rmeq (1+f_H^\rmeq)(1-f_{L \pm}^\rmeq)=(1-f_N^\rmeq) f_{L \pm}^\rmeq f_H^\rmeq,
\end{equation}
we can write
\begin{align}
  \label{eq:b14}
  c_{N \rightarrow HL\pm} &=x_N f_N^\rmeq (1+f_H^\rmeq)(1-f_{L
    \pm}^\rmeq)=
  (x_N- x_N f_N^\rmeq) f_H^\rmeq f_{L \pm}^\rmeq \, , \nonumber \\
  c_{HL\pm \rightarrow N} &= (1- x_N f_N^\rmeq) f_H^\rmeq f_{L
    \pm}^\rmeq \, ,
\end{align}
and
\begin{align}
\label{eq:158}
c_{N \rightarrow HL\pm}- c_{N \rightarrow HL\pm} &=
(x_N-1) f_H^\rmeq f_{L \pm}^\rmeq.
\end{align}
The decay densities are
\begin{align}
  \label{eq:b16}
  \gamma(N \rightarrow H L_\pm) - \gamma(HL_\pm\rightarrow N) & = \int \rmd
  \tilde{p}_N \rmd \tilde{p}_{L \pm} \rmd \tilde{p}_H (2 \pi)^4
  \delta^4(p_{N} -p_H-p_{L \pm}) \nonumber \\ 
& \times \left|\mathcal{M}_\pm^0\right|^2 (x_N-1)
  f_H^\rmeq f_{L \pm}^\rmeq \nonumber \\
& = (x_N-1) \gamma_{D \pm}^N,
\end{align}
where
\begin{align}
  \label{eq:b17}
  \gamma_{D \pm}^N=  \int \rmd
  \tilde{p}_N \rmd \tilde{p}_{L \pm} \rmd \tilde{p}_H (2 \pi)^4
  \delta^4(p_{N} -p_H-p_{L \pm})  \left|\mathcal{M}_\pm^0\right|^2
  f_H^\rmeq f_{L \pm}^\rmeq.
\end{align}
Note that $\gamma_{D \pm}^N$ is not the same as the equilibrium decay
density in equation~\eqref{eq:6}, but differs from the latter through
the thermal factor $f_H f_L$. It is an effective decay density, which
enters the Boltzmann equations. The Boltzmann equation for the
neutrinos reads
\begin{align}
  \label{eq:b18}
    \frac{\rmd n_{N}}{\rmd z}= - \frac{z}{H(M_{N})} (x_N-1)
    \gamma_D^N,
\end{align}
where $\gamma_D^N \equiv \gamma_+ + \gamma_-$, or, in analogy to
equation~\eqref{eq:88} in appendix~\ref{sec:boltzm-equat-at},
\begin{align}
  \label{eq:b19}
   \frac{\rmd n_{N}}{\rmd z}= - D^N (n_N-n_N^\rmeq),
\end{align}
where
\begin{align}
  \label{eq:b20}
  D^N=\frac{\gamma_D^N}{n_N^\rmeq} \frac{1}{H z}
\end{align}
and we have used $H(M_N)=H z^2$.  Most conveniently, the number
densities are normalised by the entropy density $s$ in order to
factorise their dependence on the expansion of the universe. The
entropy density scales as
\begin{align}
  \label{eq:66}
  s=g_* \frac{2 \pi^2}{45} T^3, 
\end{align}
where $g_*$ counts the total number of effectively massless degrees of
freedom and is defined as
\begin{align}
  \label{eq:212}
  g_* = \sum_{i={\rm bosons}} g_i \left( \frac{T_i}{T} \right)^4 +
  \frac{7}{8} \sum_{i={\rm fermions}} g_i \left( \frac{T_i}{T}
  \right)^4 \, ,
\end{align}
where $i$ denotes species with mass $m_i \ll T$ and the factor $7/8$
arises from the difference in Fermi and Bose
statistics~\cite{Kolb:EarlyU}. At the temperature of leptogenesis, all
SM particles have negligible masses, so $g_*=106.75$.  We define all
number densities in terms of the entropy density as
\begin{align}
  \label{eq:70}
  Y_i \equiv \frac{n_i}{s},
\end{align}
then
\begin{align}
  \label{eq:76}
  x_i = \frac{Y_i}{Y_i^\rmeq}.
\end{align}
The Boltzmann equation reads
\begin{align}
  \label{eq:71}
      \frac{\rmd Y_{N}}{\rmd z}= - \frac{z}{s H_1} (x_N-1)
    \gamma_D^N,
\end{align}
where
\begin{align}
  \label{eq:72}
  H_1 \equiv H(T=M_N) = \sqrt{\frac{4 \pi^3 g_*}{45}} \frac{M_N}{M_{\rm Pl}},
 \end{align}
and the Planck mass is
\begin{align}
  \label{eq:73}
  M_{\rm Pl}= 1.221 \cdot 10^{19} \, {\rm GeV}.
\end{align}

\subsubsection*{Lepton asymmetry evolution}
\label{sec:lept-antil-evol}

We set up evolution equations for the two different lepton modes
separately and define the phase space density of the lepton asymmetry
in the respective mode as
\begin{align}
  \label{eq:b21}
  f_{\mathcal{L}h}
  =f_{\ell h}-f_{\bar{\ell} h}.
\end{align}
where $h=\pm1$ denotes the helicity-over-chirality ratio of the
leptons. The final lepton asymmetry is then
$n_{\mathcal{L}}^{\rm fin}=n_{\mathcal{L}+}^{\rm fin}+n_{\mathcal{L}-}^{\rm fin}$ after evaluating
the Boltzmann equations for each mode separately. 
The Boltzmann
equations for leptons and antileptons read
 \begin{align}
   \label{eq:b22}
   \frac{\rmd n_{\ell h_1}}{\rmd z} = - \frac{z}{H(M_{N})} \Big\{
     & \gamma(\ell_{h_1} \phi \rightarrow N)
     - \gamma(N \rightarrow  \ell_{h_1} \phi) \nonumber \\
     & + \sum_{h_2} \left[ \gamma(\ell_{h_1} \phi \rightarrow
       \bar{\ell}_{h_2} \bar{\phi})- \gamma(\bar{\ell}_{h_2}
       \bar{\phi} \rightarrow \ell_{h_1} \phi) \right]
   \Big\} \, , \nonumber \\
   \frac{\rmd n_{\bar{\ell} h_1}}{\rmd z} = - \frac{z}{H(M_{N})}
   \Big\{ & \gamma(\bar{\ell}_{h_1} \bar{\phi} \rightarrow N)
     - \gamma(N \rightarrow  \bar{\ell}_{h_1} \bar{\phi}) \nonumber \\
     & + \sum_{h_2} \left[ \gamma(\bar{\ell}_{h_1} \bar{\phi}
       \rightarrow \ell_{h_2} \phi) - \gamma(\ell_{h_2} \phi
       \rightarrow \bar{\ell}_{h_1} \bar{\phi}) \right] \Big\},
\end{align}
where we have $(h_1,h_2)= \pm 1$ to account for the second lepton
involved in the scatterings and we have only included $\Delta L=2$
scatterings since the other two-by-two scatterings that involve only
$N$, $\ell$ and $\phi$ are negligible\cite{Giudice:2003jh}. For the
evolution of the lepton asymmetry, we have
\begin{align}
  \label{eq:b23}
  \frac{\rmd n_{\mathcal{L} h_1}}{\rmd z} =& - \frac{1}{H z}  \Big\{
    \gamma(\ell_{h_1} \phi \to N) - \gamma(\barell_{h_1} \barphi \to N) - \gamma(N
    \to \ell_{h_1} \phi) + \gamma(N \to \barell_{h_1} \barphi)  \nonumber \\
  &+ \sum_{h_2}  \left[ 
\gamma(\ell_{h_1} \phi\to \barell_{h_2} \barphi)
 - \gamma(\barell_{h_1} \barphi \to \ell_{h_2} \phi)
+ \gamma(\ell_{h_2}\phi\to \barell_{h_1} \barphi) 
 - \gamma(\barell_{h_2} \barphi \to \ell_{h_1} \phi) \right]
\Big\} \, .
\end{align}

\begin{figure}
  \centering
  \includegraphics{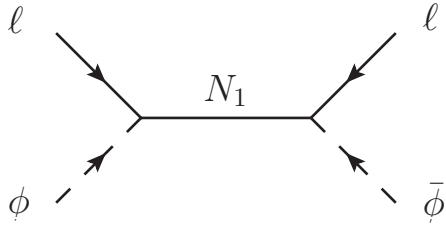}
  \caption{The $s$-channel contribution to the $\D L=2$ scattering $\ell
    \phi\to \barell \barphi$.}
  \label{fig:schannel}
\end{figure}

At leading order in the couplings, the $\D L=2$ scatterings are
computed at tree level and are consequently $CP$-conserving. However,
from these scatterings we must subtract the $CP$-violating
contribution where an on-shell $N_1$ is exchanged in the $s$ channel,
shown in figure~\ref{fig:schannel}. This is because in the Boltzmann equations
the process is already taken into account by inverse decays with
successive decays, $\ell \phi \to N \to \barell
\barphi$~\cite{Kolb:1979qa,Giudice:2003jh,Buchmuller:2004nz}.
We must therefore
replace the scattering rate by the subtracted rate,
\begin{equation}
  \label{eq:b24}
  \gamma(\ell_{h_1} \phi \to \barell_{h_2} \barphi) \to \gamma^{\rm
    sub}(\ell_{h_1} \phi\to \barell_{h_2}
  \barphi) \equiv \gamma(\ell_{h_1} \phi \to \barell_{h_2} \barphi) - \gamma^{\rm
    on-shell}(\ell_{h_1} \phi 
  \to \barell_{h_2} \barphi)
\end{equation}
and $\gamma(\barell_{h_1} \barphi \to \ell_{h_2} \phi)$ accordingly,
where $\gamma^{\rm on-shell}$ is the on-shell contribution.  The
Boltzmann equations then read
\begin{align}
  \label{eq:b30}
  \frac{\rmd n_{\mathcal{L} h_1}}{\rmd z} =& - \frac{1}{H z}  \Big\{
    \gamma(\ell_{h_1} \phi \to N) - \gamma(\barell_{h_1} \barphi \to N) - \gamma(N
    \to \ell \phi_{h_1}) + \gamma(N \to \barell_{h_1} \barphi)  \nonumber \\
  &+ \sum_{h_2}  \big[ 
\gamma^{\rm sub}(\ell_{h_1} \phi\to \barell_{h_2} \barphi)
 - \gamma^{\rm sub}(\barell_{h_1} \barphi \to \ell_{h_2} \phi)
 \nonumber \\
& + \gamma^{\rm sub}(\ell_{h_2}\phi\to \barell_{h_1} \barphi) 
 - \gamma^{\rm sub}(\barell_{h_2} \barphi \to \ell_{h_1} \phi) \big]
\Big\} \, .
\end{align}

It is convenient to define a $CP$-asymmetry in neutrino decays on
amplitude level as
\begin{equation}
  \label{eq:b26}
  \epsilon^N_{h}=\frac{\left| \mathcal{M}(N \to \phi\ell_{h}) \right|^2 - 
\left| \mathcal{M}(N \to \barphi \barell_{h}) \right|^2}
{\left| \mathcal{M}(N \to \phi\ell_{h}) \right|^2 + 
\left| \mathcal{M}(N \to \barphi \barell_{h}) \right|^2}
\end{equation}
and $\left| \mathcal{M}(N \to \phi\ell_{h}) \right|^2 + \left|
    \mathcal{M}(N \to \barphi \barell_{h}) \right|^2= \left| \mathcal{M}_{0 h}
  \right|^2$, we write
  \begin{align}
    \label{eq:b27}
    \left| \mathcal{M}(N \to \phi\ell_h) \right|^2=
\left| \mathcal{M}(\barphi \barell_h \to N) \right|^2 &= \frac{1+\epsilon_h^N}{2} 
\left| \mathcal{M}_{0 h} \right|^2, \nonumber \\
    \left| \mathcal{M}(N \to \barphi \barell_h) \right|^2=
\left| \mathcal{M}(\phi\ell_h\to N) \right|^2 &= \frac{1-\epsilon_h^N}{2} 
\left| \mathcal{M}_{0 h} \right|^2.
  \end{align}

  It is useful to write the decay rates for the above $1
  \leftrightarrow 2$ processes as in
  section~\ref{sec:neutrino-evolution},
\begin{equation}
  \label{eq:b25}
  \gamma({\rm process})=\int \prod_j \rmd \tilde{p}_j (2 \pi)^4
  \delta^4\left(\sum p_j\right) c({\rm process}),
\end{equation}
where $p_j$ denotes the relevant momenta $p_N$, $p_{\ell h}=p_{\barell h}$ and
$p_\phi=p_{\barphi}$ and $\delta^4(\sum p_j)$ the momentum conservation
$\delta^4(p_N-p_{\ell h}-p_\phi)$. The information about the specific
process is encoded in $c({\rm process})$ and we have
\begin{align}
  \label{eq:b28}
  c_{N \to \ell_h \phi} &= \left| \mathcal{M}(N \to \phi\ell_h)
\right|^2 f_N (1-f_{\ell h}) (1+f_\phi)
  \nonumber \\
  c_{\ell_h \phi \to N} &= \left| \mathcal{M}(\phi\ell_h\to N) \right|^2
 (1-f_N) f_{\ell h} f_\phi
  \nonumber \\
  c_{N \to \barell_h \barphi} &= \left| \mathcal{M}(N \to \barphi
    \barell_h) \right|^2 f_N (1-f_{\barell h})
  (1+f_{\barphi})
  \nonumber \\
  c_{\barell_h \barphi \to N} &= \left| \mathcal{M}(\barphi \barell_h \to
    N) \right|^2 (1-f_N) f_{\barell h} f_{\barphi}.
\end{align}
Since we are looking at the lepton asymmetry, the lepton distributions
have to be out of equilibrium, 
\begin{align}
  \label{eq:b29}
  f_{\ell/\barell h}&=x_{\ell/\barell h} f_{\ell h}^\rmeq, \nonumber \\
f_{\mathcal{L} h}&= x_{\mathcal{L}.h} f_{\ell h}^\rmeq, \nonumber \\
f_{\ell h}+f_{\barell h}& \approx
 2 f_{\ell h}^\rmeq,
\end{align}
while the Higgs bosons can be assumed to be in equilibrium.

As explained in appendix~\ref{sec:subtr-shell-prop}, the scattering
rates can be written as
\begin{align}
  \label{eq:b31}
  \sum_{h_f} \left[ \gamma^{\rm sub} (\ell_{h_i}\phi \right.
&
\left. \to \barell_{h_f}
    \barphi) - \gamma^{\rm sub}(\barell_{h_i} \barphi \to \ell_{h_f} \phi)
  \right] \nonumber \\
& =   \sum_{h_f} \left[ \gamma^{\rm sub} (\ell_{h_f}\phi\to \barell_{h_i}
    \barphi) - \gamma^{\rm sub}(\barell_{h_f} \barphi \to \ell_{h_i} \phi)
  \right]= \nonumber \\
& = \int \rmd \tilde{p}_N \rmd \tilde{p}_{\ell h_i} \rmd
  \tilde{p}_\phi (2 \pi)^4 \delta^4(p_N-p_{\ell h_i}-p_\phi)
  \e_{h_i}^N \left| \mathcal{M}_{h_i}^0 \right|^2 f_{\ell h_i}^\rmeq f_\phi^\rmeq
  (1-f_N^\rmeq).
\end{align}
so we define\footnote{Note that our factor $c^{\rm sub}$ differs from
  reference~\cite{HahnWoernle:2009qn}, where they have the
  out-of-equilibrium distribution $(1-f_N)$ instead of
  $(1-f_N^\rmeq)$. However, as derived in
  appendix~\ref{sec:subtr-shell-prop}, we must employ $f_N^\rmeq$,
  even if we had only one lepton mode, which also results in a
  Boltzmann equation for $(\ell - \barell)$ which is slightly
  different from reference~\cite{HahnWoernle:2009qn}.}
\begin{align}
  \label{eq:b32}
  c^{\rm sub}_h=2 \e_h^N \left| \mathcal{M}^0_h \right|^2
    f_{\ell h}^\rmeq f_\phi^\rmeq (1-f_N^\rmeq)
\end{align}
%{\color{red} Ab hier wird's jetzt unklar. Auf jeden Fall \"uberpr\"ufen und
%  rechnen!!! Wie kann die Boltzmann-Gleichung f\"ur beide Moden
%  zusammengefasst werden?!?\\}
\noindent and calculate the integrand for the right-hand side of the Boltzmann
equation~\eqref{eq:95},
\begin{align}
  \label{eq:b33}
  c(N \to \ell_h \phi)  - c(\ell_h \phi \to N) - 
 & 
c(N  \to \barell_h
    \barphi) + c(\barell_h \barphi \to N) + c^{\rm sub}_h 
= 
\nonumber \\[1ex]
& =
  x_{\mathcal{L} h} f_{\ell h}^\rmeq (f_\phi^\rmeq + x_N f_N^\rmeq )
               - 2 \e_h^N
  f_{\ell h}^\rmeq f_\phi^\rmeq \left( x_N -1 \right) \left( 1 - 2
    f_N^\rmeq \right).
\end{align}
We can easily check that this term vanishes when the neutrinos are in
equilibrium, $x_N=1$, and there is no previous lepton asymmetry,
$x_{\mathcal{L}h}=0$.

The Boltzmann equation reads now
\begin{align}
  \label{eq:b34} 
\frac{\rmd n_{\mathcal{L}h}}{\rmd z} = - \frac{1}{H
z} \left[ - \e^N_{\g h} \g_{\e h}^N \left( x_N-1 \right) +
\frac{x_{\mathcal{L}h}}{2} \left( \gamma_{W h}^N + x_N \g_{N
h}^N \right) \right] ,
\end{align}
where $\g_{W h}^N=\g_{D h}^N$ is defined in equation~\eqref{eq:b17} and
\begin{align}
  \label{eq:b35}
  \g_{\e h}^N  &= \int \rmd
  \tilde{p}_N \rmd \tilde{p}_{\ell h} \rmd \tilde{p}_\phi (2 \pi)^4
  \delta^4(p_{N} -p_\phi-p_{\ell h})  \left|\mathcal{M}_0\right|^2
  f_\phi^\rmeq f_{\ell h}^\rmeq (1 - 2 f_N^\rmeq) \nonumber \\ 
\g_{N h}^N &=  \int \rmd
  \tilde{p}_N \rmd \tilde{p}_{\ell h} \rmd \tilde{p}_\phi (2 \pi)^4
  \delta^4(p_{N} -p_\phi-p_{\ell h})  \left|\mathcal{M}_0\right|^2
  f_{\ell h}^\rmeq f_N^\rmeq, \nonumber \\
\e_{\g h}^N &=  \frac{1}{\g_{\e h}} \int \rmd
  \tilde{p}_N \rmd \tilde{p}_{\ell h} \rmd \tilde{p}_\phi (2 \pi)^4
  \delta^4(p_{N} -p_\phi-p_{\ell h}) \e_h^N \left|\mathcal{M}_0\right|^2
  f_\phi^\rmeq f_{\ell h}^\rmeq (1 - 2 f_N^\rmeq).
\end{align}
We see that the rates and the $\CP$-asymmetries that enter the
Boltzmann equations have slightly different thermal factors than the
equilibrium rate in equations~\eqref{eq:6}, which employs the factor
$f_N (1-f_\ell) (1+f_\phi)$ for $N$ decays.

We may also write
\begin{align}
  \label{eq:74}
  \frac{\rmd Y_{\mathcal{L}h}}{\rmd z} = - \frac{z}{s H_1} \left[ -
    \e^N_{\g h} \g_{\e h}^N \left( x_N-1 \right) +
    \frac{x_{\mathcal{L}h}}{2} \left( \gamma_{W h}^N + x_N \g_{N h}^N
    \right) \right]
\end{align}
or, corresponding to equation~\eqref{eq:95} in
appendix~\ref{sec:boltzm-equat-at},
\begin{align}
  \label{eq:b36}
    \frac{\rmd n_{\mathcal{L}h}}{\rmd z} = \epsilon_{\g h}^N D_{\e h}^N
    (n_N-n_N^\rmeq) - (W_{0 h}^N+ W_{N h}^N x_N)  n_{\mathcal{L}
      h},
\end{align}
where
\begin{align}
  \label{eq:b37}
  D_{\e h}^N&=\frac{1}{Hz} \frac{\g_{\e h}^N}{n_N^\rmeq} \nonumber \\
  W_{0 h}&=\frac{1}{Hz} \frac{\g_{W h}^N}{2 n_{\ell h}^\rmeq} \nonumber \\
  W_{N h} & = \frac{1}{Hz} \frac{\g_{N h}^N}{2 n_{\ell h}^\rmeq}.
\end{align}

\subsection{High temperature}
\label{sec:high-temperature}

As discussed in section~\ref{sec:decay-inverse-decay}, the neutrino
processes $N \leftrightarrow \ell \phi$ are forbidden when the thermal
masses of the Higgs bosons and leptons become too large, that is, when
$m_\phi > M_N$. However, new processes with the Higgs as single
initial or final state are then allowed, $\phi \leftrightarrow N
\ell$. These are the dominant contributions to the neutrino and lepton
evolution and they can be $CP$-violating as well, so they contribute
to generating a lepton asymmetry. We derive the Boltzmann equations
for this high temperature regime in the following.

\subsubsection*{Neutrino evolution}
\label{sec:neutrino-evolution-1}

We derive the Boltzmann equation analogously to
section~\ref{sec:neutrino-evolution},
\begin{align}
  \label{eq:b44}
    \frac{\rmd n_{N}}{\rmd z}= - \frac{1}{Hz} \sum_h \left[ 
\gamma(NL_h \rightarrow H)
- \gamma(H\rightarrow NL_h)
\right].
\end{align}
We have
\begin{align}
  \label{eq:b52}
  \gamma(NL_h \rightarrow H ) - \gamma(H\rightarrow NL_h) & = \int
  \rmd \tilde{p}_N \rmd \tilde{p}_{Lh} \rmd \tilde{p}_H (2 \pi)^4
  \delta^4 (p_H -p_N-p_{Lh}) \nonumber \\
  & \times \left[ \left|\mathcal{M}(NL_h \rightarrow H)\right|^2 f_N
    f_{Lh} (1+f_H) \right. \nonumber \\
  &- \left. \left|\mathcal{M}(H \rightarrow NL_h)\right|^2 (1-f_N)
    (1-f_{Lh}) f_H \right].
\end{align}
The tree-level matrix elements $\left| \mathcal{M}^0_h \right|^2$ are the
same at high temperature for the Higgs-processes, just the kinematics
differ. So we have
\begin{align}
  \label{eq:b53}
    \left|\mathcal{M}^0_h\right|^2 \equiv
  \left|\mathcal{M}(NL_h \rightarrow H)\right|^2 = \left|\mathcal{M}(H
    \rightarrow NL_h)\right|^2.
\end{align}
Again, we assume the Higgs bosons and leptons to be in equilibrium. We
write
\begin{align}
  \label{eq:b54}
    \left|\mathcal{M}_0\right|^2  \left[ f_N
    f_{Lh} (1+f_H)- (1-f_N) (1-f_{Lh}) f_H \right] =
  \left|\mathcal{M}_0\right|^2 \left[ c(NL_h \rightarrow H)- c(H
      \rightarrow NL_h) \right] \, .
\end{align}
Using the relation
\begin{align}
  \label{eq:b55}
  f_N^\rmeq f_{Lh}^\rmeq (1+f_H^\rmeq) = (1-f_N^\rmeq) (1-f_{Lh}^\rmeq) f_H^\rmeq,
\end{align}
we get
\begin{align}
  \label{eq:b56}
   c(NL_h \rightarrow H)- c(H
      \rightarrow NL_h) = (x_N-1) (1-f_{Lh}^\rmeq) f_H^\rmeq.
\end{align}
The Boltzmann equation then reads
\begin{align}
  \label{eq:b57}
      \frac{\rmd n_{N}}{\rmd z} = - \frac{1}{Hz} (x_N-1) 
    \gamma_{D}^\phi,
\end{align}
\begin{align}
  \label{eq:77}
    \frac{\rmd Y_{N}}{\rmd z} = - \frac{z}{s H_1} (x_N-1) 
    \gamma_{D}^\phi,
\end{align}
or
\begin{align}
  \label{eq:b60}
  \frac{\rmd n_{N}}{\rmd z} = - D^\phi (n_N-n_N^\rmeq),
\end{align}
where $\gamma_{D}^\phi= \gamma_{D+}^\phi+ \gamma_{D-}^\phi$,
\begin{align}
  \label{eq:b58}
    \gamma_{Dh}^\phi=  \int \rmd
  \tilde{p}_N \rmd \tilde{p}_{Lh} \rmd \tilde{p}_H (2 \pi)^4
  \delta^4(p_H -p_N-p_{Lh})  \left|\mathcal{M}^0_h\right|^2
  f_H^\rmeq (1-f_{Lh}^\rmeq)
\end{align}
and
\begin{align}
  \label{eq:b59}
  D^\phi=\frac{\gamma_{D}^\phi}{n_N^\rmeq} \frac{1}{H z} \, .
\end{align}

\subsubsection*{Lepton asymmetry evolution}
\label{sec:lept-antil-evol-1}

\begin{figure}
  \centering
  \includegraphics{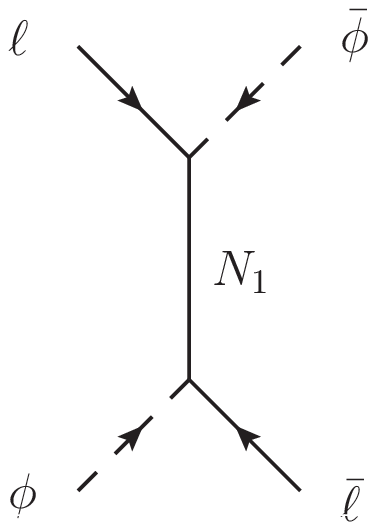}
  \caption{The $u$-channel contribution to the $\D L=2$ scattering $\ell
    \phi\to \barell \barphi$.}
  \label{fig:uchannel}
\end{figure}

The Boltzmann equations for leptons and antileptons read
\begin{align}
  \label{eq:b61}
    \frac{\rmd n_{\ell h_1}}{\rmd z}  = 
- \frac{1}{Hz} \Big\{
   & \gamma(\ell_{h_1} N \rightarrow \barphi) 
- \gamma(\barphi \rightarrow \ell_{h_1} N) \nonumber \\
  &  + \sum_{h_2} \left[\gamma(\ell_{h_1} \phi \rightarrow
      \bar{\ell}_{h_2} \bar{\phi})-
    \gamma(\bar{\ell}_{h_2} \bar{\phi} \rightarrow \ell_{h_1} \phi)
  \right] \Big\}, \\
  \frac{\rmd n_{\bar{\ell h_1}}}{\rmd z}   =
- \frac{z}{H(M_{N})}
 \Big\{
 &  \gamma(\bar{\ell}_{h_1} N \rightarrow \phi) 
- \gamma(\phi \rightarrow \barell_{h_1} N) \nonumber \\
& + \sum_{h_2} \left[ \gamma(\bar{\ell}_{h_1} \bar{\phi} \rightarrow \ell_{h_2} \phi)
 - \gamma(\ell_{h_2} \phi \rightarrow \bar{\ell}_{h_1} \bar{\phi})
\right] \Big\}, 
\end{align}
combined we get
\begin{align}
  \label{eq:b62}
  \frac{\rmd n_{\mathcal{L} h_1}}{\rmd z} =& - \frac{1}{H z} \Big\{
  \gamma(\ell_{h_1} N \to \barphi) - \gamma(\barell_{h_1} N \to
  \phi) - \gamma(\barphi
  \to \ell_{h_1} N) + \gamma(\phi \to \barell_{h_1} N)  \nonumber \\
  &+ \sum_{h_2} \left[ \gamma(\ell_{h_1} \phi\to \barell_{h_2}
    \barphi) - \gamma(\barell_{h_1} \barphi \to \ell_{h_2} \phi) +
    \gamma(\ell_{h_2}\phi\to \barell_{h_1} \barphi) -
    \gamma(\barell_{h_2} \barphi \to \ell_{h_1} \phi) \right] \Big\} \, .
\end{align}
At high temperature, there can be no on-shell neutrino in the
$s$-channel of the $\D L=2$ scatterings, but there can be an on-shell
neutrino exchange in the $u$-channel as shown in figure~\ref{fig:uchannel}.
Again, we need to subtract the $\D L=2$ rates since the $u$-channel
on-shell neutrino exchange corresponds to a Higgs decay followed by an
inverse decay, $\ell \phi \to \ell N \barell \to \barphi
\barell$. We replace
\begin{align}
  \label{eq:b63}
  \gamma(\ell_{h_1} \phi \to \barell_{h_2} \barphi) \to \gamma^{\rm
    sub}(\ell_{h_1}\phi\to \barell_{h_2} \barphi) \equiv
  \gamma(\ell_{h_1} \phi \to \barell_{h_2} \barphi) - \gamma^{\rm
    on-shell}_u(\ell_{h_1} \phi \to \barell_{h_2} \barphi)
\end{align}
and get
\begin{align}
  \label{eq:b64}
  \frac{\rmd n_{\mathcal{L} h_1}}{\rmd z} =& - \frac{1}{H z} \Big\{
  \gamma(\ell_{h_1} N \to \barphi) - \gamma(\barell_{h_1} N \to
  \phi) - \gamma(\barphi
  \to \ell_{h_1} N) + \gamma(\phi \to \barell_{h_1} N)  \nonumber \\
  &+ \sum_{h_2} \left[ \gamma^{\rm sub}(\ell_{h_1} \phi\to \barell_{h_2}
    \barphi) - \gamma^{\rm sub}(\barell_{h_1} \barphi \to \ell_{h_2} \phi) +
    \gamma^{\rm sub}(\ell_{h_2}\phi\to \barell_{h_1} \barphi) -
    \gamma^{\rm sub}(\barell_{h_2} \barphi \to \ell_{h_1} \phi) \right] \Big\} \, .
\end{align}
We define a $CP$-asymmetry in Higgs decays on amplitude level as
\begin{align}
  \label{eq:b65}
  \e_{h}^\phi \equiv \frac{\left| \mathcal{M}(\barphi \to N \ell_h) \right|^2 - 
\left| \mathcal{M}(\phi \to N \barell_h) \right|^2}
{\left| \mathcal{M}(\barphi \to N \ell_h) \right|^2 + 
\left| \mathcal{M}(\phi \to N \barell_h) \right|^2},
\end{align}
thus
\begin{align}
  \label{eq:b66}
  \left| \mathcal{M}(\barphi \to N \ell_h) \right|^2 = 
  \left| \mathcal{M}(\barell_h N \to \phi) \right|^2 =
  \frac{1+\e_{h}^\phi}{2} \left| \mathcal{M}_h \right|^2, \nonumber \\
  \left| \mathcal{M}(\phi \to N \barell_h) \right|^2 = 
  \left| \mathcal{M}(\ell_h N \to \barphi) \right|^2 =
  \frac{1-\e_{h}^\phi}{2} \left| \mathcal{M}_h \right|^2 \, .
\end{align}
As explained in appendix~\ref{sec:subtr-shell-prop}, the scattering
rates are written as
\begin{align}
  \label{eq:b67}
  \sum_{h_f} \left[ \gamma^{\rm sub} (\ell_{h_i}\phi \right.  &
  \left. \to \barell_{h_f} \barphi) - \gamma^{\rm sub}(\barell_{h_i}
    \barphi \to \ell_{h_f} \phi)
  \right] \nonumber \\
  & = \sum_{h_f} \left[ \gamma^{\rm sub} (\ell_{h_f}\phi\to
    \barell_{h_i} \barphi) - \gamma^{\rm sub}(\barell_{h_f} \barphi
    \to \ell_{h_i} \phi)
  \right] \nonumber \\
  & = \int \rmd \tilde{p}_N \rmd \tilde{p}_{\ell h_i} \rmd
  \tilde{p}_\phi (2 \pi)^4 \delta^4(p_N-p_{\ell h_i}-p_\phi) \e_h^\phi \left|
    \mathcal{M}_{h_i}^0 \right|^2 f_{\ell h_i}^\rmeq (1+f_\phi^\rmeq)
  f_N^\rmeq \, ,
\end{align}
so 
\begin{align}
  \label{eq:62}
  c_h^{\rm sub}= 2 \e^\phi_h \left| \mathcal{M}_h^0 \right|^2 f_{\ell
    h}^\rmeq (1+ f_\phi^\rmeq) f_N^\rmeq
\end{align}
and we get for the right-hand side of equation~\eqref{eq:b64}, 
\begin{align}
  \label{eq:63}
  c(\ell_{h} N \to \barphi) - & c(\barell_{h} N \to
  \phi) - 
c(\barphi
  \to \ell_{h} N) + c(\phi \to \barell_{h} N) +  c_h^{\rm sub}
\nonumber \\[1ex]
& =
  x_{\mathcal{L} h} f_{\ell h}^\rmeq (f_\phi^\rmeq + x_N f_N^\rmeq )
               + 2 \e^\phi_h
  (1-f_{\ell h}^\rmeq) f_\phi^\rmeq \left( x_N -1 \right) \left( 1 - 2
    f_N^\rmeq \right).
\end{align}

The Boltzmann equation reads now
\begin{align}
\label{eq:64}
   \frac{\rmd n_{\mathcal{L}h}}{\rmd z} = - \frac{1}{H z} \left[
    - \e_{\g h}^\phi
     \g_{\e h}^\phi \left( x_N-1 \right) + \frac{x_{\mathcal{L}h}}{2} \left(
       \gamma_{W h}^\phi + x_N \g^\phi_{N h} \right) \right] ,
\end{align}
where
\begin{align}
  \label{eq:65}
    \g^\phi_{\e h}  &= \int \rmd
  \tilde{p}_N \rmd \tilde{p}_{\ell h} \rmd \tilde{p}_\phi (2 \pi)^4
  \delta^4(p_{N} -p_\phi+p_{\ell h})  \left|\mathcal{M}_h^0\right|^2
  f_\phi^\rmeq (1-f_{\ell h}^\rmeq) (1 - 2 f_N^\rmeq) \nonumber \\ 
\g_{W h}^\phi & = \int \rmd
  \tilde{p}_N \rmd \tilde{p}_{\ell h} \rmd \tilde{p}_\phi (2 \pi)^4
  \delta^4(p_{N} -p_\phi+p_{\ell h})  \left|\mathcal{M}_h^0\right|^2
  f_\phi^\rmeq f_{\ell h}^\rmeq
\nonumber \\
\g^\phi_{N h} &=  \int \rmd
  \tilde{p}_N \rmd \tilde{p}_{\ell h} \rmd \tilde{p}_\phi (2 \pi)^4
  \delta^4(p_{N} -p_\phi+p_{\ell h})  \left|\mathcal{M}_h^0\right|^2
  f_{\ell h}^\rmeq f_N^\rmeq, \nonumber \\
\e^\phi_{\g h} &=  \frac{1}{\g^\phi_{\e h}} \int \rmd
  \tilde{p}_N \rmd \tilde{p}_{\ell h} \rmd \tilde{p}_\phi (2 \pi)^4
  \delta^4(p_{N} -p_\phi+p_{\ell h}) \e^\phi_h \left|\mathcal{M}_h^0\right|^2
  f_\phi^\rmeq (1-f_{\ell h}^\rmeq) (1 - 2 f_N^\rmeq).
\end{align}

Analogous to equation~\eqref{eq:95}, we may also write
\begin{align}
\label{eq:67}
\frac{\rmd n_{\mathcal{L}h}}{\rmd z} = - \epsilon^\phi_{\g h} D^\phi_{\e
  h} (n_N-n_N^\rmeq) - (W^\phi_{0 h}+ W^\phi_{N h} x_N) n_{\mathcal{L}
  h},
\end{align}
where
\begin{align}
\label{eq:68}
D^\phi_{\e h}&=\frac{1}{Hz} \frac{\g^\phi_{\e h}}{n_N^\rmeq} \nonumber \\
W^\phi_{0 h}&=\frac{1}{Hz} \frac{\g^\phi_{D h}}{2 n_{\ell h}^\rmeq}
\nonumber \\
W^\phi_{N h} & = \frac{1}{Hz} \frac{\g^\phi_{\mathcal{L}N h}}{2
  n_{\ell h}^\rmeq}.
\end{align}
Normalised by the entropy density, the equation reads
\begin{align}
  \label{eq:78}
     \frac{\rmd Y_{\mathcal{L}h}}{\rmd z} = - \frac{z}{s H_1} \left[ - 
     \e_{\g h}^\phi
     \g_{\e h}^\phi \left( x_N-1 \right) + \frac{x_{\mathcal{L}h}}{2} \left(
       \gamma_{W h}^\phi + x_N \g^\phi_{N h} \right) \right].
\end{align}

\subsection{Interacting Modes}
\label{sec:interacting-modes}

The Boltzmann equations in the previous sections were derived under
the assumption that the only relevant interactions in which the
leptons take part are the Yukawa interactions with Higgs bosons and
heavy neutrinos, which have very small coupling constants
$\lambda$, while it is implicitly assumed that gauge interactions keep the
leptons and the Higgs bosons in equilibrium. This scenario would imply
that the two modes only interact with each other via intermediate
neutrinos or Higgs bosons, where the distributions and also the
asymmetries in each mode are to first approximation decoupled.  In a
more realistic model, the lepton modes will couple to each other via
the $SU(2)$ and $U(1)$ gauge bosons $W^a_\mu$ and $B_\mu$ in processes
like $\ell_\pm \to \ell_\mp B$. While it is conceptually interesting
to consider the case that the two modes are completely decoupled, it
might be more realistic to study the scenario where the interactions
between the lepton modes are fast enough to keep them in chemical
equilibrium.

Chemical equilibrium implies that for species that interact via
processes $a+b \to i+j$, the corresponding chemical potentials are
related as
\begin{align}
  \label{eq:183}
  \mu_a+\mu_b=\mu_i+\mu_j \, .
\end{align}
When the processes which create or annihilate the particles and
antiparticles of some species are fast, for example via $a + \bar{a}
\to i+j$, where $i$ and $j$ are in equilibrium and their chemical
potentials vanish, then the chemical potentials of $a$ and $\bar{a}$
behave as
\begin{align}
  \label{eq:185}
  \mu_a+\mu_{\bar{a}} &= \mu_i + \mu_j = 0 \, , \nonumber \\
  \Rightarrow \mu_a &= -\mu_{\bar{a}} \, .
\end{align}

In order to derive the corresponding Boltzmann equation, we introduce
a chemical potential $\mu_h$ for the lepton mode $\ell_h$. For
simplicity, we approximate the distribution with Maxwell-Boltzmann
statistics, an approximation which is sufficient to derive the final
Boltzmann equations. The distribution functions in kinetic equilibrium
are
\begin{align}
  \label{eq:171}
  f_{\ell h}(k)&=\rme^{-\beta (\omega_h - \mu_h)} \, , \nonumber \\
  f_{\barell h}(k)&=\rme^{-\beta (\omega_h + \mu_h)} \, , \nonumber \\
  f_{\ell h}(k)-f_{\barell h}(k) &= \rme^{-\beta \omega_h}
  (\rme^{\beta \mu_h} - \rme^{-\beta \mu_h}) \approx 2 \beta \mu_h
  f_{\ell h}^{\rmeq} \, ,
\end{align}
for $\mu_h \ll T$. We assume chemical equilibrium between the plus-
and the minus-mode,
\begin{align}
  \label{eq:188}
  \mu_+ = \mu_- \equiv \mu_\ell \, .
\end{align}
Moreover, we can make the approximation that the equilibrium densities
are about the same since the thermal mass $m_\ell \approx 0.2 \, T$ is
too small to affect the momentum integration considerably in
\begin{align}
  \label{eq:189}
  n_{\ell_h}^\rmeq &= \int \frac{\rmd^3 k}{(2 \pi)^3}
  f_{\ell_h}^\rmeq(k) \, , \nonumber \\ 
\Rightarrow n_{\ell_+}^\rmeq
  &\approx n_{\ell_-}^\rmeq \approx n_{\ell, 0}^\rmeq \, ,
\end{align}
where $n_{\ell,0}^\rmeq$ is the distribution for massless leptons. With these approximations, we have
\begin{align}
  \label{eq:192}
  n_{\mathcal{L}_+} &= 2 \beta \mu_\ell n_{\ell_0}^\rmeq = n_{\mathcal{L}_-} \, , \nonumber \\
  n_{\mathcal{L}  \pm} &\equiv n_{\mathcal{L}_+} +
  n_{\mathcal{L}_-} \equiv 2 n_{\mathcal{L}_h}\, 
, \nonumber \\
  x_{\mathcal{L}  \pm} &\equiv \frac{n_{\mathcal{L} 
      \pm}}{n_{\ell 0}^\rmeq} \, ,
\end{align}
where the subscript $\pm$ indicates that we sum over the two modes,
contrary to its use in the previous sections.  We can now add the
Boltzmann equations for the two modes in equations~\eqref{eq:74} and
\eqref{eq:78} and arrive at
\begin{align}
  \label{eq:191}
       \frac{\rmd Y_{\mathcal{L} \pm}}{\rmd z} = - \frac{z}{s H_1} \left[
     \Delta \gamma_{ \pm} \left( x_N-1 \right) +
     \frac{x_{\mathcal{L} \pm}}{4} \left(
       \gamma_{W  \pm} + x_N \g_{N  \pm} \right) \right],
\end{align}
where 
\begin{align}
  \label{eq:193}
  Y_{\mathcal{L} \pm} &= Y_{\mathcal{L}+} +  Y_{\mathcal{L}-} \, , \nonumber \\
\Delta \gamma_{ \pm} &= \epsilon_{\g +} \g_{\e +} + \epsilon_{\g -} \g_{\e -} \, , \nonumber \\
\g_{W  \pm} &= \g_{W+}+\g_{W-} \, , \nonumber \\
\g_{N  \pm} &= \g_{N+}+\g_{N-} \, .
\end{align}
The factor $1/4$ comes from the fact that $x_{\mathcal{L}h} =
x_{\mathcal{L} \pm}/2$. Depending on the temperature regime, we either
have to employ the Higgs boson or the neutrino rates in the Boltzmann
equations.

\subsection{One-Mode Approximation}
\label{sec:one-mode-appr}

As we did in section~\ref{sec:hard-thermal-loop} for the decay rates
and the $\CP$ asymmetries, we also employ the one-mode approach for
the Boltzmann equations. The equations are derived in analogy to
sections~\ref{sec:low-temperature} and~\ref{sec:high-temperature} and
read
\begin{align}
  \label{eq:194}
  \frac{\rmd Y_{N}}{\rmd z}&= - \frac{z}{s H_1} (x_N-1)
  \gamma_{D m_\ell}, \nonumber \\
  \frac{\rmd Y_{\mathcal{L}}}{\rmd z} &= - \frac{z}{s H_1} \left[ - \D
    \g_{m_\ell} \left( x_{N}-1 \right) + \frac{x_{\mathcal{L}}}{2}
    \left( \gamma_{W m_\ell} + x_N \g_{N m_\ell} \right) \right],
\end{align}
where $\gamma_{D m_\ell}$, $\gamma_{W m_\ell}$, $\gamma_{N m_\ell}$
and $\Delta \gamma_{m_\ell}$ are the same as the rates defined in
equations~\eqref{eq:b17}, \eqref{eq:b35}, \eqref{eq:b58} and
\eqref{eq:65} and one has to make the appropriate replacements for the
matrix elements and the lepton dispersion relations of the one-mode
approach for $m_\ell$ and $\sqrt{2} \, m_\ell$.

\subsection{Evaluation of the Boltzmann Equations}
\label{sec:eval-boltzm-equat}

We solve the Boltzmann equations for five different scenarios:
\begin{enumerate}
\item
 the
zero temperature case with Maxwell-Boltzmann statistics,
\item the
two-lepton-mode approach where the two modes do not interact with each
other, 
\item the two-mode approach where the modes couple strongly to each
  other,
\item the one-mode approach for a thermal mass $m_\ell$,
\item and the one-mode approach for an asymptotic thermal mass
  $\sqrt{2} \, m_\ell$. 
\end{enumerate}
In the decoupled case, the lepton asymmetries for the plus- and the
minus-mode evolve separately from each other. When solving the
equations, one has to specify the initial conditions for the neutrino
abundance and the lepton asymmetry. We assume a vanishing initial
lepton asymmetry and distinguish between three cases for the neutrino
abundances:
\begin{enumerate}
\item Zero initial abundance: this is the case, for example, when an
  inflaton field decays mostly into SM particles and not into the
  heavy neutrinos.
\item Thermal initial abundance: this can be realised when some
  additional interactions keep the neutrinos in equilibrium at $T \gg
  M_1$, for example via a heavy $Z'$ boson related to $SO(10)$
  unification~\cite{Plumacher:1996kc}.
\item Dominant initial abundance: this is the case, for example, when
  an inflaton decays predominantly into $N_1$.
\end{enumerate}
The coupling $(\lambda^\dagger \lambda)_{11}$, which enters the
neutrino decay rate, is parameterised by the so-called decay parameter
$K$, defined as
\begin{align}
  \label{eq:195}
  K \equiv \frac{\widetilde{m}_1}{m^*} \, ,
\end{align}
where 
\begin{align}
  \label{eq:11}
  \widetilde{m}_1 = \frac{(\l^\dagger \l)_{11} v^2}{M_1}
\end{align}
is the conveniently defined effective neutrino mass, which is of the
order of the light neutrino mass scale, and 
\begin{align}
  \label{eq:13}
  m^* =  \left. 8 \pi \frac{v^2}{M_1^2} H \right |_{T=M_1} \simeq 1.1 \times 10^{-3} \,{\rm eV}\,
\end{align}
is called equilibrium neutrino mass. In the language of these masses,
the out-of-equilibrium condition, $\G < H$, corresponds to $K>1$ and
is called strong washout regime. The case $K<1$ is called weak washout regime.

We want to analyse the evolution of the neutrino abundance and lepton
asymmetries for the weak and strong washout regimes and different
initial abundances. To this end, we write the Boltzmann equations for
the different scenarios in the form of equations~\eqref{eq:b19},
\eqref{eq:b36}, \eqref{eq:b60} and \eqref{eq:67},
\begin{align}
  \label{eq:168}
  \frac{\rmd Y_N}{\rmd z} &= - D (Y_N - Y_N^\rmeq) \, . \nonumber \\
\frac{\rmd Y_{\mathcal{L}}}{\rmd z} &= \epsilon_0 D_\epsilon (Y_N - Y_N^\rmeq) -
 (W + W_N x_N) Y_{\mathcal{L}} \, ,
\end{align}
where
\begin{align}
  \label{eq:198}
  D &= \frac{z}{H_1} \frac{\g_D}{s Y_N^\rmeq} \, ,
  & D_\epsilon &= \frac{z}{H_1} \frac{\D \g}{\epsilon_0 s Y_N^\rmeq}\, , \nonumber \\
  W & = \frac{z}{H_1} \frac{\g_W}{2 s Y_{\mathcal{L}}^\rmeq} \, ,
  \quad & W_N & = \frac{z}{H_1} \frac{\g_N}{2 s Y_{\mathcal{L}}^\rmeq}
  \, .
\end{align}
One usually refers to $D_\e (Y_N-Y_N^\rmeq)$ as source term since this
term is responsible for the production of a lepton asymmetry. The term
$(W+W_N x_N) Y_\ml^\rmeq$ is called washout term since it usually has
the opposite sign as the source term and reduces the production of the
lepton asymmetry. The terms $D$, $D_\e$, $W$ and $W_N$ are different
for the different scenarios. Note that for the finite temperature
cases, $D_\epsilon$ is not the same as $D$ and there is an additional
washout term $W_N$ due to the quantum statistics. Our analysis closely
follows the arguments and explanations in
reference~\cite{Buchmuller:2004nz} and the interested reader will find
a comprehensive explanation of leptogenesis dynamics in the vacuum
case therein.

\subsubsection*{Weak washout for zero initial abundance}
\label{sec:weak-washout}

\begin{figure}
  \centering
  \includegraphics[width=\textwidth]{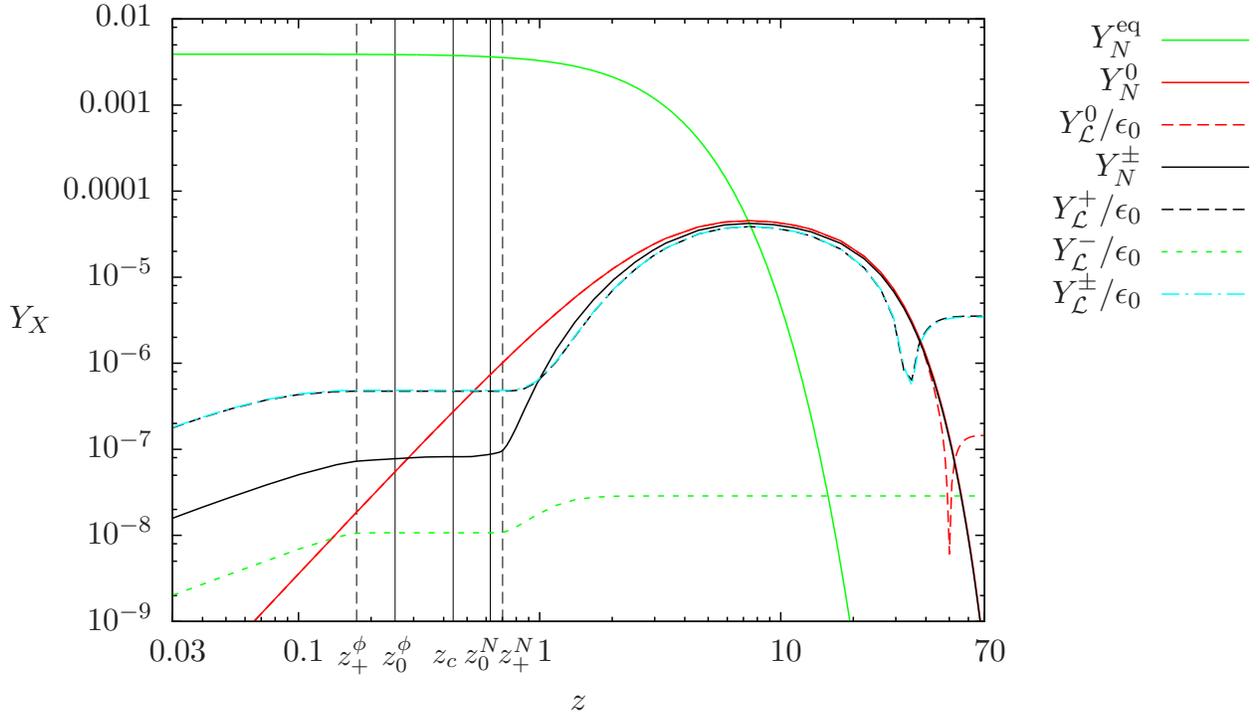}
  \caption[$Y_N(z)$ and $Y_\ml(z)$ for weak washout, zero initial
  $N$ abundance and two-mode cases.]{Evolution of neutrino abundance $Y_N(z)$ and
    lepton asymmetry $Y_\ml(z)$ for $K=0.005$ and zero initial
    neutrino abundance. We show the the two-mode cases and the vacuum
    case.}
  \label{fig:nell2k0.005ztd0}
\end{figure}

\begin{figure}
  \centering
  \includegraphics[width=\textwidth]{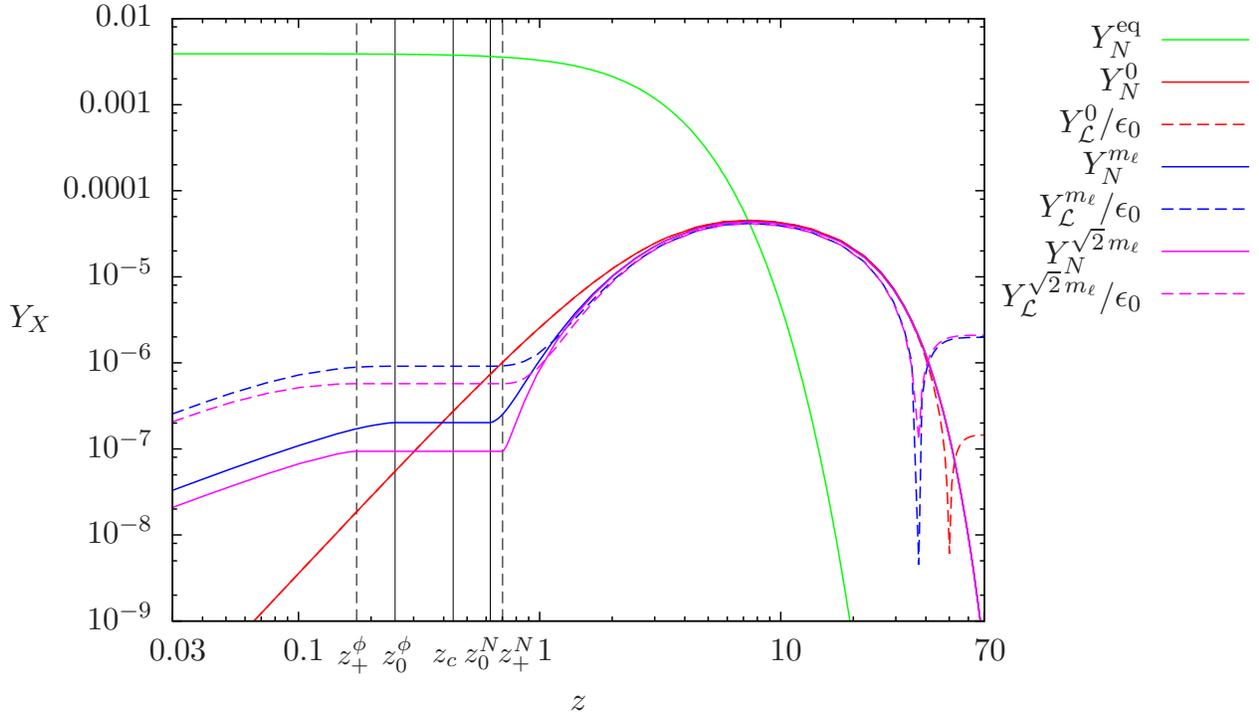}
  \caption[Same as above; weak washout, zero $N$ abundance,
  one-mode cases]{Evolution of neutrino density and lepton asymmetry
    for $K=0.005$ and zero initial neutrino abundance. We show the
    one-mode cases and the vacuum case.}
  \label{fig:nell1k0.005ztd0}
\end{figure}

Let us start with the weak washout regime and zero initial
abundance.  We define a value $z_\rmeq$ by the condition
\begin{align}
  \label{eq:200}
  Y_N(z_\rmeq) = Y_N^\rmeq(z_\rmeq) \, .
\end{align}
For $z \ll 1$, the neutrino abundance is negligible compared to
$Y_N^\rmeq$,
\begin{align}
  \label{eq:199}
  \frac{\rmd Y_N}{\rmd z} \simeq D Y_N^\rmeq \, ,
\end{align}
where $Y_N^\rmeq$ is approximately constant for $z \ll 1$. The entropy
density $s$ is proportional to $z^{-3}$ and $\gamma_D$ is proportional
to $z^{-2}$ in vacuum and $z^{-4}$ for the Higgs boson decays at high
temperature in the finite temperature cases. Thus $D \sim z^2$ in the
vacuum case and $D \sim {\rm const.}$ at finite
temperature. Neglecting $Y_{N}^{\rm initial}$ and $z^{\rm initial}$,
the integration yields $Y_N \simeq z D(z)/3 \sim z^3 $ for the vacuum
case and $Y_N \simeq z D \sim z$ for the finite temperature cases. We
show the numerical results for $K=0.005$ and zero initial abundance in
figures~\ref{fig:nell2k0.005ztd0} and~\ref{fig:nell1k0.005ztd0}, where
these power laws for $Y_N(z)$ can be observed for $z \lesssim 0.1$.
We also see that $Y_N^{m_\ell} > Y_N^{\sqrt{2} \, m_\ell} > Y_N^\pm
\gg Y_N^0$, which reflects $\gamma_D^{m_\ell} > \gamma_D^{\sqrt{2} \,
  m_\ell} > \g_D^\pm \gg \g_0$. Between the thresholds $z_+^\phi$ and
$z_+^N$, the finite-temperature abundances do not evolve much, which
reflects that the decay rates are very low or vanishing in this
regime. The neutrino abundance for the asymptotic mass $\sqrt{2} \,
m_\ell$ does not rise at all at high temperature, while $Y_N^{m_\ell}$
rises slightly between $z_0^{\phi/N}$ and $z_+^{\phi/ N}$, where the
rate is non-zero. The two-mode rate $\g_\pm$ is, though very
suppressed, present over the whole threshold range between $z_+^\phi$
and $z_+^N$ due to the minus-modes, so $Y_N^\pm$ rises slightly. At
low temperature, $z > z_\rmeq$, the neutrino abundances of the
different scenarios are very close to each other, since for $z \gtrsim
2$, the rates are very close to the vacuum rate, $\gamma_{D,W,N}^{T>0}
\simeq \D \g^{T>0} / \e_0 \simeq \g_0$. In this regime, $Y_N$ is much
larger than $Y_N^\rmeq$ since the coupling is too small to keep the
abundance close to equilibrium.

Having outlined the evolution of the neutrino abundance at high
temperature, we can understand the evolution of the lepton asymmetry
at high temperature. The washout term in the
Boltzmann equations proportional to $Y_\ml$ is much smaller than the
source term $D_\e (Y_N-Y_N^\rmeq)$, since $Y_\ml/\e_0 \ll Y_N^\rmeq$,
and we have for $z \ll z_\rmeq$,
\begin{align}
  \label{eq:203}
  Y_\ml^{T>0} &\simeq - z \e_0 D_\e \, \nonumber \\
  Y_\ml^{T=0} &\simeq - z \e_0 \frac{D_\e(z)}{3} \, .
\end{align}
In the high temperature regime, the lepton asymmetry is negative and
follows the neutrino abundance in its absolute value,
\begin{align}
  \label{eq:204}
  Y_\ml(z) \simeq - \e_0 \frac{D_\e}{D} Y_N(z) = -\frac{\D \g}{\g_D}
  Y_N(z) \, .
\end{align}
For the vacuum case, $D_\e \equiv D$ and $Y_\ml/\e_0\simeq -Y_N$,
while for the finite temperature cases, $D_\e/D = \D \g/(\e_0 \g_D)
\sim \mathcal{O}(10^1)$.\footnote{We saw in section~\ref{sec:cp-asymm}
  that $\D \g / (\e_0 \g_D) \sim \mathcal{O}(10^2)$ at high
  temperature instead of $\mathcal{O}(10^1)$. The discrepancy is due
  to rates and asymmetries that occur in the Boltzmann equations and
  that are slightly different from the usual rates and
  $\CP$-asymmetries in section~\ref{sec:hard-thermal-loop} due to the
  different statistical factors they employ, for example $1$ in
  equation~\eqref{eq:b58} or $(1-2 f_N^\rmeq)$ in
  equation~\eqref{eq:65} instead of the usual factor $(1-f_N^\rmeq)$
  in equation~\eqref{eq:7}.}  Both behaviours can be observed in
figures~\ref{fig:nell2k0.005ztd0} and~\ref{fig:nell1k0.005ztd0}. The
lepton asymmetry in the minus-mode obeys the same power law as the
other finite-temperature modes, but is about a factor $100$ lower due
to the lower rates. The combined ($\pm$)-abundance closely follows the
plus-abundance since the influence of the minus-mode rates is very
suppressed and also the different washout term with factor $1/4$
instead of $1/2$ can be neglected.

Before turning to the intermediate temperatures $z \sim z_\rmeq$, let
us discuss the low temperature regime. For $z > z_\rmeq$, $Y_N \gg
Y_N^\rmeq$, so the source term dominates and washout can be neglected.
Since in this regime, $D_\e^{T>0} \simeq D^{T>0} \simeq D^{T=0}$, we
can write
\begin{align}
  \label{eq:205}
  \frac{\rmd Y_\ml}{\rmd z} \simeq \e_0D (Y_N- Y_N^\rmeq) = - \e
  \frac{\rmd Y_N}{\rmd z} \, .
\end{align}
To first order, the negative lepton asymmetry created below $z_\rmeq$
and the positive contribution from above $z_\rmeq$ have the same
magnitude and cancel each other. For the remaining asymmetry that did
not cancel, the washout contribution up to $z_\rmeq$ and the exact
behaviour of the abundances around $z_\rmeq$ are crucial.

Assuming that $Y_N(z=\infty) = 0$, we get 
\begin{align}
  \label{eq:206}
  Y_\ml^{\rm fin} \simeq \e_0 Y_N(z_\rmeq) - |Y_\ml(z_\rmeq)| \, ,
\end{align}
so we see that the evolution of the difference $\D Y(z) \equiv Y_N(z)
- |Y_\ml(z)|/\e_0$ below $z_\rmeq$ is crucial for the final lepton
asymmetry. For the regime $1 \lesssim z \lesssim z_\rmeq$, we can
write
\begin{align}
  \label{eq:207}
  \frac{\rmd \D Y}{\rmd z} &\simeq Y_N^\rmeq (D -D_\e) \, , \nonumber
  \\
  D-D_\e &= \frac{z}{H_1} \frac{1}{s Y_N^\rmeq} \left(\g_D-\frac{\D
      \g}{\e}\right) \, .
\end{align}
For the finite-temperature cases, the $\CP$-asymmetry in the decay
rates, $\D \g/\e_0$, is considerably smaller than the decay rate
$\g_D$ in the range $z_0^N \lesssim z \lesssim 2$, which can be seen
in figure~\ref{fig:cpn0}. Moreover, for the one-mode cases, $\D
\g_m/\e_0$ approaches $\g_m$ faster than $\D \g_{++}$ approaches
$\g_+$ for the plus-mode. Above $z_+^N$, the ratio $\D \g/(\e_0 \g_D)$
for the two one-mode cases is about the same. Thus, the difference $\D
D \equiv D-D_\e$ is largest for the two-mode approach, smaller and
about the same for the two one-mode approaches and vanishing for the
vacuum approach, $\D D^+ > \D D^{m_\ell} \simeq \D D^{\sqrt{2} \,
  m_\ell}$ and $\D D^0=0$. As a result, $\D Y^+ \gtrsim \D Y^{m_\ell}
\simeq \D Y^{\sqrt{2} \, m_\ell}$ at $z_\rmeq$ and therefore the final
asymmetries are related as $Y_\ml^+ > Y_\ml^{m_\ell} \simeq
Y_\ml^{\sqrt{2}\, m_\ell} \gg Y_\ml^0$. Note that the final asymmetry
is non-vanishing for the vacuum case, since the washout at higher
temperature is larger than at lower temperature due to the larger
decay rate. For the finite temperature cases, the difference $D-D_\e$
in the crucial regime $z \simeq z_\rmeq$ governs the final asymmetry.

The evolution of the decoupled minus-mode at low temperature is very
different but not hard to understand. The asymmetry rises at $z
\gtrsim z_+^N$, because this is the regime above $z_c$ where $\Delta
\g_{-+}$ is maximal and therefore $D_\e^-$ is maximal as well. Between
the thresholds $z_+^\phi$ and $z_+^N$, $\D \g_{-+}$ is suppressed by
the internal plus-lepton and $\D \g_{--}$ is suppressed by the residue
of the internal minus-lepton, so $D_\e$ is negligible and $Y_\ml^-$
does not rise. Above $z \sim 1$, $\D \g_{-+}$ falls due to the residue
of the external minus-mode and $Y_\ml^-$ does not change. The final
lepton asymmetry therefore does not change its sign above $z \gtrsim
z_\rmeq$ and keeps the value it achieves at around $1 \lesssim z
\lesssim 2$ where the $\CP$-asymmetry $\D \g_{-+}$ becomes small.

\begin{figure}
  \centering
  \includegraphics[width=\textwidth]{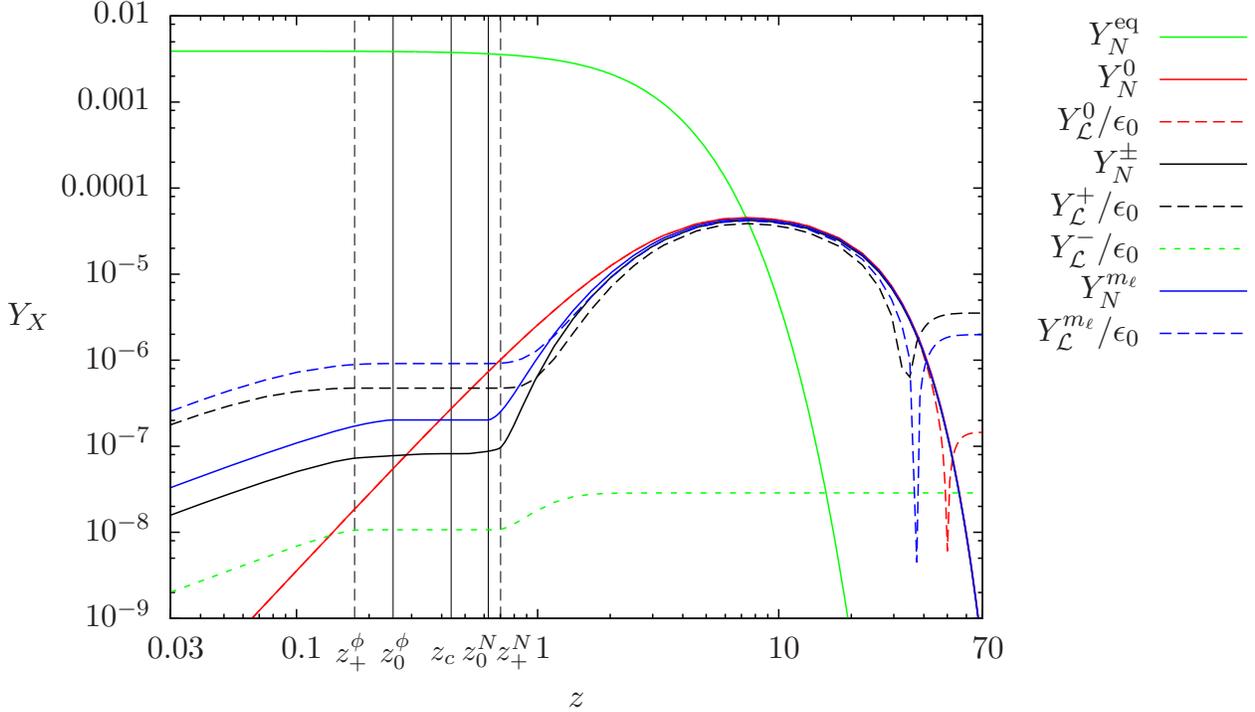}
  \caption[Same as above; weak washout, zero $N$ abundance, ($+$)-,
  ($-$)-, $m_\ell$-cases]{Evolution of neutrino density and lepton
    asymmetry for $K=0.005$ and zero initial neutrino abundance. We
    display the four modes from figures~\ref{fig:nell2k0.005ztd0}
    and~\ref{fig:nell1k0.005ztd0} that give different final lepton
    asymmetries, that is, the plus-mode the minus-mode, the
    $m_\ell$-mode and the vacuum case.}
  \label{fig:nellpmmlk0.005ztd0}
\end{figure}

The combined ($\pm$)-mode does not evolve differently from the
plus-mode since the influence from the $\g_-$ rates can be neglected
and also the washout term with the additional factor $1/2$ is not
noticeable since washout is very small in all temperature
regimes. Summarising, there are four differing lepton asymmetries in
this regime: the vacuum case, the $m_\ell$-case, the ($+$)-case and
the ($-$)-case. The $\sqrt{2} \, m_\ell$-case yields the same
asymmetry as the $m_\ell$-case and the ($\pm$)-case yields the same
asymmetry as the ($+$)-case. We show the four differing lepton
asymmetries together in figure~\ref{fig:nellpmmlk0.005ztd0}.

\subsubsection*{Strong and intermediate washout for zero initial abundance}
\label{sec:strong-washout}

\begin{figure}
  \centering
  \includegraphics[width=\textwidth]{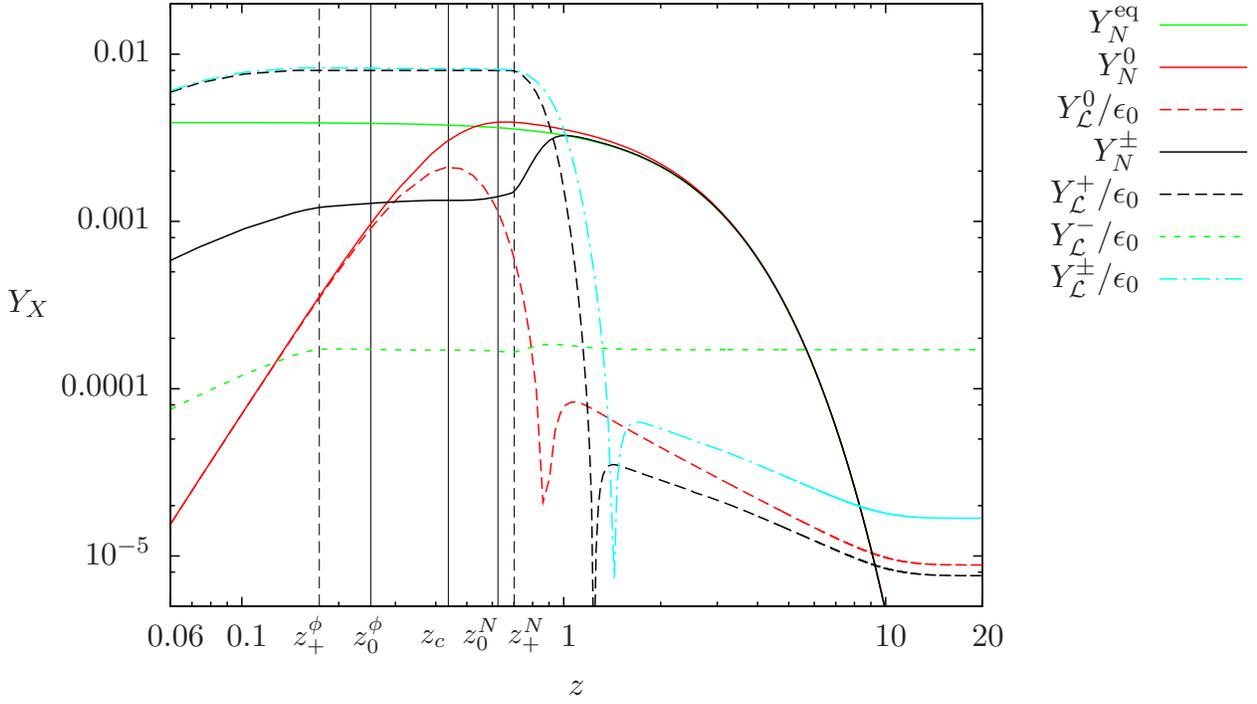}
  \caption [Same as above; strong washout, zero $N$ abundance, two-mode
  cases]{Evolution of neutrino abundance $Y_N(z)$ and lepton asymmetry
    $Y_\ml(z)$ for $K=100$ and zero initial neutrino abundance. We
    show the two-mode cases and the vacuum case.}
  \label{fig:nell2k100.000ztd0}
\end{figure}
\begin{figure}
  \centering
  \includegraphics[width=\textwidth]{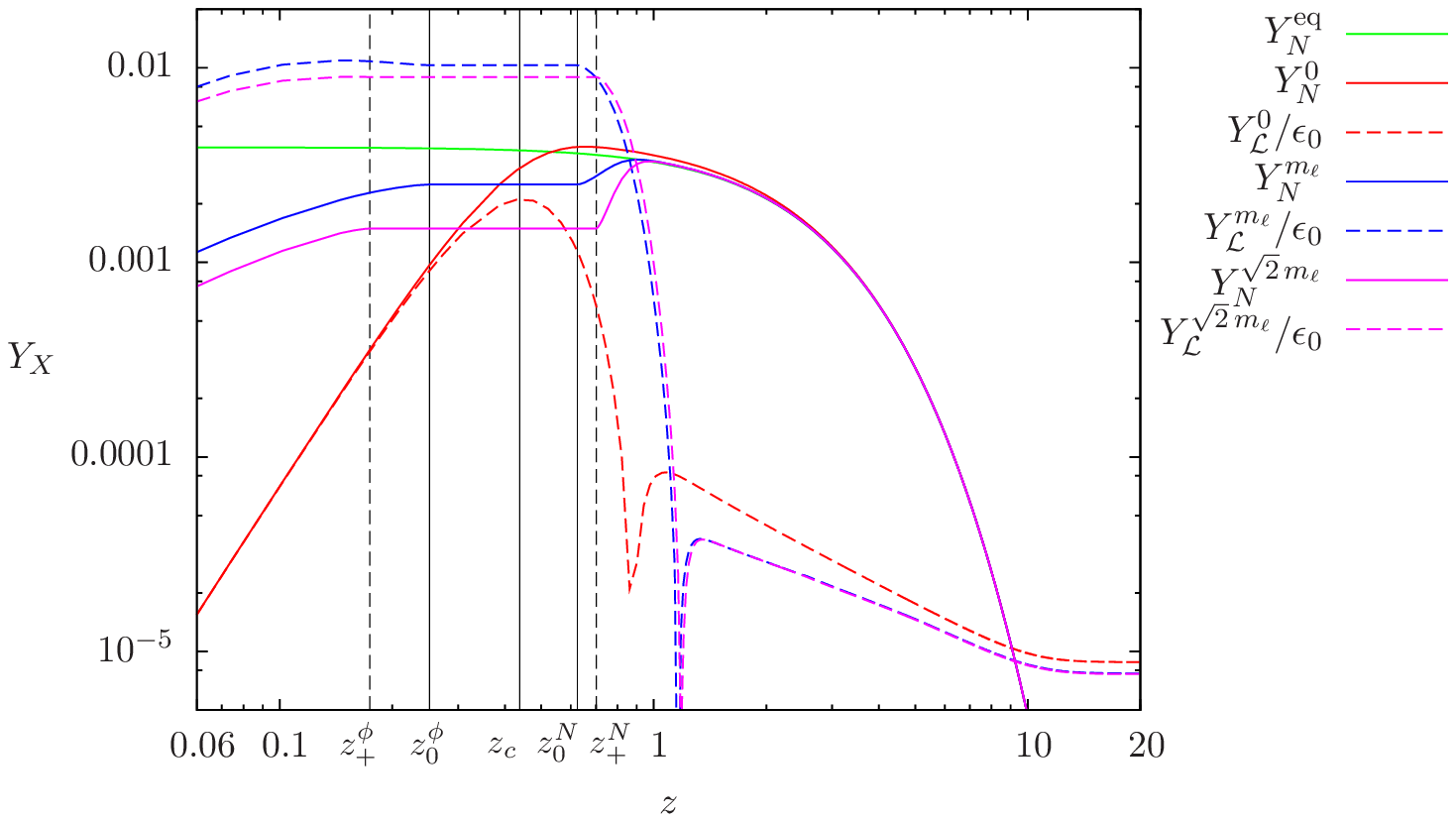}
  \caption[Same as above; strong washout, zero $N$ abundance, one-mode
  cases]{Evolution of neutrino abundance $Y_N(z)$ and
    lepton asymmetry $Y_\ml(z)$ for $K=100$ and zero initial
    neutrino abundance. We show the one-mode cases and the vacuum
    case.}
  \label{fig:nell1k100.000ztd0}
\end{figure}
For strong washout, the evolution of the neutrino abundance is
analogous to the weak washout regime, with $Y_N^T \sim z$ and $Y_N^0 \sim
z^3$, as shown in figures~\ref{fig:nell2k100.000ztd0} and~\ref{fig:nell1k100.000ztd0}. 
The couplings are stronger, therefore the abundances rise faster and
meet $Y_N^\rmeq$ earlier at $z_\rmeq \sim 1$. For larger $z$, the
couplings are strong enough to keep $Y_N$ close to equilibrium. The
evolution of the lepton asymmetry is nicely explained in
reference~\cite{Buchmuller:2004nz} for the vacuum case with some
rather accurate analytical approximations. In this work, we are only
interested in the difference of the vacuum case to the finite
temperature case. In the strong washout regime, the lepton asymmetries
rise rather fast and the washout term, which competes with the source
term, becomes larger than the latter at some temperature $z_{\rm
  min}$, where the lepton asymmetry reaches its most negative
value. The source term becomes small when $Y_N$ approaches its
equilibrium value, so the washout term drives the asymmetry evolution
back to zero. At $z \gtrsim z_\rmeq$, the neutrino abundance slightly
overshoots $Y_N^\rmeq$, so the source term changes sign and adds to
the washout term until the lepton asymmetry becomes positive, where
the washout term changes its sign as well and is competing again. At
low temperature, source term and washout term have the same magnitude
when the lepton asymmetry reaches a maximum at $z_{\rm max}$. Above
$z_{\rm max}$, the lepton asymmetry is again driven to zero by the
larger washout. At very low temperature, washout and source term
become very small and do not influence the asymmetry further, which
settles at a final value $Y_\ml^{\rm fin}$. We see that in the strong
washout regime, the dynamics are governed by the washout term. The
evolution of the finite temperature lepton asymmetries is analogous to
the vacuum case, but they settle to a different final value. For the
finite temperature cases, the equilibrium density of leptons is
smaller than in the vacuum case due to the thermal mass $m_\ell \sim
0.2 \, T$. The washout term is effectively larger than for the vacuum
case and competes with the source term in a stronger way. Therefore,
the asymmetry evolution appears slightly damped compared to the vacuum
case and the final asymmetry is marginally lower. The evolution of the
minus-mode is analogous to the evolution at weak washout, rises fast
below $z_+^\phi$ and does not change above the thresholds since $Y_N
\sim Y_N^\rmeq$ in this regime. The combined ($\pm$)-mode tracks the
plus-mode until washout becomes relevant at $z \gtrsim z_\rmeq$. The
washout term for the ($\pm$)-mode is always about a factor two smaller
than for the plus-mode, since we add the minus- and plus-washout
rates, where the minus-rate is always negligible compared to the
plus-rate. Thus, the ($\pm$)-abundance is less affected by washout,
so the dynamics are affected by the source term in a stronger way and the
final asymmetry is larger. We can view this behaviour as always
distributing half the asymmetry in a mode $\ell_-$ which couples
strongly to $\ell_+$ and is not affected by washout. The final
asymmetry is about a factor two larger than for the other scenarios in
the strong washout regime.

\begin{figure}
  \centering
  \includegraphics[width=\textwidth]{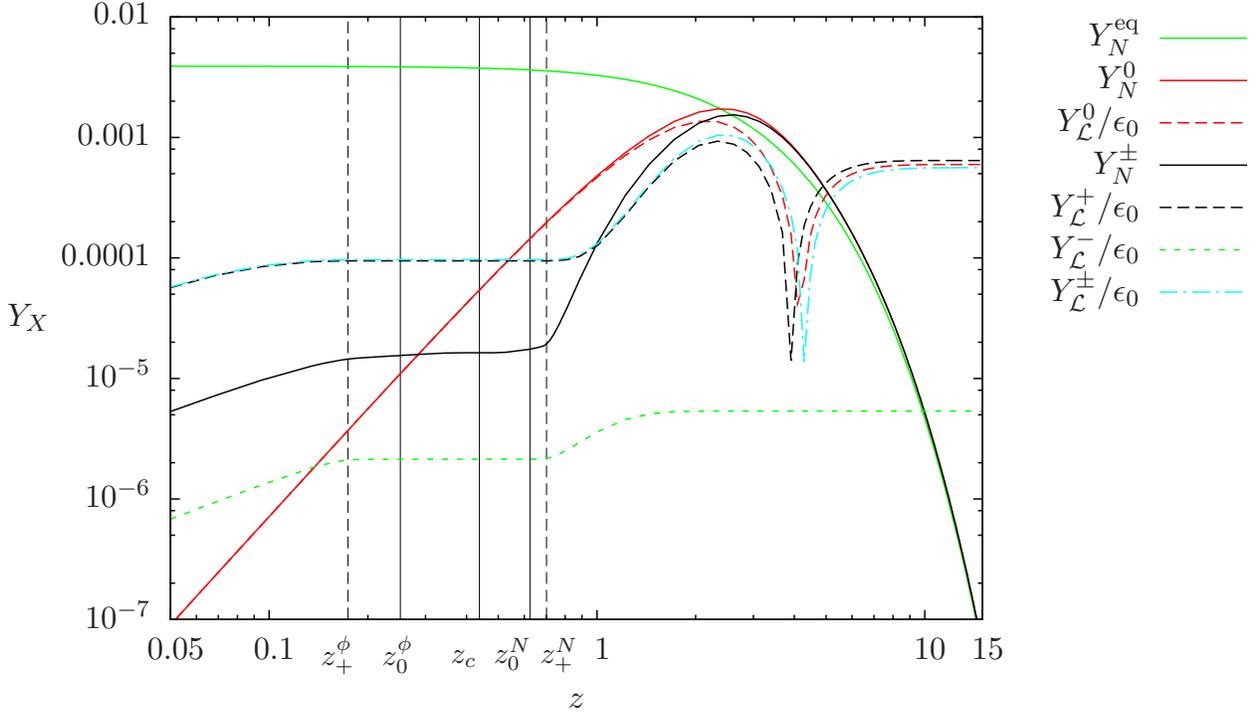}
  \caption[Same as above; intermediate washout, zero $N$ abundance, two-mode
  cases]{Evolution of neutrino abundance $Y_N(z)$ and
    lepton asymmetry $Y_\ml(z)$ for $K=1$ and zero initial
    neutrino abundance. We show the two-mode cases and the vacuum
    case.}
  \label{fig:nell2k1.000ztd0}
\end{figure}

The case of intermediate washout is shown in
figure~\ref{fig:nell2k1.000ztd0}, where we only show the two-mode
cases since in this regime, the final lepton asymmetries of the one-mode cases are the
same as for the plus-mode. The dynamics can be viewed as an
interpolation between the strong and weak washout regimes and the
final asymmetries are very similar to each other.

\subsubsection*{Non-zero initial abundance}
\label{sec:non-zero-initial}

\begin{figure}
  \centering
  \includegraphics[width=\textwidth]{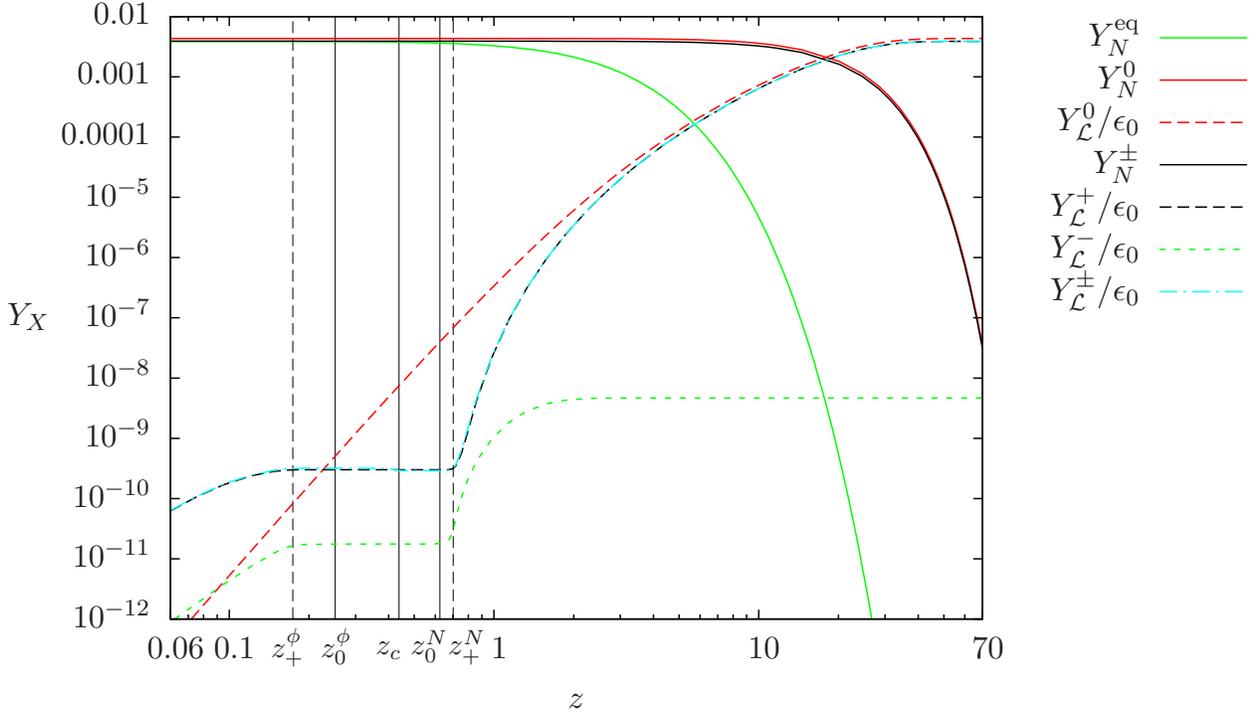}
  \caption[Same as above; weak washout, thermal $N$ abundance, two-mode
  cases]{Evolution of neutrino abundance $Y_N(z)$ and
    lepton asymmetry $Y_\ml(z)$ for $K=0.005$ and thermal initial
    neutrino abundance. We show the two-mode cases and the vacuum
    case.}
  \label{fig:nell2k0.005ztd1}
\end{figure}

\begin{figure}
  \centering
  \includegraphics[width=\textwidth]{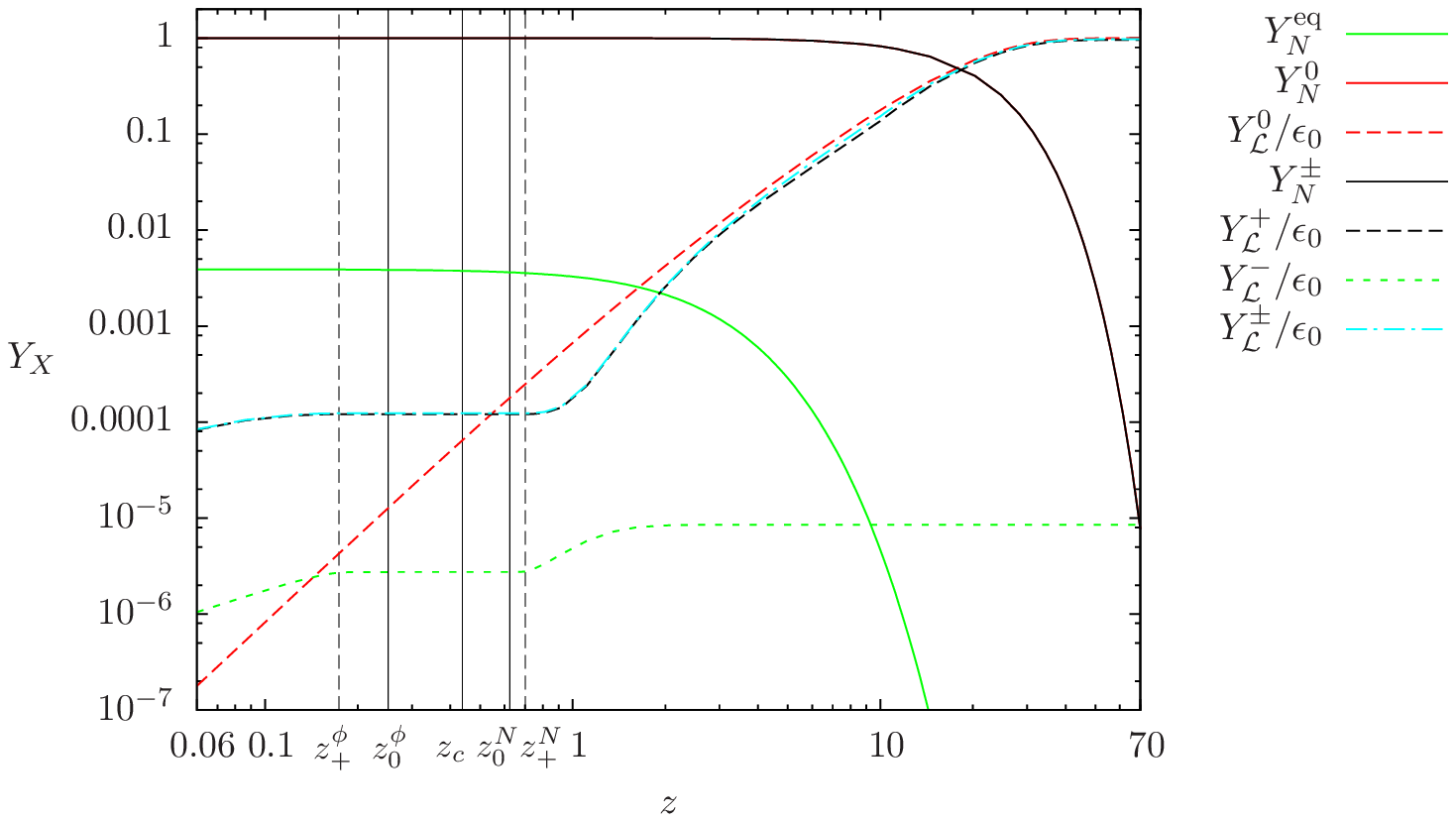}
  \caption[Same as above; weak washout, dominant $N$ abundance, two-mode
  cases]{Evolution of neutrino abundance $Y_N(z)$ and
    lepton asymmetry $Y_\ml(z)$ for $K=0.005$ and dominant initial
    neutrino abundance. We show the two-mode cases and the vacuum
    case.}
  \label{fig:nell2k0.005ztd2}
\end{figure}

\begin{figure}
  \centering
  \includegraphics[width=\textwidth]{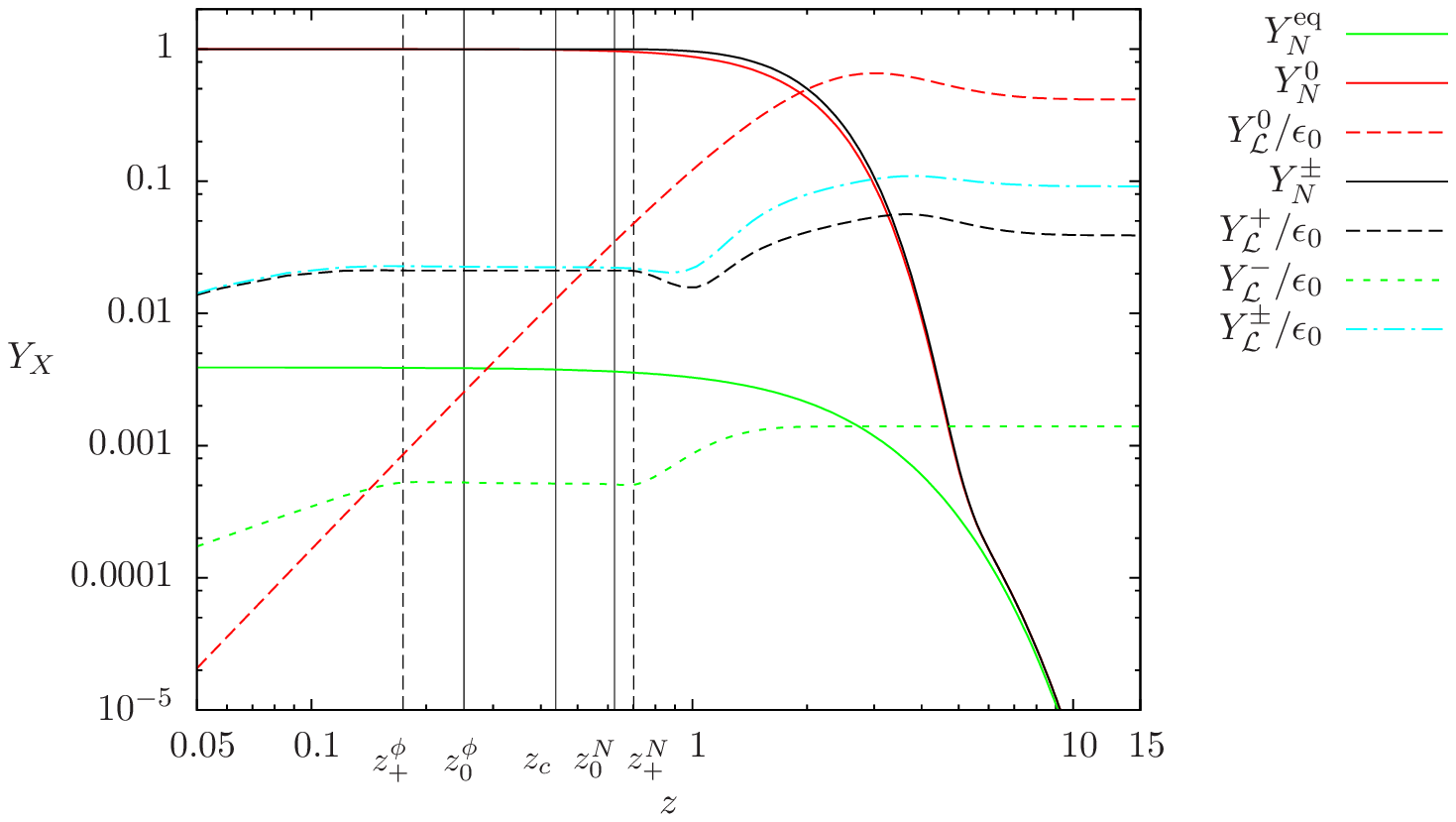}
  \caption[Same as above; intermediate washout, dominant $N$ abundance, two-mode
  cases]{Evolution of neutrino abundance $Y_N(z)$ and
    lepton asymmetry $Y_\ml(z)$ for $K=1$ and dominant initial
    neutrino abundance. We show the two-mode cases and the vacuum
    case.}
  \label{fig:nell2k1.000ztd2}
\end{figure}

\begin{figure}
  \centering
  \includegraphics[width=\textwidth]{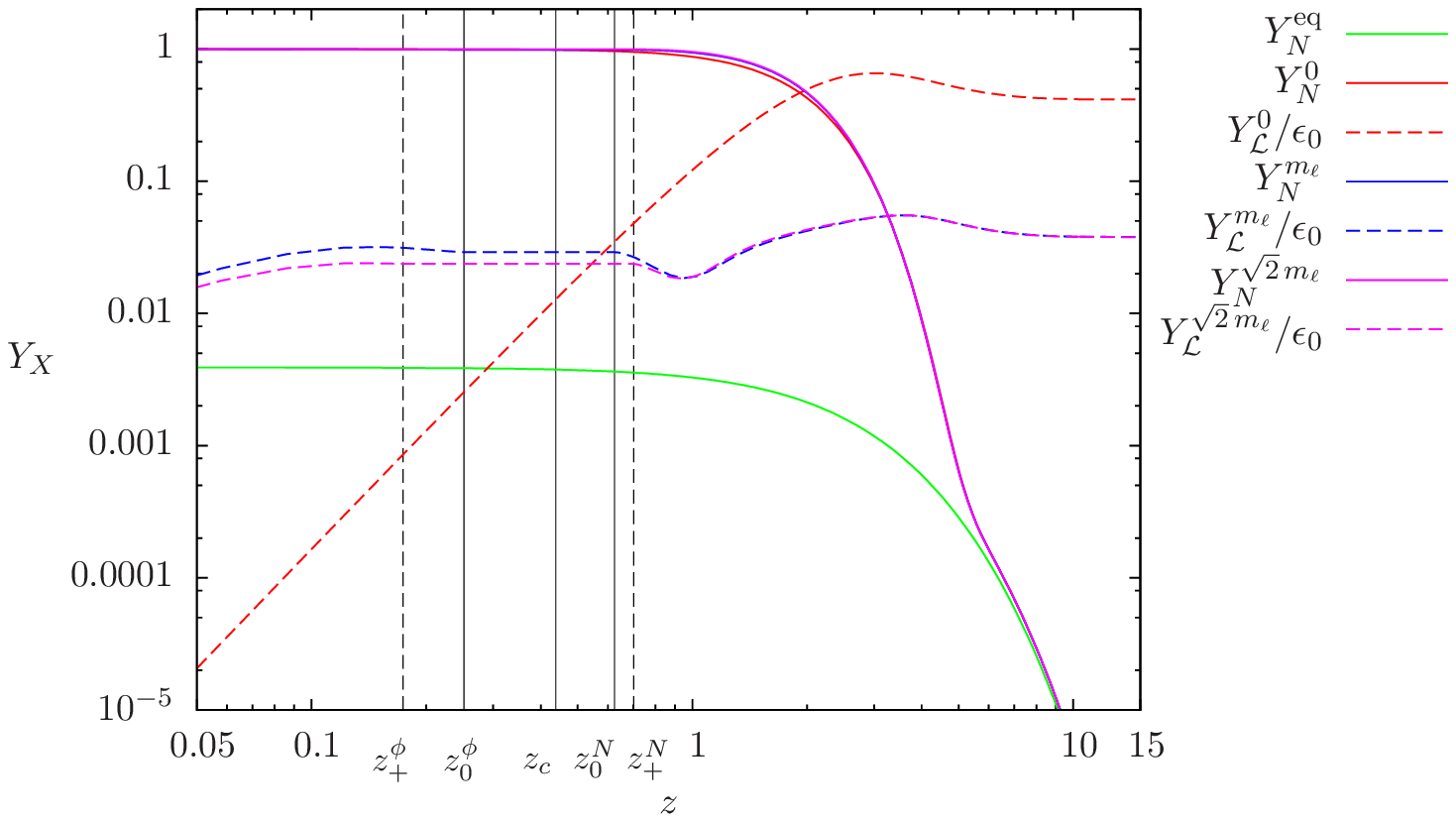}
  \caption[Same as above; intermediate washout, dominant $N$ abundance, one-mode
  cases]{Evolution of neutrino abundance $Y_N(z)$ and
    lepton asymmetry $Y_\ml(z)$ for $K=1$ and dominant initial
    neutrino abundance. We show the one-mode cases and the vacuum
    case.}
  \label{fig:nell1k1.000ztd2}
\end{figure}

\begin{figure}
  \centering
  \includegraphics[width=\textwidth]{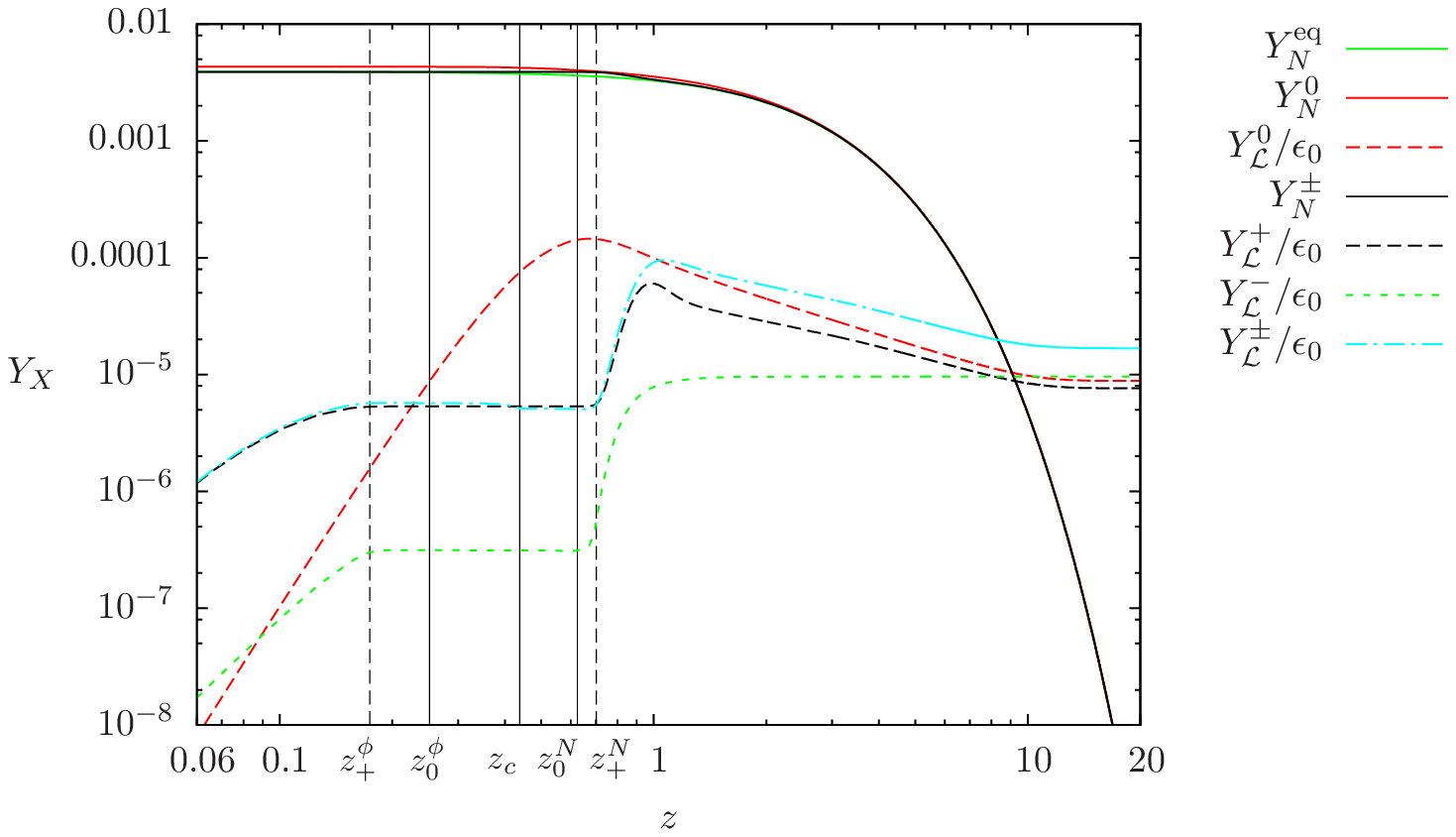}
  \caption[Same as above; strong washout, thermal $N$ abundance, two-mode
  cases]{Evolution of neutrino abundance $Y_N(z)$ and
    lepton asymmetry $Y_\ml(z)$ for $K=100$ and thermal initial
    neutrino abundance. We show the two-mode cases and the vacuum
    case.}
  \label{fig:nell2k100.000ztd1}
\end{figure}

\begin{figure}
  \centering
  \includegraphics[width=\textwidth]{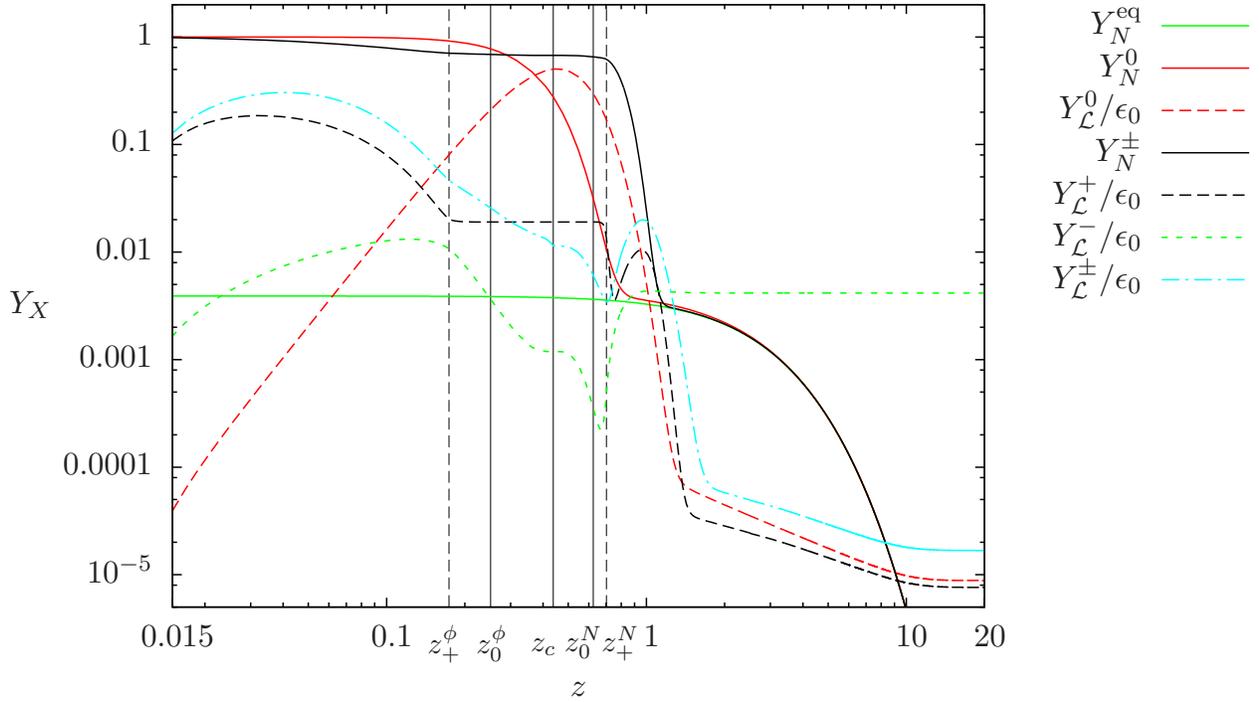}
  \caption[Same as above; strong washout, dominant $N$ abundance, two-mode
  cases]{Evolution of neutrino abundance $Y_N(z)$ and
    lepton asymmetry $Y_\ml(z)$ for $K=100$ and dominant initial
    neutrino abundance. We show the two-mode cases and the vacuum
    case.}
  \label{fig:nell2k100.000ztd2}
\end{figure}

We also present the dynamics for thermal and dominant initial
abundance in
figures~\ref{fig:nell2k0.005ztd1}--\ref{fig:nell2k100.000ztd2}.
For weak washout and thermal initial abundance, $Y_N \gg |Y_\ml|$ for
low temperatures, and according to equation~\eqref{eq:206},
$Y_\ml^{\rm final}/\e_0 \sim Y_N^{\rm initial}$. For weak washout and
dominant initial abundance, this equation holds as well, as can be
seen in figures~\ref{fig:nell2k0.005ztd1}
and~\ref{fig:nell2k0.005ztd2}.  For intermediate washout $K \sim 1$
and dominant abundance, shown in figures~\ref{fig:nell2k1.000ztd2} and
\ref{fig:nell1k1.000ztd2}, the lepton asymmetry production is stopped
between the thresholds for the thermal cases and the production above
$z \sim 1$ does not succed in producing an asymmetry as high as in the
vacuum case.  For the ($\pm$)-case, the asymmetry production is larger
since it is not as much affected by washout.

For the strong washout regime and large initial neutrino abundances,
the dynamics at high temperature are interesting, as shown in
figures~\ref{fig:nell2k100.000ztd1} and~\ref{fig:nell2k100.000ztd2},
but the interplay between source term and washout term at low
temperature governs the final asymmetry as in the zero-abundance
case. We reproduce the well-known fact that the initial conditions do
not influence the final asymmetry in the strong washout regime, while
the arguments concerning the equilibrium distribution of leptons with
thermal mass and the reduced washout of the ($\pm$)-mode still hold
and lead to the same lepton asymmetry as for zero initial
abundance. The decoupled minus-mode is very much affected by the
coupling, that is the decay parameter $K$, and the initial conditions,
since the final lepton asymmetry is produced at high temperatures. The
stronger the coupling, the larger the asymmetry production of the
minus-mode at high temperatures and the larger the final
value. Moreover, the larger the initial deviation of the neutrino
abundance from equilibrium, the larger the asymmetry production and
the final asymmetry. The final lepton asymmetry in this mode is thus
lowest for neutrinos with thermal initial abundance.

\subsubsection*{Final lepton asymmetries}
\label{sec:final-asymmetries}

\begin{figure}
  \centering
  \includegraphics[width=\textwidth]{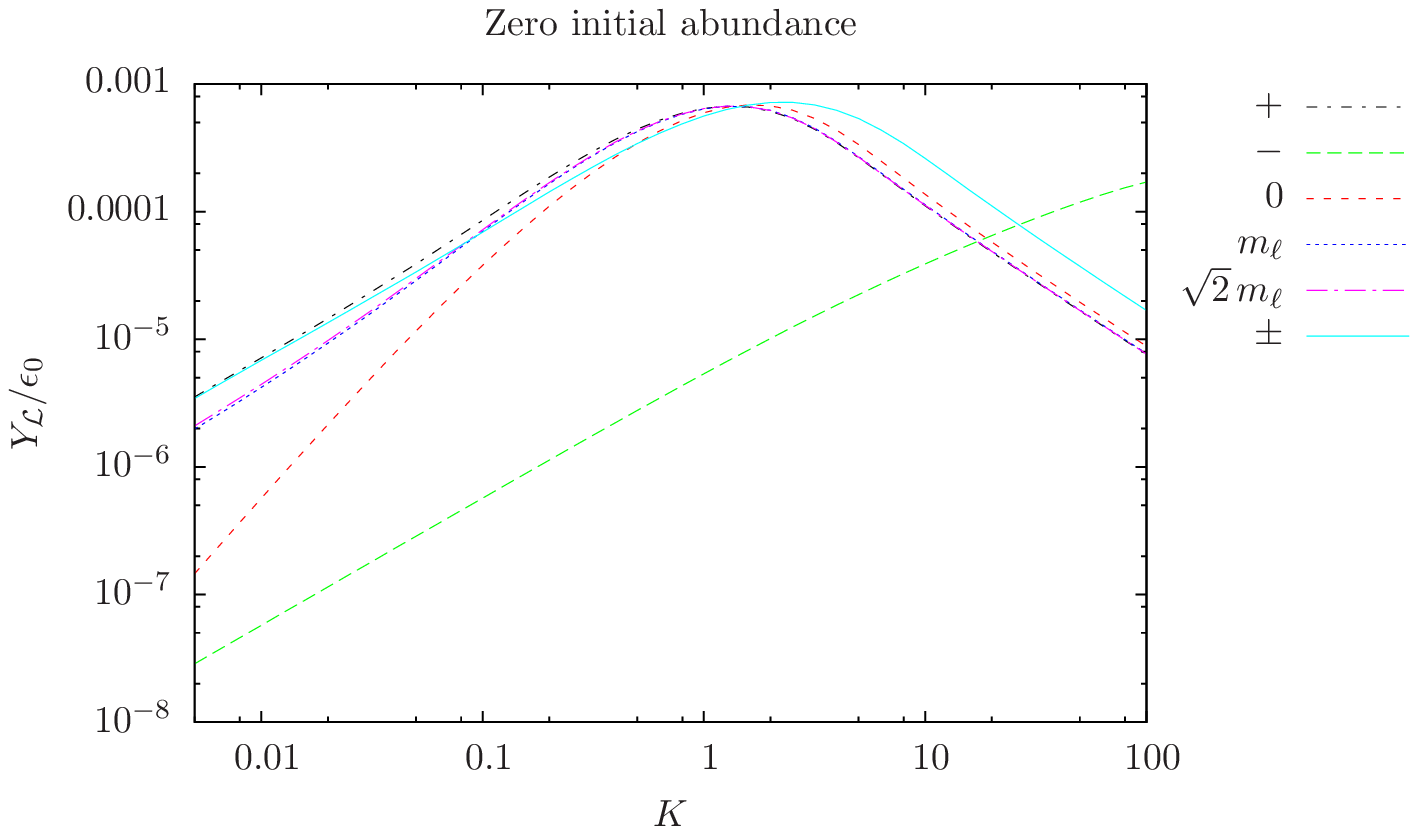}
  \caption[$K$-dependence of $|Y_\ml^{\rm fin}|$ for zero initial
  neutrino abundance]{Final value of the lepton asymmetry for
    different values of $K$ for zero initial neutrino abundance.}
  \label{fig:final0}
\end{figure}

\begin{figure}
  \centering
  \includegraphics[width=0.98\textwidth]{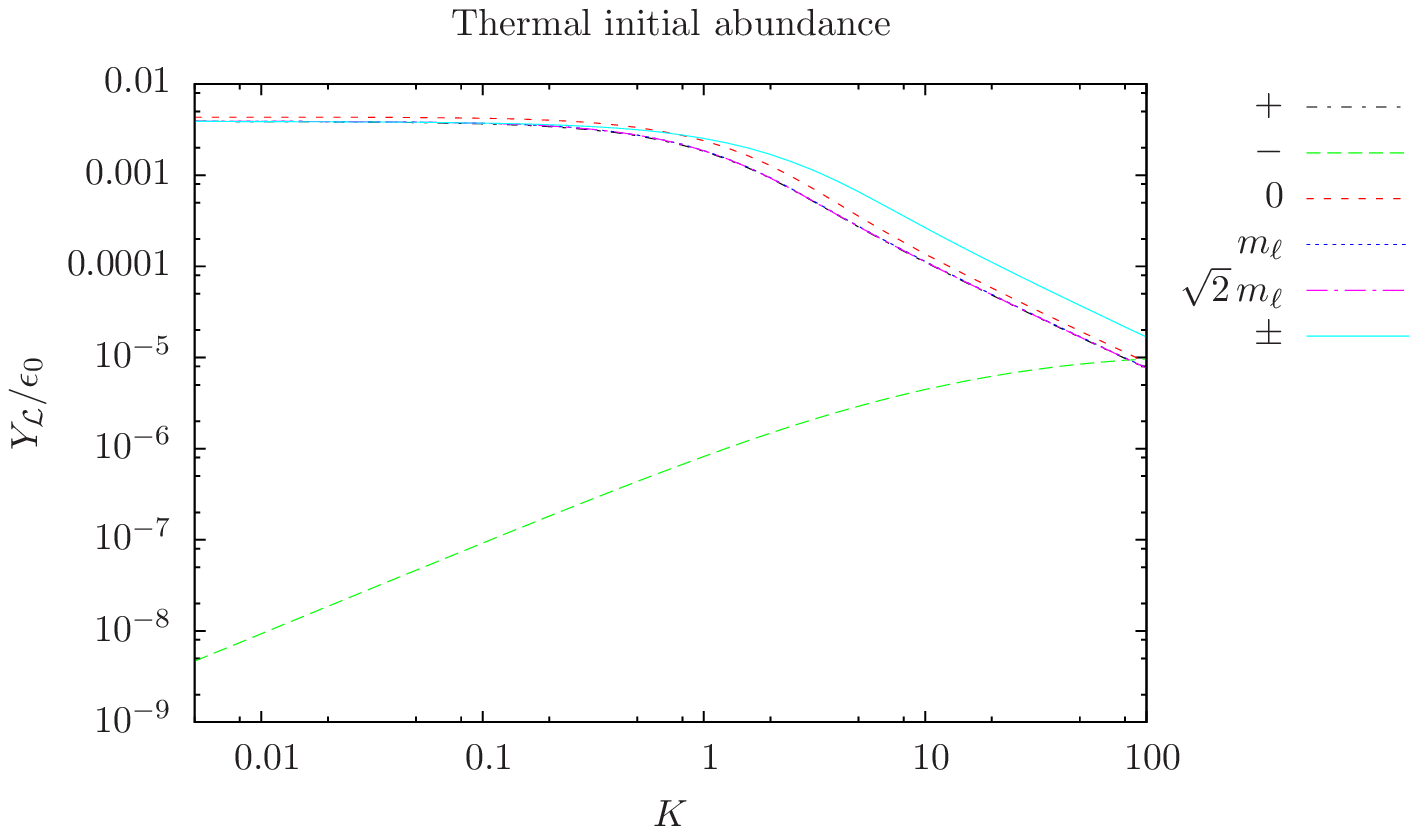}
  \caption[$K$-dependence of $|Y_\ml^{\rm fin}|$ for thermal initial
  neutrino abundance]{Final value of the lepton asymmetry for
    different values of $K$ for thermal initial neutrino abundance.}
  \label{fig:final1}
\end{figure}

The values of the final asymmetries are shown in
figures~\ref{fig:final0}--\ref{fig:final2} for different initial
abundances.
For zero initial abundance and weak washout, shown in
figure~\ref{fig:final0}, the asymmetries for the finite-temperature
cases are larger than for the vacuum case by about one order of
magnitude due to the difference of the thermal rates $\g_D$ and the
$\CP$-asymmetries $\D \g$ at $z \gtrsim z_+^N$. The lepton asymmetry
for the plus-mode is also slightly larger than for the one-mode cases
due to a suppression of the $\CP$ asymmetry compared to the one-mode
approaches. For strong washout, the asymmetry production in the vacuum
case is marginally more efficient than in the thermal cases due to a
smaller lepton equilibrium distribution, while the lepton asymmetry in
the ($\pm$)-approach is by a factor two larger than in the other cases
since half of the asymmetry is stored in the $\ell_-$-modes and not
affected by washout. The minus-mode case is completely decoupled, the
lepton asymmetry bears the opposite sign as the other lepton
asymmetries and rises with stronger couplings, that is with larger
decay parameter $K$. As discussed in
section~\ref{sec:interacting-modes}, this scenario might not be
realistic since the modes will couple to each other via gauge bosons,
so an evolution similar to the ($\pm$)-case seems more likely.
{\begin{figure}
  \centering
  \includegraphics[width=0.95\textwidth]{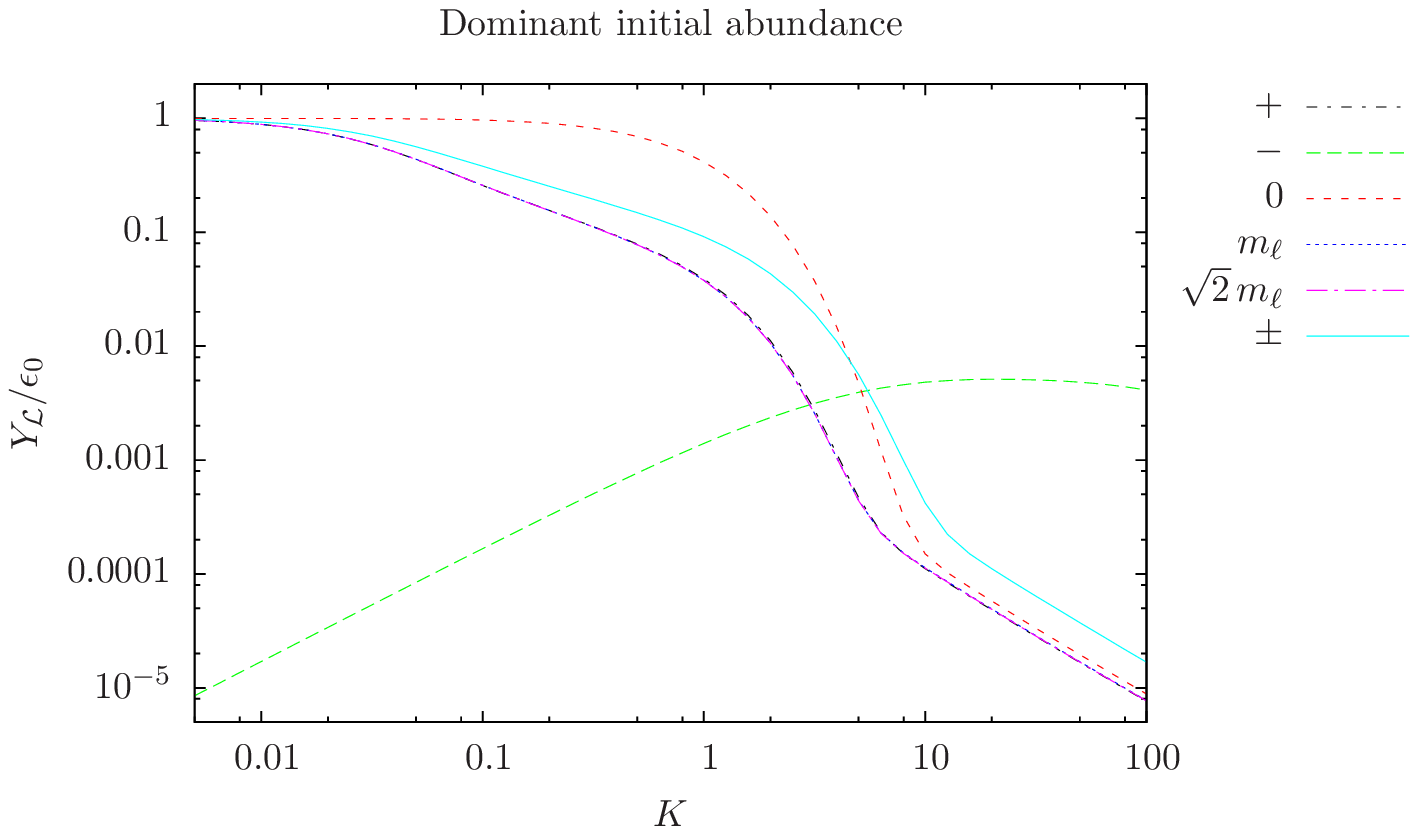}
  \caption[$K$-dependence of $|Y_\ml^{\rm fin}|$ for dominant initial
  neutrino abundance]{Final value of the lepton asymmetry for different values of
    $K$ for thermal initial neutrino abundance.}
  \label{fig:final2}
\end{figure}
\begin{figure}
  \centering
  \includegraphics[width=0.95\textwidth]{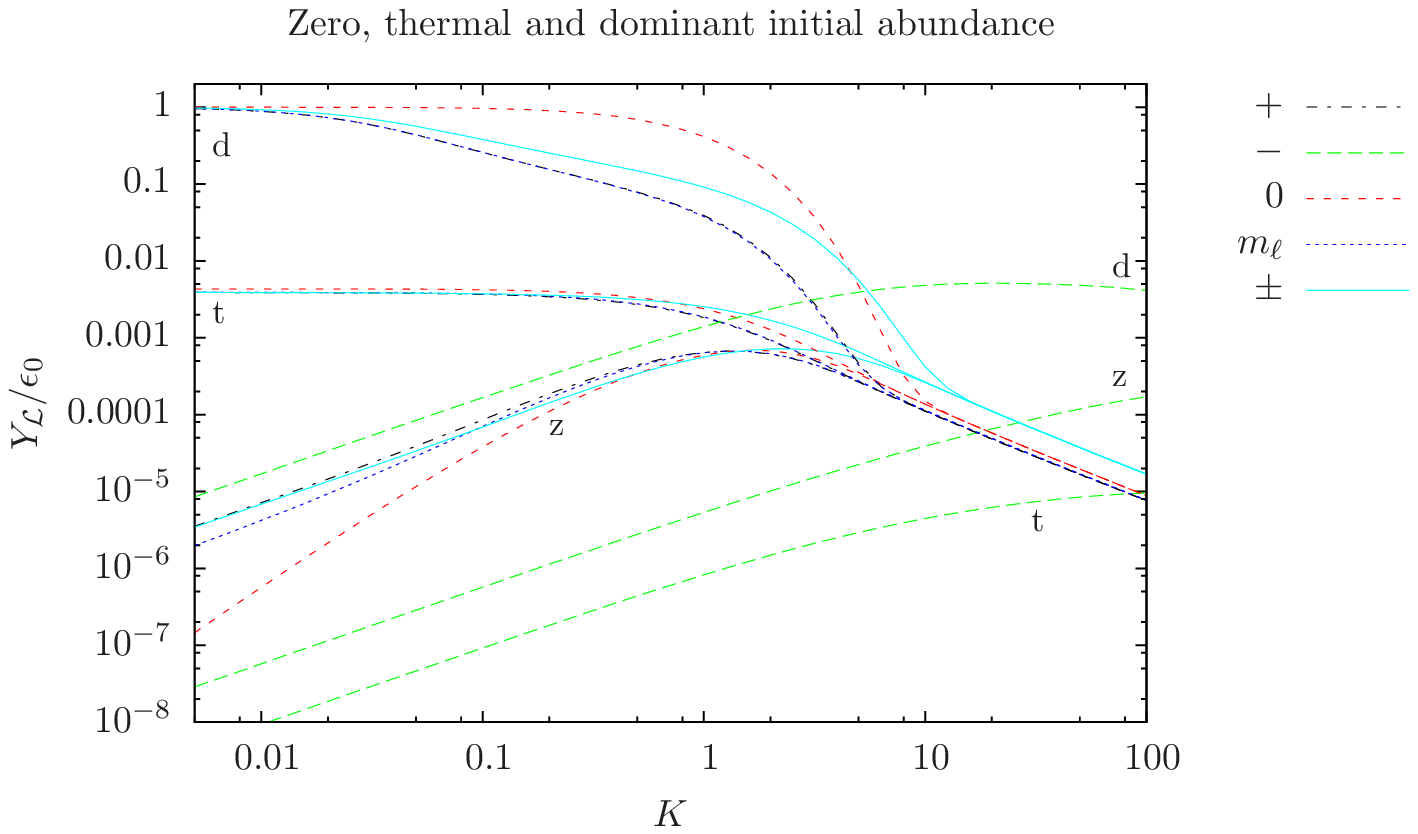}
  \caption[$K$-dependence of $|Y_\ml^{\rm fin}|$ for different initial
  neutrino abundances]{Final value of the lepton asymmetry for
    different values of $K$ for zero, thermal and dominant initial
    neutrino abundance. The letters z, t and d denote the curves for
    zero, thermal and dominant abundance. Note that the final
    asymmetry of the minus-mode has opposite sign for zero initial
    neutrino abundance, compared to the asymmetries of all other
    cases.}
  \label{fig:finaltot}
\end{figure}}

For thermal initial abundance and weak washout, shown in
figure~\ref{fig:final1}, the final asymmetry equals the equilibrium
abundance $Y_\ml/\e_0$, while in the strong washout regime, it shows
the same behaviour as in the case of zero initial neutrino
abundance. The minus-mode asymmetry is very low for thermal initial
neutrino abundance since the neutrinos are close to equilibrium at
high temperatures. Contrary to the zero initial abundance case, it
bears the same sign as the lepton asymmetries of the other scenarios.

For dominant initial abundance, shown in figure~\ref{fig:final2}, the
final lepton asymmetries assume their maximal value in the weak washout
regime, when the coupling is weak enough not to wash them out at low
temperature. For larger couplings $K \sim 1$, the thresholds lead to a
halted asymmetry production for the finite temperature cases and not
as much asymmetry can be produced as for the vacuum case. The
($\pm$)-case shows a larger asymmetry compared to the other thermal
cases due to the weaker washout. At strong coupling $K \gg 1$, the
asymmetries are the same as for thermal and zero initial neutrino
abundance.
%the dependence on the initial conditions vanishes. 
The minus-mode asymmetry is large in all washout regimes, since the
neutrinos are far from equilibrium at high temperatures when the
$\ell_-$-asymmetry is produced.

A summary of the several initial conditions can be seen in
figure~\ref{fig:finaltot}, where we have omitted the $\sqrt{2} \,
m_\ell$ case since it is very close to the $m_\ell$ case in all
scenarios.  In the weak washout regime, the case with zero initial
abundance is most strongly affected by thermal corrections which
amount to one order of magnitude, and the plus-mode asymmetry is
additionally enhanced by a factor of about two. In the intermediate
regime, the dominant-initial-abundance case is influenced very much by
thermal corrections. Therefore, a production mechanism for dominant
neutrino abundance has to take into account such thermal effects. In
the strong washout regime, one would naturally expect that thermal
corrections can be neglected. We see that this is not the case since
for strongly interacting leptonic quasiparticles, a part of the lepton
asymmetry can be hidden in the $\ell_-$-mode which is unaffected by
washout, thus producing an asymmetry by up to a factor of two larger
than at zero temperature. The effect of the thermal lepton mass on the
equilibrium distribution of the leptons is an interesting feature, but
very small and can be neglected for all practical purposes.

\section{Conclusions}
\label{sec:conclusions}

For a minimal and self-consistent toy model of leptogenesis, which
consists only of neutrinos, leptons and Higgs bosons, we have
performed an extensive analysis of the effects of HTL
corrections. This implies capturing the effects of thermal masses,
modified dispersion relations and modified helicity structures. We put
special emphasis on the influence of the two fermionic quasiparticles,
which show a different behaviour than particles in vacuum, notably
through their dispersion relations, but also the helicity structure of
their interactions.  Our work is thus similar to the work done in
reference~\cite{Giudice:2003jh}, where the authors of the latter work
did not include the effects of fermionic quasiparticles and get a
different result for the $\CP$-asymmetries, which are crucial for the
evolution of the lepton asymmetry. Our toy model produces two lepton
asymmetries stored in the two different lepton modes without the
possibility of an equilibration of these asymmetries by SM
processes. Since we expect the lepton modes to interact via gauge
bosons in the bath, we examine a second case where the modes are
strongly coupled to each other. As a third and fourth case, we
approximate the lepton propagators by zero temperature propagators
with the zero temperature mass replaced by the thermal lepton mass in
one case and the asymptotic mass in the other case. We refer to these
cases as one-mode approach. All four thermal cases are compared to the
zero-temperature case.

We have calculated interaction rates and $C\!P$-asymmetries in
references~\cite{Kiessig:2010pr},~\cite{Kiessig:2010zz}
and~\cite{Kiessig:2011fw}, where a detailed analysis can be found. We
present the rates and $C\!P$-asymmetries shortly in
section~\ref{sec:hard-thermal-loop}. Neglecting the zero-temperature
fermion mass, the resummation of HTL fermion self-energies results in
an effective fermion propagator that does not break chiral invariance
and is split up into two helicity modes. The external fermion states
therefore behave conceptually different from the ones with
chirality-breaking thermal masses that have been inserted in the
kinematics by hand. Moreover, one has to take care of one additional
mode, which has implications for the Boltzmann equations.

We derive and evaluate the Boltzmann equations in
section~\ref{sec:boltzmann-equations}, performing the crucial
subtraction of on-shell intermediate states in
appendix~\ref{sec:subtr-shell-prop}\footnote{Reference~\cite{HahnWoernle:2009qn}
  uses a thermal factor $(1-f_N)$ without explicitly deriving this
  factor. We show in this appendix that we have to use the equilibrium
  distribution for the neutrinos in $(1-f_N^\rmeq)$ instead.}. We
compare the results of the Boltzmann equations for our five cases,
that is, decoupled lepton modes, strongly coupled lepton modes, the
one-mode approach with $m_\ell$, the one-mode approach with $\sqrt{2}
\, m_\ell$, and the vacuum case. We assume three different initial
values for the abundance of neutrinos: zero, thermal and dominant
abundance, motivated by different scenarios for the production of
heavy neutrinos after inflation~\cite{Giudice:2003jh
%,Antusch:2010mv
}. In the weak
washout regime, we find that using thermal masses enhances the final
lepton asymmetry by about one order of magnitude for zero initial
neutrino abundance. This is due to the fact that the $\CP$-asymmetry
and the decay rate evolve differently at $z \gtrsim 1$ when using
thermal masses, since the $\CP$-asymmetry suffers from an additional
suppression by thermal masses through the leptons and Higgs bosons in
the loop. Due to the helicity structure of the modes, the
$\CP$-asymmetry of the plus-mode is additionally suppressed, which
results in an additional enhancement of the plus-mode lepton asymmetry
compared to the final asymmetries of the one-mode approaches. The
enhancement we find is similar to the one found in
reference~\cite{Giudice:2003jh} in this regime, but hard to compare
quantitatively due to their different approach, which includes
scatterings, and the discrepancy in the $\CP$-asymmetry.

In the strong washout regime, thermal masses do not show an influence,
as expected\footnote{There is a slight suppression of the lepton
  asymmetry for thermal masses, since the thermal mass suppresses the
  equilibrium distribution of the leptons somewhat and thereby
  enhances the washout term.}. However, when we couple the plus- and
minus-mode strongly, we observe an enhancement of the lepton asymmetry
by a factor of about two, since we stored half of the asymmetry in a
mode that essentially does not interact with the neutrinos and is
therefore not affected by washout. For intermediate washout, that is
$K \sim 1$, we find that the lepton asymmetries with thermal masses
are about one magnitude lower than in the vacuum case when we assume
dominant initial neutrino abundance. This is due to the fact that the
lower $\CP$-asymmetry does not succeed in producing as much lepton
asymmetry at $z \gtrsim 1$ when using thermal masses.  

A decoupled minus-mode would show a behaviour completely different
from the other thermal cases and the vacuum case for all initial
values of the neutrino abundance. The lepton asymmetry in such a
decoupled mode is produced mainly at high temperature and only
slightly affected by the development at $z \gtrsim 1$, where it
decouples from the evolution of the other abundances. Therefore, the
washout parameter $K$, which determines the coupling strength and
thereby the asymmetry production at high temperatures, is crucial for
the final value of the lepton asymmetry in this mode, as is the
initial neutrino abundance.

Summarising, we argue that for an accurate description of medium
effects on leptogenesis, the influence of thermal quasiparticles,
notably the effects of the two fermionic modes, cannot be
neglected. Similar to reference~\cite{Giudice:2003jh}, our study shows
that thermal masses have a strong effect in the weak washout regime,
while the effect of fermionic modes has an additional influence on the
final lepton asymmetry in this regime. We also showed that notably in
the strong washout regime, the presence of a quasi-sterile lepton mode
that is not affected by washout can have a non-negligible effect on
the final lepton asymmetry. Future studies should clarify the dynamics
of the interaction between the two fermionic modes and determine
whether the evolution of the asymmetries in the two modes is closer to
the decoupled or the strongly coupled case.

Another important aspect that might be studied in future works is the
influence of the finite width of the fermionic
modes~\cite{Drewes:2010yy}, notably the minus-mode. Such effects could
be studied using formalism that takes into account non-equilibrium
quantum effects, such as Kadanoff-Baym
equations~\cite{Anisimov:2008dz, Garny:2009rv, Garny:2009qn,
  Anisimov:2010aq, Garny:2010nj, Beneke:2010wd, Garny:2010nz,
  Anisimov:2010dk, Beneke:2010dz, Drewes:2010pf, Drewes:2010zz}. In
the quest for a unified description of finite-temperature effects on
leptogenesis, it is important to include SM interactions in the
Kadanoff-Baym studies that are under way. To this end, quasiparticle
excitations of fermions and gauge-bosons should be taken into
account. Last but not least, the fermionic modes might have an
influence on other related dynamics in the early universe that involve
fermions, such as thermal production of axions, axinos or gravitinos,
which could be studied in future works.

\subsection*{Acknowledgements}

We would like to thank Mathias Garny, Georg Raffelt, Michael
A.~Schmidt and Markus Thoma for their support and comments in this
project. Thanks also to Denis Besak, Dietrich B\"odeker, Wilfried
Buchm\"uller, Valerie Domcke, Marco Drewes, Andreas Hohenegger,
Alexander Kartavtsev and Christoph Weniger for fruitful and inspiring
discussions.

\appendix

\section{Particle Kinematics}
\label{sec:particle-kinematics}

The Boltzmann equations describe the time evolution of the
distribution function of a particle species $\psi$. We assume an
isotropic and spatially homogeneous universe described by the
Friedmann-Lemaitre-Robertson-Walker (FLRW) metric~\cite{Kolb:EarlyU},
\begin{equation}
  \label{eq:b1}
 \rmd s^2 = \rmd t^2 - a(t)^2 \left \{ \frac{\rmd r^2}{1-k r^2} + r^2
  \rmd \theta^2 + r^2 \sin^2 \theta \rmd \phi^2 \right \} ,
\end{equation}
where $a(t)$ is the cosmic scale factor, which describes the expansion
of the universe, $k=\pm1,0$ specifies the curvature, and
$(t,r,\theta,\phi)$ are the comoving coordinates.

The trajectory of a particle $\psi$ with mass $m_{\psi}\geq0$ moving
in a gravitational field is given by the geodesic equations of
motion~\cite{Weinberg:Gravitation}:
%~\footnote{For simplicity we
%  neglect the index $\psi$ at the momentum and space--time
 % coordinates.}:
\begin{align}
  \label{eq:eqm1}
  \frac{\rmd p_{\psi}^{\mu}}{\rmd \tau}+\Gamma^{\mu}_{\nu\alpha}
\,p_{\psi}^{\nu}\,p_{\psi}^{\alpha}&=0, \\
  \label{eq:eqm2}
  \frac{\rmd x_{\psi}^{\mu}}{\rmd \tau}&=p_{\psi}^{\mu}.
\end{align}
Since $s=m_{\psi}\tau$ is the eigen-time of the particle, $\tau$ is
fixed and $p^{\mu}$ is the momentum of a particle $\psi$.

In the FLRW metric the $\mu=0$ component of
Eq~(\ref{eq:eqm1}) is given as
\begin{align}
  \frac{\rmd p_{\psi}^{0}}{\rmd \tau}+ \frac{\dot{a}}{a}
  \mathbf{p}_{\psi}^{2}=0,\ \qquad \textrm{with} \quad \dot{a}=
  \frac{ \partial a}{ \partial t}.
\end{align}
Writing $p_{\psi}^{0}\,\rmd p_{\psi}^{0}=\vert\mathbf{p}_{\psi}\vert
\,\rmd \vert\mathbf{p}_{\psi}\vert$, this leads to:

\begin{align}
  \label{eq:scaling}
  \hphantom{\Leftrightarrow}&\vert\dot{\mathbf{p}}_{\psi}\vert
  a+\dot{a}\vert\mathbf{p}_{\psi}\vert=0
  \nonumber\\
  \Leftrightarrow&\frac{\rmd}{\rmd t}(\vert\mathbf{p}_{\psi}\vert a)=0
  \nonumber \\
  \Leftrightarrow &\vert\mathbf{p}_{\psi}\vert=
  {\rm const.}\times\frac{1}{a}.
\end{align}
Therefore the 3-momentum scales as $1/a$.

In general, the Liouville operator describing the evolution of a point
particle's phase space in a gravitational field is given by
\begin{align}
  \label{eq:liouville}
  L = p^{\alpha} \frac{\partial}{\partial x^{\alpha}} -
  \Gamma{^\alpha_{\beta \gamma}}p^{\beta}p^{\gamma}
  \frac{\partial}{\partial p^{\alpha}}.
\end{align}
With this operator the equations of motion~(\ref{eq:eqm1})
and~(\ref{eq:eqm2}) can be written for the momentum as
\begin{align}
  \label{eq:eqm-liouville1}
  \frac{\rmd p^{\mu}}{\rmd \tau} = L \left[p^{\mu} \right],
\end{align}
and for the space-time as
\begin{align}
  \label{eq:eqm-liouville2}
  \frac{\rmd x^{\mu}}{\rmd \tau} = L \left[x^{\mu} \right].
\end{align}
Furthermore, the time derivative of the phase space
distribution of a non-interacting gas vanishes,
\begin{align}
  \label{eq:derivative-phase}
  \frac{\rmd f(x,p)}{\rmd \tau} = 0.
\end{align}
Using the equations of motion for the particle we obtain the
Boltzmann equations for the non-interacting particle species $\psi$,
\begin{align}
  \label{eq:derivative-liouville}
  L \left[f_{\psi}(x,p) \right] =0.
\end{align}
Since we assume a Robertson--Walker universe which is isotropic and
homogeneous, the distribution function $f_{\psi}$ depends only on $t$
and $\vert \mathbf{p}_{\psi} \vert$. Therefore, the Boltzmann equation
can be written as~\cite{Kolb:EarlyU}
\begin{align}
  \label{eq:boltzmann-without}
  L\left[f_{\psi} \right] = E_{\psi} \frac{\partial f_{\psi}}
  {\partial t} - H \vert \mathbf{p}_{\psi} \vert^{2} \frac{
    \partial f_{\psi}}{\partial E_{\psi}} = 0,
\end{align}
where we have omitted arguments for the sake of notational clarity.

Since $p_{\psi}^{2}=m_{\psi}^{2}$ and because of the spatial isotropy
of the Robertson--Walker--Metric, we have
\begin{align}
  \label{eq:E-p-relation}
 \vert \mathbf{p}_{\psi} \vert^{2}\,\frac{
   \partial f_{\psi}}{\partial E_{\psi}} =E_{\psi}\, \vert
 \mathbf{p}_{\psi} \vert\, \frac{\partial f_{\psi}}{\partial \vert
   \mathbf{p}_{\psi} \vert}.
\end{align}
After dividing by $E_{\psi}$, equation~(\ref{eq:boltzmann-without}) has the form
\begin{align}
  \label{eq:boltzmann-without-2}
  L'\left[f_{\psi} \right] = \frac{\partial f_{\psi}}{\partial t}-H\, \vert
  \mathbf{p}_{\psi} \vert \,\frac{\partial f_{\psi}}{\partial \vert
    \mathbf{p}_{\psi}\vert}.
 \end{align}
Interactions are introduced on the right-hand side 
by a collision term $C \left[f_{\psi} \right]$, which drives
the distribution function towards its equilibrium value.
The complete Boltzmann equation reads
\begin{align}
  \label{eq:boltzmann-with}
   L' \left[f_{\psi} \right] =  \frac{\partial f_{\psi}}
  {\partial t} - H\, \vert \mathbf{p}_{\psi} \vert\, \frac{
    \partial f_{\psi}}{\partial \vert \mathbf{p}_{\psi}\vert} = C
  \left[f_{\psi} \right].
\end{align}
Thus, the Boltzmann equation in a Robertson--Walker universe has the form
of a partial differential equation. However, in the radiation
dominated phase of the universe, in which leptogenesis takes place,
equation~(\ref{eq:boltzmann-with}) can be written as an ordinary
differential equation by transforming to the dimensionless coordinates
$z= m_{\psi}/T$ and $y_{\psi}=\vert \mathbf{p}_{\psi}\vert/T$. Using
the relation $\rmd T/\rmd t = - H T$, the differential operator
$\partial_t-\vert \mathbf{p}_{\psi} \vert H \partial_{\vert
  \mathbf{p}_i\vert}$ is written as $z H \partial_z$, and
consequently~\cite{Kawasaki:1992kg}
\begin{align}
  \label{eq:boltzmann-with-2}
  \frac{\partial f_{\psi}(z,y)}{\partial z} = \frac{z}{H(m_{\psi})}\,
  C_{D}\left[ f_{\psi}(z,y)\right]
\end{align}
with $H \left(m_{\psi}\right)= H \left|_{T=m_\psi} \right.$. 
%given in equation~(\ref{eq:exp-rate}). 
In this form, the Boltzmann equation can be solved numerically on a
grid for specific rescaled momenta $y$. For the right hand side, we
have to sum over the collision terms of all processes which involve
the particle $\psi$ and change the phase space distribution. The
collision term for a process $\psi + a + \cdots \leftrightarrow i + j
+ \cdots$ is given by\cite{Kolb:EarlyU}\footnote{We have chosen a
  normalisation different from Kolb and Turner, so
  $C_\textrm{here}=\frac{1}{2 E_\psi} C_{\rm KT}$}
\begin{align}
  \label{eq:b3}
g_\j  \, C[\psi + a + \cdots \leftrightarrow i + j + \cdots]& = - \frac{1}{2
    E_\psi} \int \prod_\alpha \rmd \tilde{p}_\alpha (2 \pi)^4
  \delta^4(p_\psi+p_a + \cdots - p_i - p_j -\cdots) \nonumber \\
& \times \left[ \left| \mathcal{M}
    (\psi + a + \cdots \rightarrow i + j + \cdots) \right|^2 f_\psi
  f_a \cdots (1\pm f_i)(1\pm f_j) \cdots \right. \nonumber \\
& - \left. \left| \mathcal{M}
 (i + j + \cdots \rightarrow \psi + a + \cdots) \right|^2 f_i
  f_j \cdots (1\pm f_\psi)(1\pm f_a) \cdots \right],
\end{align}
where $\alpha=(a, \cdots,i,j,\cdots)$,
\begin{equation}
  \label{eq:b4}
\rmd \tilde{p}_\alpha= \frac{\rmd^3 p_\alpha}{(2 \pi)^3 2 E_\alpha}.
\end{equation}
The terms $(1\pm f_i)$ hold for fermions ($-$) and bosons ($+$) and are
interpreted as Fermi-blocking ($-$) and Bose-enhancement ($+$). In
practice, we will only look at processes which involve three or four
particles, that is, decays, inverse decays and scatterings. We have
included the internal degrees of freedom, $g_\psi, g_a, \cdots, g_i, g_j,
\cdots$, in the matrix elements, therefore we need to put $g_\psi$ in
front of the collision term since it is not included in the
phase-space density $f_\psi$.

%We do not want
%to solve the Boltzmann equations for each mode, but rather make
%simplifying assumptions for the particle distributions and solve
%integrated equations. 
We integrate
equation~\eqref{eq:boltzmann-with-2} over the phase space of the
incoming particle with $g_\psi \int \rmd^3 p_\psi/(2 \pi)^3$ and arrive at
\begin{equation}
  \label{eq:b2}
  \frac{\rmd n_\psi}{\rmd z}= - \frac{z}{H(m_\psi)} \sum_{\rm processes} \left[ 
\gamma(\psi + a + \cdots \rightarrow i + j + \cdots)
- \gamma(i + j + \cdots \rightarrow \psi + a + \cdots)
\right],
\end{equation}
where
\begin{align}
  \label{eq:b5}
  \gamma(\psi + a + \cdots \rightarrow i + j + \cdots) &= - g_\psi \int
  \frac{\rmd^3 p_\psi}{(2 \pi)^3} C[\psi + a + \cdots \rightarrow i +
  j + \cdots] \nonumber \\
& =   \int \prod_\beta \rmd \tilde{p}_\beta (2 \pi)^4
  \delta^4(p_\psi+p_a + \cdots - p_i - p_j -\cdots) \nonumber \\
& \times \left| \mathcal{M}
    (\psi + a + \cdots \rightarrow i + j + \cdots) \right|^2 f_\psi
  f_a \cdots (1\pm f_i)(1\pm f_j) \cdots,
\end{align}
where we now integrate over $p_\psi$ as well, that is,
$\beta=(\psi,a,\cdots,i,j,\cdots)$. The analogous equation holds for
$\gamma (i + j + \cdots \rightarrow \psi + a + \cdots)$.

\section{Boltzmann Equations at Zero Temperature}
\label{sec:boltzm-equat-at}

We can derive the Boltzmann equations for the neutrino and lepton
evolution at zero temperature, approximating the phase-space densities
with Maxwell-Boltzmann distributions,
\begin{align}
  \label{eq:75}
  f_i(E_i) = \exp(- E_i \beta) \, ,
\end{align}
where energy conservation in scatterings and decays implies
\begin{align}
  \label{eq:79}
  f_N &= f_L f_H \,
\end{align}
and there are no Higgs decays at high temperature. For the neutrino
evolution, we get, analogous to equation~\eqref{eq:71},
\begin{align}
  \label{eq:80}
  \frac{\rmd Y_{N}}{\rmd z}= - \frac{z}{s H_1} (x_N-1) \gamma_0,
\end{align}
where 
\begin{align}
  \label{eq:81}
  \gamma_0 =  \int \rmd
  \tilde{p}_N \rmd \tilde{p}_{L} \rmd \tilde{p}_H (2 \pi)^4
  \delta^4(p_{N} -p_H-p_{L})  \left|\mathcal{M}_0\right|^2
  f_N^\rmeq.
\end{align}
The matrix element evaluated at zero temperature reads
\begin{align}
  \label{eq:82}
  \left| \mathcal{M}_0 \right|^2 = 4 \times 2 \, P_N \cdot P_L,
\end{align}
where the factor $4$ originates from summing over $\ell$ and
$\barell$, as well as over the doublets $(e^-,\phi^+)$ and $(\nu,
\phi^0)$.

We can express the decay rate $\gamma_0$ in terms of the total decay width
$\G_{\rm rf}^{\rm tot}$ in the rest-frame of the neutrino,
\begin{align}
  \label{eq:83}
  \gamma_0  =  g_N \int \frac{\rmd p_N^3}{(2 \pi)^3} \frac{M}{E_N} \G_{\rm rf}^{\rm tot} f_N^\rmeq,
\end{align}
where $g_N=2$ accounts for the internal degrees of freedom of the
neutrino, the two spins, and
\begin{align}
  \label{eq:84}
  \G_{\rm rf}^{\rm tot}(N \to H L)= \frac{(\lambda^\dagger \lambda)_{11} M_1}{4 \pi g_N}
\end{align}
describes the decay of a neutrino with a definite spin into $(\phi \ell)$ and $(\barphi \barell)$.

Evaluating equation~\eqref{eq:83}, we get
\begin{align}
  \label{eq:85}
  \gamma_0= g_N \frac{M^2}{2 \pi^2} T \, \G_{\rm rf}^{\rm tot} K_1(z),
\end{align}
where $z=M/T$ and $K_1(z)$ is a Bessel function of second kind.
For the equilibrium density of the neutrinos, we get
\begin{align}
  \label{eq:86}
  n_N^\rmeq = g_N \int \frac{d^3 p_N}{(2 \pi)^3} f_N^\rmeq = g_N
  \frac{M^2}{2 \pi^2} T K_2(z), 
\end{align}
so that
\begin{align}
  \label{eq:87}
  \frac{\gamma_0}{n_N^\rmeq}= \G_{\rm rf}^{\rm tot} \frac{K_1(z)}{K_2(z)}.
\end{align}
We can write the Boltzmann equation as
\begin{align}
  \label{eq:88}
  Y_N' = -D (Y_N-Y_N^\rmeq),
\end{align}
where
\begin{align}
  \label{eq:91}
  Y_X' \equiv \frac{\rmd Y_X}{\rmd z},
\end{align}
\begin{align}
  \label{eq:89}
  D= \frac{z}{H_1} \frac{\gamma_0}{n_N^\rmeq} = z K \frac{K_1(z)}{K_2(z)} \, ,
\end{align}
\begin{align}
  \label{eq:99}
  Y_N^\rmeq= \frac{45}{4 \pi^4} \frac{g_N}{g_*} z^2 K_2(z)
\end{align}
and
\begin{align}
  \label{eq:90}
  K = \frac{\G_{\rm rf}^{\rm tot}}{H_1}= \frac{\tilde{m}}{m^*}
\end{align}
is called decay parameter.

For the lepton evolution, the subtraction of on-shell propagators can
be performed analogously to the finite temperature case in
appendix~\ref{sec:subtr-shell-prop}, so that
\begin{align}
  \label{eq:92}
  \gamma^{\rm sub}(\ell \phi \to \barell \barphi) - \gamma^{\rm
    sub}(\barell \barphi \to \ell \phi) = \epsilon_0 \gamma_0,
\end{align}
where 
\begin{align}
  \label{eq:93}
  \epsilon_0 \equiv \frac{\Gamma(N \to \ell \phi)- \Gamma(N \to
    \barell \barphi)}{\Gamma(N \to \ell \phi) + \Gamma(N \to
    \barell \barphi)}
\end{align}
is the $C\!P$-asymmetry. 
Analogous to equation~\eqref{eq:74}, we get
\begin{align}
  \label{eq:94}
    \frac{\rmd Y_{\mathcal{L}}}{\rmd z} = - \frac{z}{s H_1} \left( -
    \e_0 \left( x_N-1 \right) +
    \frac{x_{\mathcal{L}}}{2} \right) \gamma_0,
\end{align}
which can be rewritten as
\begin{align}
  \label{eq:95}
  Y_{\mathcal{L}}'=\epsilon_0 D (Y_N-Y_N^\rmeq) -W Y_{\mathcal{L}},
\end{align}
where 
\begin{align}
  \label{eq:96}
  W \equiv \frac{z}{H_1} \frac{\gamma_0}{2 n_{\ell}^\rmeq}.
\end{align}
We have
\begin{align}
  \label{eq:97}
  n_\ell^\rmeq= g_\ell \int \frac{d^3 p_\ell}{(2 \pi)^3} f_\ell^\rmeq = g_\ell
  \frac{T^3}{\pi^2},
\end{align}
where $g_\ell=2$ accounts for the lepton doublet
components, so we get
\begin{align}
  \label{eq:98}
  W = \frac{1}{4} \frac{g_N}{g_\ell} z^3 K K_2(z).
\end{align}

\section{Subtraction of On-Shell Propagators}
\label{sec:subtr-shell-prop}

\subsection{Low Temperature}
\label{sec:low-temperature-2}

We verify the relation in equation~\eqref{eq:b31}. The scattering rate
$\gamma(\ell \phi \to \barell \barphi)$ can be split up into four
scatterings with different kinematics, corresponding to the four
possibilities of combining the in- and outgoing lepton modes. The
scattering rates read
\begin{align}
  \label{eq:b38}
  \g(\ell_{h_i} \phi \to \barell_{h_f} \barphi)= \int \rmd
  \tilde{p}_{\ell h_i} \rmd \tilde{p}_\phi \rmd
  \tilde{p}_{\barell h_f} \tilde{p}_{\barphi} & (2 \pi)^4 \delta^4
  (p_{\ell h_i} + p_\phi- p_{\barell h_f} -p_{\barphi})
  \nonumber \\
  & \times \left| \mathcal{M}(\ell_{h_i} \phi \to \barell_{h_f}
      \barphi) \right|^2 f_{\ell h_i} f_\phi (1- f_{\barell h_f}) (1+
    f_{\barphi})  ,
\end{align}
where $(h_i, h_f) = \pm 1$ denote the helicity-to-chirality ratio of
the initial- and final-state leptons (or antileptons). We will drop
the subscript for this appendix part, unless it is necessary, and all
equations are valid for one specific mode for each involved lepton,
unless otherwise noted. With this simplified notation, each of the
four matrix elements is evaluated as\\
% {\color{red} The matrix elements are already summed over spins by
%   definition. They should also be summed over lepton and Higgs degrees
%   of freedom, so a factor two in the initial state and a factor two in
%   the final state... Check this!}
\begin{align}
  \label{eq:b39}
  \sum_{s_\ell,s_{\barell}} \left| \mathcal{M}(\ell_{h_i} \phi \to \barell_{h_f}
    \barphi) \right|^2 = \left[(\l^\dagger \l)_{11} \right]^2
 \left| D_N
  \right|^2 2 \left[2 (p_N \cdot p_{\ell h_i}) (p_N \cdot p_{\barell h_f}) -
    (p_N \cdot p_N) (p_{\ell h_i} \cdot p_{\barell h_f}) \right],
\end{align}
where we sum over the lepton spins $s_\ell$ and $s_{\barell}$ and the
lepton flavours and $D_N=1/[P_N^2-M_N^2+\rmi \, p_N^0 \Gamma_N(p_N^0)]$
is the neutrino propagator in the narrow-width approximation and
$\Gamma_N(p_N^0)$ the total width of the neutrino, which equals the
total interaction rate, including both lepton modes. Putting the
propagator on its mass shell, $P_N^2=M_N^2$, we get
\begin{align}
  \label{eq:b40}
  \sum_{s_\ell,s_{\barell}} \left| \mathcal{M}^{\rm os}(\ell_{h_i}
    \phi \to \barell_{h_f} \barphi) \right|^2 = \left[(\l^\dagger
    \l)_{11} \right]^2 \left| D_N^{\rm os} \right|^2 2 \left[2 (p_N
    \cdot p_{\ell h_i}) (p_N \cdot p_{\barell h_f}) - M_N^2 (p_{\ell h_i} \cdot
    p_{\barell h_f}) \right],
\end{align}
where
\begin{align}
  \label{eq:b42}
  \left| D_N^{\rm os} \right|^2= \frac{ \pi
    \delta(P_N^2-M^2)}{p_N^0 \Gamma_N(p_N^0)}
\end{align}
In vacuum without thermal masses, this reads
\begin{align}
  \label{eq:b41}
   \sum_{s_\ell,s_{\barell}} \left| \mathcal{M}^{\rm os}(\ell \phi \to \barell
    \barphi) \right|^2 = \left[(\l^\dagger \l)_{11} \right]^2
  \left| D_N^{\rm os} \right|^2
 2 \left[
\frac{M_N^4}{4} (1+\eta)
\right],   
\end{align}
where the dependence on the angle $\eta$ between the external leptons
cancels out in the integration for symmetry reasons, so we can neglect
it and write
\begin{align}
  \label{eq:b43}
   \sum_{s_\ell,s_{\barell}} \left| \mathcal{M}^{\rm os}(\ell \phi \to \barell
    \barphi) \right|^2 =
   \sum_{s_\ell,s_{\barell}}
 \left| \mathcal{M}(\ell \phi \to N) \right|^2 
  \left| D_N^{\rm os} \right|^2
 \left| \mathcal{M}(N \to \barell \barphi) \right|^2. 
\end{align}
At finite temperature with quasiparticle dispersion relations, we can
not derive equation~\eqref{eq:b43} accurately, but in the narrow-width
approximation\cite{Giudice:2003jh}, one assumes that the influence of
the angle between the external particles is negligible and
equation~\eqref{eq:b43} holds.

Using the relations in equation~\eqref{eq:b27}, we derive
\begin{align}
  \label{eq:b45}
  &\left| \mathcal{M}^{\rm os}(\ell_{h_i} \phi_i \to \barell_{h_f}
    \barphi_f) \right|^2 f_{\ell h_i} f_{\phi,i} (1- f_{\barell h_f})
  (1+ f_{\barphi,f}) - \left| \mathcal{M}^{\rm os}(\barell_{h_i}
    \barphi_i \to \ell_{h_f} \phi_f) \right|^2
  f_{\barell h_i} f_{\barphi,i} (1- f_{\ell h_f}) (1+ f_{\phi,f}) \nonumber \\
  = &\left| D_N^{\rm os} \right|^2 \frac{1}{4} \left|
    \mathcal{M}_{h_i}^0 \right|^2 \left| \mathcal{M}_{h_f}^0 \right|^2
  \left[ f_{\mathcal{L}h_i} (1- f_{\ell h_f}^\rmeq) +f_{\ell
      h_i}^\rmeq f_{\mathcal{L} h_f}-4 \e^N_h f_{\ell h_i}^\rmeq (1-f_{\ell
      h_f}^\rmeq) \right] f_{\phi,i}^\rmeq (1-f_{\phi,f}^\rmeq),
\end{align}
where we have neglected terms of order $\e^2$ and $x^2_{\mathcal{L}}$
and added the subscripts $i$ and $f$ in the Higgs boson
distributions to clarify which momentum to use,
\begin{align}
  \label{eq:b70}
  f_{\phi,i}=f_{\phi}(\omega_{\phi,i})=f_{\phi}(\omega_N-\omega_{\ell h_i})
\end{align}
and likewise for $f_{\phi,f}$.

For the tree-level, $CP$-conserving amplitude, we have
\begin{align}
  \label{eq:b46}
  \left| \mathcal{M}^{\rm tree}(\ell_{h_i} \phi \to \barell_{h_f} \barphi)
  \right|^2
=   \left| \mathcal{M}^{\rm tree}(\barell_{h_i} \barphi \to \ell_{h_f} \phi)
  \right|^2 \equiv   \left| \mathcal{M}_{\D L=2} \right|^2_{h_i h_f}.
\end{align}
For the full amplitude $\left| \mathcal{M}_{\D L=2} \right|^2$, the
on-shell part is also dominant. Since it is $CP$-conserving, we write
\begin{align}
  \label{eq:b48}
  \left| \mathcal{M}_{\D L=2} \right|^2_{h_i h_f} \approx 
  \left| \mathcal{M}^{\rm os}_{\D L=2} \right|^2_{h_i h_f} = 
\left| D_N^{\rm os}
  \right|^2 \frac{1}{4} \left| \mathcal{M}_{h_i}^0 \right|^2
 \left| \mathcal{M}_{h_f}^0 \right|^2
\end{align}
and we get
\begin{align}
  \label{eq:b47}
  \left| \mathcal{M}^{\rm tree}(\ell_{h_i} \phi \right. 
& \left. 
 \to \barell_{h_f} \barphi)
  \right|^2 f_{\ell h_i} f_\phi (1- f_{\barell h_f}) (1+ f_{\barphi})
  \nonumber \\
&-  \left|
    \mathcal{M}^{\rm tree}(\barell_{h_i} \barphi \to \ell_{h_f} \phi) \right|^2
  f_{\barell h_i} f_{\barphi} (1- f_{\ell h_f}) (1+ f_{\phi}) \nonumber \\
%\nonumber \\
% & 
&  = \left| \mathcal{M}_{\D L=2} \right|^2_{h_i h_f} \left[
    f_{\mathcal{L}, h_i} (1 - f_{\ell h_f}^\rmeq) + f_{\ell h_i}^\rmeq
    f_{\mathcal{L} h_f} \right].
\end{align}
Subtracting equations~\eqref{eq:b45} and~\eqref{eq:b47}, we derive
\begin{align}
  \label{eq:b49}
  \gamma^{\rm sub} (\ell_{h_i}\phi\to \barell_{h_f} \barphi) -
  \gamma^{\rm sub}(\barell_{h_i} \barphi \to \ell_{h_f} \phi) = \int &
  \rmd \tilde{p}_{\ell h_i} \rmd \tilde{p}_\phi \rmd
  \tilde{p}_{\barell h_f} \tilde{p}_{\barphi} (2 \pi)^4 \delta^4
  (p_{\ell h_i} + p_\phi- p_{\barell h_f} -p_{\barphi})
  \nonumber \\
  & \times \e_h^N \left| D_N^{\rm os} \right|^2 \left|
    \mathcal{M}_{h_i}^0 \right|^2 \left| \mathcal{M}_{h_f}^0 \right|^2
  f^\rmeq_{\ell h_i} f^\rmeq_\phi (1- f^\rmeq_{\ell h_f}) (1+
  f^\rmeq_{\phi}) \nonumber \\
  \equiv & \e^N_h \gamma^{\rm os}_\rmeq(L_{h_i} H \to L_{h_f} H)
\end{align}
Using the relations
\begin{align}
  \label{eq:b71}
  (1-f_{\ell h}^\rmeq)(1+f_\phi^\rmeq) & = (1-f_N^\rmeq)
  (1-f_{\ell h}^\rmeq+f_\phi^\rmeq),
  \\
  f_{\ell h}^\rmeq f_\phi^\rmeq & = f_N^\rmeq
  (1-f_{\ell h}^\rmeq+f_\phi^\rmeq)
%\text{ and} 
  \\
  \text{and } f_\phi^\rmeq f_{\ell h}^\rmeq (1-f_N^\rmeq) & =
  (1+f_\phi^\rmeq) (1-f_{\ell h}^\rmeq) f_N^\rmeq,
\end{align}
which hold for $\omega_N=\omega_{\ell h}+\omega_\phi$, it is
straightforward to derive
\begin{align}
  \label{eq:b72}
    \gamma^{\rm sub} (\ell_{h_i}\phi\to \barell_{h_f} \barphi) - \gamma^{\rm
    sub}(\barell_{h_i} \barphi \to \ell_{h_f} \phi) 
&=   \gamma^{\rm sub} (\ell_{h_f}\phi\to \barell_{h_i} \barphi) - \gamma^{\rm
    sub}(\barell_{h_f} \barphi \to \ell_{h_i} \phi), \nonumber \\
\gamma^{\rm os}_\rmeq(L_{h_i} H \to L_{h_f} H)
& = \gamma^{\rm os}_\rmeq(L_{h_f} H \to L_{h_i} H)
\end{align}

Inserting $1=\int \rmd^4 p_N/(2 \pi)^4 \delta^4(p_N-p_{\ell h_i}
-p_\phi)$ into equation~\eqref{eq:b72}, again using the first relation
from equations~\eqref{eq:b49} and the expression for the total neutrino
width,
\begin{align}
  \label{eq:b51}
  \Gamma_N(p_N^0)=\frac{1}{2 p_N^0} \sum_{h_f=\pm 1} \int \rmd \tilde{p}_{L h_f}
  \tilde{p}_H (2 \pi)^4 \delta^4 (p_N-p_{L h_f}-p_H)
  \left| \mathcal{M}_{h_f}^0 \right|^2 (1-f^\rmeq_{L h_f} +f^\rmeq_\phi),
\end{align}
we arrive at equation~\eqref{eq:b31},
\begin{align}
  \sum_{h_f} \left[ \gamma^{\rm sub} (\ell_{h_i}\phi \right.
&
\left. \to \barell_{h_f}
    \barphi) - \gamma^{\rm sub}(\barell_{h_i} \barphi \to \ell_{h_f} \phi)
  \right] \nonumber \\
& =   \sum_{h_f} \left[ \gamma^{\rm sub} (\ell_{h_f}\phi\to \barell_{h_i}
    \barphi) - \gamma^{\rm sub}(\barell_{h_f} \barphi \to \ell_{h_i} \phi)
  \right] \nonumber \\
& = \int \rmd \tilde{p}_N \rmd \tilde{p}_{\ell h_i} \rmd
  \tilde{p}_\phi (2 \pi)^4 \delta^4(p_N-p_{\ell h_i}-p_\phi)
  \e_h^N \left| \mathcal{M}_{h_i}^0 \right|^2 f_{\ell h_i}^\rmeq f_\phi^\rmeq
  (1-f_N^\rmeq) \nonumber \\
& \equiv \e_h^N \gamma_\rmeq(L_{h_i} H \to N).
\end{align}

\subsection{High Temperature}
\label{sec:high-temperature-2}

For the $u$-channel resonance at high temperature when Higgs bosons
decay into neutrinos and leptons while the neutrinos are stable, we
can derive a relation similar to equation~\eqref{eq:b31}. The width in
the on-shell neutrino propagator is then not the decay rate but an
interaction rate which accounts for the processes where the neutrino
interacts with the medium, that is, $H \to NL$ and $NL \to H$. This
width acts as a regulator of the $u$-channel resonance.

In the narrow-width approximation, the on-shell amplitude reads
\begin{align}
  \label{eq:55}
  \sum_{s_{\ell},s_{\barell}} \left| \mathcal{M}^{\rm os}(\ell_{h_i}
    \phi \to \barell_{h_f} \barphi) \right|^2 = \sum_{s_\ell,s_{\barell}}
  \left| \mathcal{M}(\phi \to N \barell_{h_f}) \right|^2 \left| D_N^{\rm os}
  \right|^2 \left| \mathcal{M}(N \ell_{h_i}\to \barphi) \right|^2,
\end{align}
where the on-shell propagator is the same as in
equation~\eqref{eq:b42}, but the width $\Gamma_N$ is given by the kinematically
allowed processes, $H \to NL$ and $NL \to H$. 

Using the relations in equation~\eqref{eq:b66}, we derive
\begin{align}
  \label{eq:56}
    &\left| \mathcal{M}^{\rm os}(\ell_{h_i} \phi_i \to \barell_{h_f}
    \barphi_f) \right|^2 f_{\ell h_i} f_{\phi,i} (1- f_{\barell h_f})
  (1+ f_{\barphi,f}) - \left| \mathcal{M}^{\rm os}(\barell_{h_i}
    \barphi_i \to \ell_{h_f} \phi_f) \right|^2
  f_{\barell h_i} f_{\barphi,i} (1- f_{\ell h_f}) (1+ f_{\phi,f}) \nonumber \\
  = &\left| D_N^{\rm os} \right|^2 \frac{1}{4} \left|
    \mathcal{M}_{h_i}^0 \right|^2 \left| \mathcal{M}_{h_f}^0 \right|^2
  \left[ f_{\mathcal{L}h_i} (1- f_{\ell h_f}^\rmeq) +f_{\ell
      h_i}^\rmeq f_{\mathcal{L} h_f}+4 \e_h^\phi f_{\ell h_i}^\rmeq (1-f_{\ell
      h_f}^\rmeq) \right] f_{\phi,i}^\rmeq (1-f_{\phi,f}^\rmeq),
\end{align}
Analogous to equation~\eqref{eq:b47}, we derive
\begin{align}
  \label{eq:57}
    \left| \mathcal{M}^{\rm tree}(\ell_{h_i} \phi \right. 
& \left. 
 \to \barell_{h_f} \barphi)
  \right|^2 f_{\ell h_i} f_\phi (1- f_{\barell h_f}) (1+ f_{\barphi})
  \nonumber \\
&-  \left|
    \mathcal{M}^{\rm tree}(\barell_{h_i} \barphi \to \ell_{h_f} \phi) \right|^2
  f_{\barell h_i} f_{\barphi} (1- f_{\ell h_f}) (1+ f_{\phi}) \nonumber \\
%\nonumber \\
% & 
&  = \left| \mathcal{M}_{\D L=2} \right|^2_{h_i h_f} \left[
    f_{\mathcal{L}, h_i} (1 - f_{\ell h_f}^\rmeq) + f_{\ell h_i}^\rmeq
    f_{\mathcal{L} h_f} \right],
\end{align}
so that
\begin{align}
  \label{eq:58}
    \gamma^{\rm sub} (\ell_{h_i}\phi\to \barell_{h_f} \barphi) - \gamma^{\rm
    sub}(\barell_{h_i} \barphi \to \ell_{h_f} \phi) =
- \e_h^\phi \gamma^{\rm os}_\rmeq(L_{h_i} H \to L_{h_f} H) \, .
\end{align}

Using the relations
\begin{align}
\label{eq:59}
  (1-f_{\ell h}^\rmeq)f_\phi^\rmeq & = f_N^\rmeq
  (f_{\ell h}^\rmeq+f_\phi^\rmeq),
  \\
  f_{\ell h}^\rmeq (1+f_\phi^\rmeq) & = (1-f_N^\rmeq)
  (f_{\ell h}^\rmeq+f_\phi^\rmeq)
%\text{ and} 
  \\
  \text{and } f_\phi^\rmeq (1-f_{\ell h}^\rmeq) (1-f_n^\rmeq) & =
  (1+f_\phi^\rmeq) f_{\ell h}^\rmeq f_N^\rmeq,
\end{align}
which hold for $\omega_\phi=\omega_{\ell h}+\omega_N$, it is
straightforward to derive
\begin{align}
\label{eq:61}
    \gamma^{\rm sub} (\ell_{h_i}\phi\to \barell_{h_f} \barphi) - \gamma^{\rm
    sub}(\barell_{h_i} \barphi \to \ell_{h_f} \phi) 
&=   \gamma^{\rm sub} (\ell_{h_f}\phi\to \barell_{h_i} \barphi) - \gamma^{\rm
    sub}(\barell_{h_f} \barphi \to \ell_{h_i} \phi), \nonumber \\
\gamma^{\rm os}_\rmeq(L_{h_i} H \to L_{h_f} H)
& = \gamma^{\rm os}_\rmeq(L_{h_f} H \to L_{h_i} H).
\end{align}

Inserting $1=\int \rmd^4 p_N/(2 \pi)^4 \delta^4(p_\phi-p_{\ell h_i}
-p_N)$ into equation~\eqref{eq:61}, again using the first relation
from equations~\eqref{eq:59} and the expression for the total neutrino
width at high temperature, 
\begin{align}
\label{eq:60}
  \Gamma_N(p_N^0)=\frac{1}{2 p_N^0} \sum_{h_f=\pm 1} \int \rmd \tilde{p}_{L h_f}
  \tilde{p}_H (2 \pi)^4 \delta^4 (p_H-p_{L h_f}-p_N)
  \left| \mathcal{M}_{h_f}^0 \right|^2 (f^\rmeq_{L h_f} +f^\rmeq_H),
\end{align}
we arrive at equation~\eqref{eq:b67} ,
\begin{align}
  \sum_{h_f} \left[ \gamma^{\rm sub} (\ell_{h_i}\phi \right.
&
\left. \to \barell_{h_f}
    \barphi) - \gamma^{\rm sub}(\barell_{h_i} \barphi \to \ell_{h_f} \phi)
  \right] \nonumber \\
& =   \sum_{h_f} \left[ \gamma^{\rm sub} (\ell_{h_f}\phi\to \barell_{h_i}
    \barphi) - \gamma^{\rm sub}(\barell_{h_f} \barphi \to \ell_{h_i} \phi)
  \right] \nonumber \\
& = - \int \rmd \tilde{p}_N \rmd \tilde{p}_{\ell h_i} \rmd
  \tilde{p}_\phi (2 \pi)^4 \delta^4(p_N-p_{\ell h_i}-p_\phi)
  \e_h^\phi \left| \mathcal{M}_{h_i}^0 \right|^2 f_{\ell h_i}^\rmeq (1+f_\phi^\rmeq)
  f_N^\rmeq \, .
\end{align}

\bibliographystyle{JHEPMS}
\bibliography{../dissertation/main}
%\bibliography{/Users/beamer/dissertation/main}
%\bibliography{main}
%\bibliography{boltz}

\end{document}